\pdfoutput=1

\providecommand{\ifmain}[1]{#1}

\documentclass[10pt]{amsart}
\newcommand{\tpath}{.}

\usepackage{
  amssymb,
  latexsym, 
  graphpap, 
  graphics, 
  verbatim, 
  harvard, 
  layout, 
  amsbsy, 
  calc, 
  pstricks, 
  pst-plot, 
  pst-tree,
  pst-pdf, 
  pstricks-add, 
  accents,
  }
\usepackage[bottom]{footmisc} 
\usepackage[inline]{enumitem}

\newcommand{\ifdraft}[1]{} 
\newcommand{\iffinal}[1]{#1}

\newcommand{\eput}{ \ifmain{\showbib \end{document}} \endinput}

\newtheorem{lemma}{\indent Lemma}
\newtheorem{nthm}[lemma]{\indent Theorem}
\newtheorem{ncrly}[lemma]{\indent Corollary}
\newtheorem{prop}[lemma]{\indent Proposition}
\theoremstyle{definition}
\newtheorem{pfa}[lemma]{\indent Proof}
\newenvironment{npf}
  { \begin{pfa} }
  { \hfill $\Box$ \end{pfa} }
\theoremstyle{remark}
\newtheorem*{pfb}{\indent Proof}
\newenvironment{pf} 
  { \begin{pfb} }
  { \hfill $\Box$ \end{pfb}  }

\newcommand{\tlistindent}{0mm} %
\newcounter{tlistno} %
\newenvironment{tlist} %
  { \begin{list} %
      {\indent(\alph{tlistno}) } %
      { \setlength{\leftmargin}{0mm} %
        \setlength{\rightmargin}{0mm} %
        \setlength{\itemindent}{\tlistindent} 
        \setlength{\labelwidth}{0mm} %
        \setlength{\labelsep}{0mm} %
        \setlength{\itemsep}{0mm} %
        \setlength{\topsep}{0mm} %
        \usecounter{tlistno} } } %
  { \end{list} }

\newcommand{\yl}[1]{\item\label{#1}}

\allowdisplaybreaks[1] 
\newcommand{\zz}{}
\newcommand{\nt}{\notag \\ }
\newcommand{\miz}{ \left. \begin{matrix} }
\newcommand{\moz}{ \end{matrix} \right. }
\newcommand{\mpi}{ \begin{minipage} }
\newcommand{\mpo}{ \end{minipage} }

\newcommand{\nichts}[1]{}
\newcommand{\showbib}{ \bibliographystyle{econometrica} 
  \bibliography{\tpath/bibtex/streuf,\tpath/bibtex/others,References}}  
\newcommand{\markb}[1]{\markboth{#1}{#1}}
\providecommand{\ifssec}[1]{}

\newcommand{\ssec}[1]{\vspace{0.9mm}\subsection{}{\scshape #1}\vspace{1.6mm}} 
  
\newcommand{\myskip}{\vspace{3mm}}

\newcommand{\lstep}[1]{\vspace{1mm} {\em #1}}
\newcommand{\rf}[1]{\ref{#1}}

\allowdisplaybreaks

\newcommand{\nnexists}{\raisebox{0.15ex}{/}\hspace{-1.15ex}\exists}

\usepackage[utf8]{inputenc}
\usepackage{amssymb}
\usepackage{newunicodechar}
\newcommand{\nuc}[2]{\newunicodechar{#1}{#2}} 

\nuc{α}{{ \alpha}} 
\nuc{β}{{ \beta}}
\nuc{γ}{{ \gamma}} \nuc{Γ}{{ \varGamma}}
\nuc{δ}{{ \delta}} \nuc{Δ}{{ \varDelta}} 
\nuc{ε}{{ \varepsilon}}
\nuc{ζ}{{ \zeta}}
\nuc{η}{{ \eta}}
\nuc{θ}{{ \theta}} \nuc{Θ}{{ \varTheta}}
\nuc{ι}{{ \iota}}
\nuc{κ}{{ \kappa}}
\nuc{λ}{{ \lambda}} \nuc{Λ}{{ \varLambda}}
\nuc{μ}{{ \mu}}
\nuc{ν}{{ \nu}}
\nuc{ξ}{{ \xi}} \nuc{Ξ}{{ \varXi}}
\nuc{π}{{ \pi}} \nuc{Π}{{ \varPi}}
\nuc{ρ}{{ \rho}} 
\nuc{σ}{{ \sigma}} \nuc{Σ}{{ \varSigma}}
\nuc{τ}{{ \tau}}
\nuc{υ}{{ \upsilon}} \nuc{Υ}{{ \varUpsilon}}
\nuc{φ}{{ \phi}} \nuc{Φ}{{ \varPhi}}
\nuc{χ}{{ \chi}} 
\nuc{ψ}{{ \psi}} \nuc{Ψ}{{ \varPsi}}
\nuc{ω}{{ \omega}} \nuc{Ω}{{ \varOmega}}

\nuc{≠}{{ \ne}}
\nuc{∾}{{ \sim}} 
\nuc{∈}{{ \in}} \nuc{∉}{{ \notin}}
\nuc{∋}{{ \ni}} \nuc{∌}{{ \not\ni}}   
\nuc{⊊}{{ \subset}} \nuc{⊆}{{ \subseteq}} 
\nuc{⊋}{{ \supset}} \nuc{⊇}{{ \supseteq}} 
\nuc{≤}{{ \leq}}
\nuc{≥}{{ \geq}}
\nuc{≺}{{ \prec}} \nuc{≼}{{ \preccurlyeq}}
\nuc{≻}{{ \succ}} \nuc{≽}{{ \succcurlyeq}}

\nuc{∑}{{ \Sigma}} 
\nuc{±}{{ \pm}}
\nuc{×}{{ \times}} \nuc{∏}{{ \Pi}}
\nuc{÷}{{ \div}} 
\nuc{○}{{ \circ}}
\nuc{∩}{{ \cap}} \nuc{⋂}{{ \textstyle\bigcap\nolimits}}
\nuc{∪}{{ \cup}} \nuc{⋃}{{ \textstyle \bigcup\nolimits}}
\nuc{⧷}{{ \smallsetminus}} 
\nuc{±}{{ \oplus}} 
\nuc{⊗}{{ \otimes}}
\nuc{∗}{{ \star}} 
\nuc{·}{{ \cdot}}

\nuc{→}{{ \rightarrow}}  
\nuc{⇉}{{ \rightrightarrows}} 
\nuc{←}{{ \leftarrow}}
\nuc{⟷}{{ \leftrightarrow}}
\nuc{⇒}{{ \Rightarrow}}  
\nuc{⇐}{{ \Leftarrow}} 
\nuc{⟺}{{ \Leftrightarrow}}
\nuc{⇡}{{ \uparrow}}
\nuc{⇣}{{ \downarrow}}

\nuc{ℓ}{{ \ell}}
\nuc{∞}{{ \infty}}
\nuc{‴}{^{ \prime\prime\prime}} 
\nuc{″}{^{ \prime\prime}} 
\nuc{′}{^{ \prime}}
\nuc{¬}{{ \neg}}
\nuc{∀}{{ \forall}}
\nuc{∃}{{ \exists}}
\nuc{∄}{{ \nnexists}} 
\nuc{∅}{{ \varnothing}}

\nuc{⎨}{{ \lbrace}}
\nuc{⎬}{{ \rbrace}}
\nuc{Ṛ}{{ \mathbb R}}
\nuc{Ẓ}{{ \mathbb Z}}
\nuc{⋅}{{ \ }} 

\newcommand{\capp}{}

\newcounter{cllistno} %
\newenvironment{cllist} %
  { \begin{list} %
      {\emph{Claim \arabic{cllistno}:} } 
      { \setlength{\leftmargin}{0mm} %
        \setlength{\rightmargin}{0mm} %
        \setlength{\itemindent}{\parindent} %
        \setlength{\labelwidth}{0mm} %
        \setlength{\labelsep}{0mm} %
        \setlength{\itemsep}{1.2mm}%{\parskip} % 
        \setlength{\topsep}{1.2mm}% 
        \usecounter{cllistno} } } %
  { \end{list} }

\newcommand{\qedup}{\vspace{-5.2mm}}
\newcommand{\unskipcl}{\qedup} 

\newcommand{\ili}{\begin{enumerate*}[label=$\lbrack$\alph*$\rbrack$]}
\newcommand{\ilr}{\begin{enumerate*}[resume*]}
\newcommand{\ilo}{\end{enumerate*}}
\newcommand{\ilitem}[1]{\item\label{#1}\ilo\hspace{-1mm}\ifdraft{\hspace{10mm}}}
\newcommand{\ilc}[1]{\ili\ilitem{#1}}
\newcommand{\il}[1]{\ilr\ilitem{#1}}

\newcommand{\lii}{\begin{enumerate*}[label=$\lbrack$\arabic*$\rbrack$]}
\newcommand{\lir}{\begin{enumerate*}[resume*]}
\newcommand{\lio}{\end{enumerate*}}
\newcommand{\liitem}[1]{\item\label{#1}\lio\hspace{-1mm}\ifdraft{\hspace{10mm}}}
\newcommand{\lic}[1]{\lii\liitem{#1}}
\newcommand{\li}[1]{\lir\liitem{#1}}

\newcommand{\ttc}[2]{\begin{enumerate*}[label={#1}]\item\label{#2}
  \hspace{-1mm}\end{enumerate*}\ifdraft{\hspace{8mm}}}
\newcommand{\ttt}[2]{\tag*{#1}\label{#2}}

\newcommand{\mpic}{\begin{minipage}{4cm}\begin{center}}
\newcommand{\mpoc}{\end{center}\end{minipage}}

\newenvironment{maneq}{%
  \setlength{\arraycolsep}{.1ex}
  
  \begin{array}{lc} \rule{5ex}{0ex} & \rule{79ex}{0ex} \\[-3ex] }{%
  \end{array}}
\newcommand{\mqi}{\begin{maneq}}
\newcommand{\mqo}{\end{maneq}}

  \newcommand{\ct}[1]{\ensuremath{\mathbf{#1}}}
  
  \newcommand{\ud}[1]{{\text{\d{$#1$}}}}
  \newcommand{\ie}[1]{\ensuremath{\dot{#1}}}

  \newcommand{\xin}[1]{#1}

  \newcommand{\mapsfrom}{\text{\reflectbox{$\mapsto$}}}

  \newcommand{\notenclosedby}{\ensuremath{{\not\mspace{-5mu}\ud{→}}}}
  \newcommand{\sencloses}{\ensuremath{\ud{←}\mspace{0mu}{\notenclosedby}}}

  \newcommand{\py}[1]{\ensuremath{\mathit{P}\mathit{#1}}}
  \newcommand{\dra}{\rotatebox[origin=c]{180}{$\Lsh$}} 
  \newcommand{\sj}{|^{\mathsf{sj}}}
  \renewcommand{\max}{\text{\normalfont max}\,} 
  \renewcommand{\min}{\text{\normalfont min}\,}

  \newcommand{\sh}{^{\sharp}}
  \newcommand{\fl}{^{\flat}}
  \newcommand{\ps}{^{\prime\sharp}}
  \newcommand{\po}{^{\prime o}}
  \newcommand{\T}{^{\mathsf{T}\!}}
  \newcommand{\pA}{^{\prime A}}
  \newcommand{\pB}{^{\prime B}}
  \newcommand{\pgr}{^{\prime\,\mathsf{gr}}}

  \newcommand{\dF}{\bar{F}}
  \newcommand{\dT}{\bar{T}} 
  \newcommand{\dU}{\bar{U}}
  \newcommand{\dv}{\bar{v}}
  \newcommand{\dd}[1]{\bar{#1}} 
  \newcommand{\dtau}{\bar{\tau}}

  \newcommand{\rt}{\dot{t}}
  \newcommand{\rV}{\dot{V}}
  \newcommand{\rZ}{\dot{Z}}

  \newcommand{\ga}{{\tilde{a}}}

  \newcommand{\zC}{{\check{C}}}
  \newcommand{\zc}{\check{c}}
  \newcommand{\zI}{{\check{I}}}
  \newcommand{\zi}{{\check{i}}}
  \newcommand{\zT}{\check{T}}
  \newcommand{\zt}{\check{t}}
  \newcommand{\zU}{\check{U}}
  
  \newcommand{\ztau}{\check{\tau}}
  \newcommand{\zZZ}{\check{\ZZ}}

  \newcommand{\HH}{{ \mathcal H}}
  \newcommand{\PP}{{ \mathcal P}} 
  \newcommand{\Ss}{{ \mathcal S}}
  \newcommand{\WW}{{ \mathcal W}}
  \newcommand{\WWt}{\WW_{T⧷⎨t^o⎬}}  
  \newcommand{\WWf}{\WW_{T⧷X}} 
   
  \newcommand{\YY}{{ \mathcal Y}}
  \newcommand{\ZZ}{{ \mathcal Z}}
  \newcommand{\ZZf}{ \mathcal Z_{\mathsf{ft}}}
  \newcommand{\ZZi}{\mathcal Z_{\mathsf{inft}}}

  \newcommand{\tNa}{{{ \mathbb N}_1}}
  \newcommand{\tNz}{{{ \mathbb N}_0}}

  \newcommand{\vb}{\boldsymbol{\beta}}

  \newcommand{\src}{\mathsf{src}}
  \newcommand{\trg}{\mathsf{trg}}
  \newcommand{\id}{\mathsf{id}}
  \newcommand{\gr}{^{\mspace{1mu}\mathsf{gr}}}
  \newcommand{\inc}{\mathsf{inc}}

  \newcommand{\fa}{{\mathsf{a}}}
  \newcommand{\fb}{{\mathsf{b}}}
  \newcommand{\fc}{{\mathsf{c}}}
  \newcommand{\fd}{{\mathsf{d}}}
  \newcommand{\fe}{{\mathsf{e}}}
  \newcommand{\ff}{{\mathsf{f}}}
  \newcommand{\fg}{{\mathsf{g}}}
  \newcommand{\f}[1]{\mathsf{#1}} 

  \newcommand{\FB}{\mathsf{F}}
  \newcommand{\FO}{\mathsf{F_0}}
  \newcommand{\FA}{\mathsf{F_1}}

  \newcommand{\PB}{\mathsf{P}}
  
  \newcommand{\PA}{\mathsf{P_1}}
  \newcommand{\TB}{\mathsf{T}}
  \newcommand{\Ef}{E_{\mathsf{ft}}}

\providecommand{\forpreamble}{} \forpreamble
\hypersetup{colorlinks=true,linkcolor=black}
\begin{document}

\numberwithin{lemma}{section}
\numberwithin{figure}{section}
\numberwithin{table}{section}

\title{The Category of \\ Node-and-Choice Extensive-Form Games}
\date{\today. This version makes only expositional changes to arXiv 2004.11196v1. {\em Keywords:} isomorphic inclusion, isomorphic equivalence, choice-sequence, choice-set. {\em Classifications:} JEL C73, MSC 91A70. {\em Contact information:} pstreuf@uwo.ca, 519-661-2111x85384, Economics Department, University of Western Ontario, London, Ontario, N6A 5C2, Canada.}  
\thanks{} 

\maketitle 

\iffinal{\vspace{-5mm}}
\begin{centering}
Peter A. Streufert \\[-.5mm]
Economics Department \\[-.5mm]
Western University \\
\end{centering} 
\vspace{2mm}

\begin{abstract} This paper develops the category \ct{NCG}.  Its
  objects are node-and-choice games, which include essentially all
  extensive-form games.  Its morphisms allow arbitrary transformations
  of a game's nodes, choices, and players, as well as monotonic
  transformations of the utility functions of the game's players.
  Among the morphisms are subgame inclusions.  Several
  characterizations and numerous properties of the isomorphisms are
  derived.  For example, it is shown that the isomorphisms preserve the game-theoretic concepts of no-absentmindedness, perfect-information, and (pure-strategy) Nash-equilibrium.  Finally, full subcategories are defined for choice-sequence games and choice-set games, and relationships among these two subcategories and \ct{NCG} itself are expressed and derived via isomorphic inclusions and equivalences.
\end{abstract}

\section{Introduction}\label{9274}\markb{\sc \rf{9274}. Introduction}

\ssec{A foundational question}\label{9270}

Extensive-form games can be specified in many different styles, and some of these styles are reviewed in Section~\rf{9265} below.  Game theorists informally understand that any fundamental concept in one style should have the same meaning in any other style, and that any fundamental result in one style should also hold in any another style.  

This informal understanding might be formally developed.  In particular, when a fundamental concept or result is translated from one style to another, how would we define the sense in which the translation itself was correct or incorrect?  A good answer to this broad question promises to let us more efficiently identify and manipulate fundamental game-theoretic concepts and results.

This paper is the third in a series of papers which aim to answer this and related questions by means of category theory.  The first two papers are Streufert 2018 (henceforth ``SP'') and Streufert 2020 (henceforth ``SF'').\nocite{ncp-1810-as-SP}\nocite{ncf-2001-as-SF}  The present paper's role is discussed in Section~\rf{9266} below. 

\ssec{Category Theory and Game Theory}\label{9735}

Before proceeding, note that both category theory and game theory are to be used, and that readers with this combination of interests are presently rare.  So the series tries to help category theorists learning game theory, and game theorists learning category theory.

To those learning game theory, the series offers a self-contained formal introduction to extensive-form games.  In particular, this paper's Section 2 summarizes the game theory in the previous two papers, and notes \rf{9721}, \rf{9722}, \rf{9723}, \rf{9724}, and \rf{9725}, with Figures \rf{9718}, \rf{9719}, and \rf{9720}, work through an introductory example.  Game-theory textbooks offer more discussion.  Mas-Colell, Whinston, and Green 1995, Section 7.C aligns relatively well with the formulation here.\nocite{MWG95}  Other starting places are Fudenberg and Tirole 1991, Section 3.3;\nocite{FuTi91b} Myerson 1991, Section 2.1;\nocite{Myers91} Osborne and Rubinstein 1994, Section 11.1;\nocite{OsRu94} and Shoham and Leyton-Brown 2009, Section 5.2. 

To those learning category theory, this paper requires only the basic concepts of category, isomorphism, functor, and full subcategory.  These fundamental definitions can be found in the early pages of Simmons 2011,\nocite{Simm11} Awodey 2010,\nocite{Awod10} and Mac Lane 1998 (arranged in decreasing order of accessibility).  Further, the paper tries to avoid the notational shortcuts used by expert category theorists.  As a result, the notation is unusually explicit.  This is discussed further in Section 2.1 below.

\ssec{Specification styles}\label{9265}

Figure~\rf{9259} depicts two games, both resembling Selten 1975, Figure 1.  The two games have the same underlying ``form'', that is, the same configuration of nodes, choices, and information sets, and the same assignment of information sets to players.  Briefly, in both games, nodes are numbers, choices are letters, the information set $⎨\f0⎬$ is assigned to player $\mathit{P1}$, the information set $⎨\f1⎬$ is assigned to player $\mathit{P2}$, and the information set $⎨\f3,\f4⎬$ is assigned to player $\mathit{P3}$.

At the same time, the two games differ in how they assign utilities.  For example, consider the the play $⎨\f0,\f3,\f5⎬$, that is, the play through nodes $\f0$, $\f3$, and $\f5$.  In game~(a), player $\mathit{P1}$ gets utility 1 from this play (shown as the first entry in the vector beneath node $\f5$).  Meanwhile in game (b), player $\mathit{P1}$ gets utility 3 from this play.  Nonetheless, in the context of this paper, the utilities in the two games have the same meaning in the sense that they have the same ordinal content.\footnote{\label{9268}The utilities in the two games would not have the same meaning if mixed strategies were allowed and the specified utilities were used to construct expected utilities.  Expected utilities attach more meaning to the specified utilities, and this additional meaning can be embodied by an \ct{NCG} subcategory that admits only affine utility transformations.  The construction of this subcategory is left for future research.} For example, in both games, player $\mathit{P1}$ most desires play $⎨\f0,\f3,\f5⎬$, least desires play $⎨\f0,\f1,\f4,\f8⎬$, and regards the remaining three plays as equally desirable.\footnote{\label{9263}To tell a story matching these two games, suppose a student (called $\mathit{P1}$) must decide between the acceptable choice of doing her homework (called $\fa$) and the bad choice of not doing her homework (called $\fb$).  Knowing that the homework has been finished (node $\f1$), a goat ($\mathit{P2}$) must decide between the good choice of taking a nap ($\fg$) and the dumb choice of eating the homework ($\fd$).  Finally, without knowing whether the student played bad (node $\f3$) or the student played acceptable and the goat played dumb (node $\f4$), the teacher ($\mathit{P3}$) must choose between excusing the student ($\fe$) and failing the student ($\ff$).  The student most prefers being excused without doing the homework (play $⎨\f0,\f3,\f5⎬$), and least prefers failing after doing the homework (play $⎨\f0,\f1,\f4,\f8⎬$).  The goat likes eating homework (going through node $\f4$).  The teacher does not want to excuse a badly behaving student (play $⎨\f0,\f3,\f5⎬$) or to fail an acceptably behaving student (play $⎨\f0,\f1,\f4,\f8⎬$).}

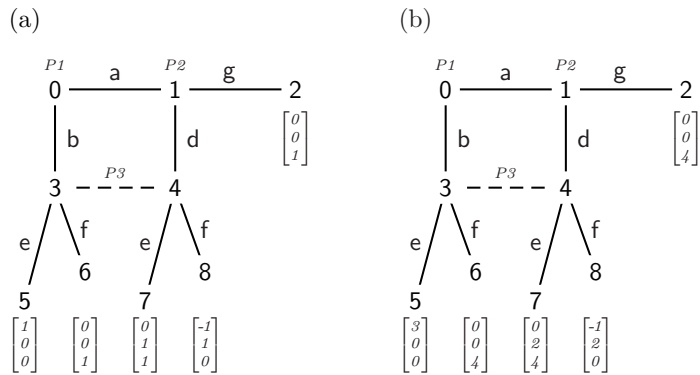
\renewcommand{\capp}{(a) is a node-and-choice game (later called an ``\ct{NCG} game'').  (b) is another.  In both games, player $\mathit{P3}$ selects choice $\fe$ or choice $\ff$ without knowing whether she is at node $\f3$ or node $\f4$.  In the context of this paper, the games' utilities have the same meaning.$^{\rf{9268}}$}
\begin{figure}[t]
  \newcommand{\hgth}{135}  
  \begin{picture}(0,\hgth)
  \put(-163,-8){\scalebox{.96}{ 
    \begin{pspicture}(4,2)(17,-6)  
      \end{pspicture}
    }} \end{picture}
  \caption{\small \capp} \label{9259} 
  \end{figure} 

In addition to the different ways of assigning utilities, there are many styles in which to specify the underlying form of a game.\footnote{\label{9269}The next four paragraphs draw heavily from SF (Streufert 2020) Section 1.2.}  All form styles must specify [a] nodes, which are variously called ``histories'', ``vertices'', or ``states'', and [b] choices, which are variously called ``actions'', ``alternatives'', ``labels'', or ``programs''.  The following paragraphs arrange these form styles into five broad groups according to how the form styles specify nodes and choices.  Representatives of these five groups are arranged in a spectrum by Streufert 2019, Figure~2, and that paper's Section 7 explains how each has its own advantages and disadvantages.

\newcommand{\notelts}{\footnote{Some labeled transition systems and process graphs have recursive transitions.  These do not support extensive-form games because extensive-form games require trees.  (Similarly, all stochastic games (Mertens 2002)\nocite{Merte02} have recursive transitions, and these too are not extensive-form games.)}}

\vspace{.8mm}{\em Group 1}. Some form styles specify nodes and choices abstractly without restriction.  Classic examples from economics include the style of Kuhn 1953\nocite{Kuhn5397} and the style of Selten 1975\nocite{Selt75}.  Examples from computer science and/or logic include the ``labeled transition system'' style in Blackburn, de Rijke, and Venema 2001, page 3 (and elsewhere);\nocite{BlackRV01} the style of Shoham and Leyton-Brown 2009, page 125;\nocite{ShohLeyt09} and the ``epistemic process graph'' style of van Benthem 2014, page 70.{\notelts}  A final example is the ``node-and-choice'' style of this paper (which has already appeared in Figure~\rf{9259}).  Because each of these form styles specifies nodes and choices abstractly without restriction, each can be understood to encompass essentially all other form styles as special cases.\footnote{\label{8282}Accordingly, this paper's node-and-choice games include essentially all extensive-form games.  Several aspects of this claim should be clarified.  [a] A node-and-choice game is discrete in the sense that each node has a finite number of predecessors.  This assumption excludes non-discrete extensive-form games such as those of Dockner, J{\o}rgensen, Long, and Sorger 2000, \nocite{DocknJLS00} and Al\'os-Ferrer and Ritzberger 2016.  [b] A node-and-choice game assumes that information sets do not share alternatives.  This assumption is insubstantial in the sense of Streufert 2020, note 19.  [c] A node-and-choice game assumes that exactly one player moves at each information set.  Accordingly, simultaneous moves by several players are specified by several information sets, as in Osborne and Rubinstein, 1994, page 202.  [d] A node-and-choice game specifies payoffs by utility functions.  Alternatively, payoffs could be specified by preference relations.  A corresponding category is left for future research.}

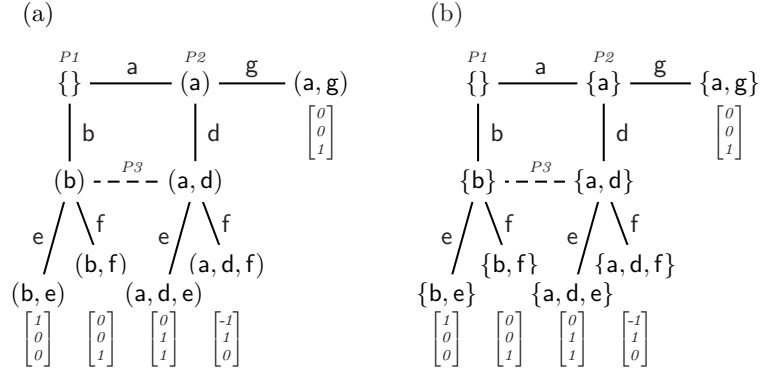
\renewcommand{\capp}{(a) A choice-sequence game (later called a ``\ct{CsqG} game'').  (b) A choice-set game (later called a ``\ct{CsetG} game'').  This paper develops subcategories for these special kinds of node-and-choice games.}
\begin{figure}[t]
  \newcommand{\hgth}{135}  
  \begin{picture}(0,\hgth) 
  \put(-172,-8){\scalebox{.96}{ 
    \begin{pspicture}(4,2)(17,-6)
      \end{pspicture}
    }} \end{picture}
  \caption{\small \capp} \label{9261} 
  \end{figure}  

\vspace{.8mm}{\em Group 2}. Other form styles specify nodes as sequences of choices.  Examples from economics include the style of Harris 1985\nocite{Harris85} and the style of Osborne and Rubinstein 1994, page 200.\nocite{OsRu94}  Examples from logic include the ``logical game'' style of Hodges 2013, Section 2,\nocite{Hodge13} and the ``epistemic forest model'' style of van Benthem 2014, page 130.\nocite{Benth14}  Examples from computer science include the ``protocol'' style of Parikh and Ramanujam 1985,\nocite{PariRama85} the ``history-based multi-agent structure'' style of Pacuit 2007,\nocite{Pacu07} and the ``sequence-form representation'' style of Shoham and Leyton-Brown 2009, page 129.  A final example is the ``choice-sequence'' style of this paper (see Figure~\rf{9261}(a)).  

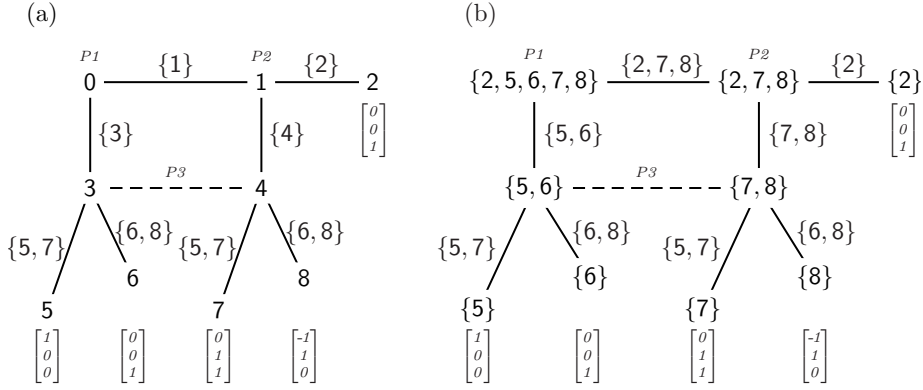
\renewcommand{\capp}{In (a), choices are node sets.  In (b), both nodes and choices are outcome sets.  This paper does not develop subcategories for these special kinds of node-and-choice games.}
\begin{figure}[t]
  \newcommand{\hgth}{140}  
  \begin{picture}(0,\hgth) 
  \put(-178,-8){\scalebox{.94}{ 
    \begin{pspicture}(0,2)(17,-6.5)
      \end{pspicture}
    }} \end{picture}
  \caption{\small \capp} \label{9262} 
\end{figure} 

\vspace{.8mm}{\em Group 3}. Some new form styles specify nodes as sets of choices.  These are the ``choice-set'' style of Streufert 2019, and the closely related ``choice-set'' style of this paper (see Figure~\rf{9261}(b)).\nocite{five-1903}

\vspace{.8mm}Finally, two more groups remain.  While subcategories corresponding to groups 2 and 3 are developed in this paper, subcategories corresponding to groups 4 and 5 are left for future research (see note~\rf{9439}).  {\em Group 4}. Some form styles specify choices as sets of nodes, as in the ``simple'' style of Al\'{o}s-Ferrer and Ritzberger 2016, Section~6.3 (see Figure~\rf{9262}(a)).\nocite{AlRi16}  {\em Group 5}. Other form styles express both nodes and choices as sets of outcomes, as in the style of von Neumann and Morgenstern 1944, Section~10,\nocite{vNMo53} and the style of Al\'{o}s-Ferrer and Ritzberger 2016, Section~6.2 (see Figure~\rf{9262}(b)).

\ssec{Summary and Motivation}\label{9266}

This section discusses five topics, and lastly, related literature.  The paper's main contributions within each topic are summarized and motivated. 

\vspace{.8mm}{\em Translating Games}. Intuitively, the six games in Section~\rf{9265} should be equivalent.  Formally, Section~\rf{9272} develops an appealing category in which these six games are isomorphic.  The category is called ``\ct{NCG}'', for the ``category of node-and-choice games''.  Each of its objects is a game consisting of [A] a form that specifies nodes, choices, and players, and [B] a utility function for each player that assigns a utility number to each of the form's plays.  Each of the category's morphisms consists of [A] a form morphism that transforms nodes, choices, and players, and [B] a monotonic transformation of each player's utility function.  

The two part [A]'s in the two previous sentences show that games and game morphisms are built on forms and form morphisms.  Relatedly, Theorem~\rf{8647} constructs a ``forgetful'' functor from \ct{NCG} to \ct{NCF}, which is the category of node-and-choice forms in SF (Streufert 2020).  SF, in turn, constructed a ``forgetful'' functor from \ct{NCF} to \ct{NCP}, which is the category of node-and-choice preforms in SP (Streufert 2018).  In this fashion, the results in the predecessor papers are made formally accessible.  In addition, Theorem~\rf{8248} and Corollary~\rf{8636} provide four characterizations of \ct{NCG} isomorphisms.  Finally, Proposition~\rf{9334} and its preceding paragraph derive more than 25 fundamental properties of \ct{NCG} isomorphisms. 

\myskip{\em Translating Styles}.  Section~\rf{9087} defines two \ct{NCG} subcategories for the two styles of Figure~\rf{9261}.  In particular, \ct{CsqG} is the subcategory for choice-sequence games like that of Figure~\rf{9261}(a), and \ct{CsetG} is the subcategory for choice-set games like that of Figure~\rf{9261}(b).

Theorem~\rf{8270} compares \ct{NCG} and \ct{CsqG} via the concept of isomorphic inclusion (i.e.\ isomorphic enclosure) from SF (Streufert 2020).  In particular, the theorem shows that the entire category \ct{NCG} is ``isomorphically included'' in \ct{CsqG} (symbolically \ct{NCG} \ie{⊆} \ct{CsqG}) in the sense that each \ct{NCG} game is (\ct{NCG}) isomorphic to a \ct{CsqG} game.  Thus, since the converse (obviously) holds, Corollary~\rf{8271} states that \ct{NCG} and \ct{CsqG} are ``isomorphically equivalent'' (symbolically \ct{NCG} \ie{\simeq} \ct{CsqG}).   Similar results are then derived for \ct{CsetG}.  In particular,  Corollary~\rf{8310} shows that \ct{NCG_\ga} \ie{\simeq} \ct{CsetG}, where \ct{NCG_\ga} is the subcategory for (general style) games with no-absentmindedness.\footnote{\label{9439}These isomorphic equivalences unify and extend earlier ad hoc equivalences by Kline and Luckraz 2016\nocite{KlinLuck16} and Streufert 2019.  Similar categorical results promise to extend the ad hoc equivalences in Al\'{o}s-Ferrer and Ritzberger 2016, Section 6.3.  This future research will concern the two games in Figure~\rf{9262}.}  

\vspace{.8mm}{\em Translating Concepts}.  Intuitively, fundamental game-theoretic concepts like Nash-equilibrium, no-absentmindedness, and perfect-information should mean the same thing in two equivalent games.  Formally, Proposition \rf{8261}, Corollary~\rf{8268}, and Corollary~\rf{8269} show how these three concepts are preserved by \ct{NCG} isomorphisms.  This starts to address Section~\rf{9270}'s broad question: Games and styles can be ``translated'' by isomorphisms, and from this perspective, the ``fundamental'' (i.e.\ ``translatable'') concepts are the ones that are isomorphically preserved.

Further, Proposition~\rf{9109} shows that, conveniently, any isomorphic inclusion or equivalence can be restricted by any invariant property.  Thus, for example, the aforementioned Corollary~\rf{8271} implies that \ct{NCG_\ga} \ie{\simeq} \ct{CsqG_\ga}, where \ct{CsqG_\ga} is the subcategory for choice-sequence games with no-absentmindedness.  So, naturally, the aforementioned Corollary~\rf{8310} implies that \ct{NCG_\ga}, \ct{CsqG_\ga}, and \ct{CsetG} are all isomorphically equivalent.  This convenience and naturalness suggest that Section~\rf{9270}'s broad question is being addressed in a useful way.

\vspace{.8mm}{\em Translating Results}. In accord with Section~\rf{9270}, the above suggests a larger agenda, namely, the systematic translation of results across different styles.  Such an overarching translation system promises conceptual benefits.  Foremost in the author's mind is the formal synthesis of results and questions from the many disciplines and subdisciplines which are each studying some version of game theory.  There seems to be much to gain because there is so much diversity.

In addition, the author has been made aware of another benefit, namely, that categorical translations between games may allow for syntactic translations between the logical languages that are interpreted in those games.  This would accord with the correspondence theory of van Benthem 2001, and Conradie, Ghilardi, and Palmigiano 2014.\nocite{Benth01} \nocite{ConraGP14}

\vspace{.8mm}{\em Categorical Reformulations}.  Finally, in a different direction, it appears that some game-theoretic concepts can be reformulated in categorical terms, and that such categorical reformulations could conceivably lead to new game-theoretic results in the future.  One example may be Section~\rf{9444}'s reformulation of the game-theoretic concept of ``subgame'' in terms of a special \ct{NCG} morphism called a ``subgame inclusion''.  Another example may be that monotonic transformations are naturally built into the definition of an \ct{NCG} morphism (and that affine transformations can naturally define an \ct{NCG} subcategory as suggested in note~\rf{9268}).

\newcommand{\NonCat}{\footnote{In addition, other papers study equivalences between games without using category theory.  These include the references in note~\rf{9439}, as well as 
McKinsey 1950\nocite{Mckin50},  
Thompson 1952\nocite{Thomp5297}, 
Dalkey 1953\nocite{Dalke53}, 
Kohlberg and Mertens 1986\nocite{KohlbM86},
Bonanno 1992\nocite{Bonan92},  
Elmes and Reny 1994\nocite{ElmesR94}, and 
van Benthem 2014 (pages 43--51).}}

\newcommand{\BoltHZ}{\footnote{This sentence may require explanation because the focus of their paper is elsewhere.  In particular, the larger contribution of their paper is to systematically compose game fragments (called ``open games'') by regarding these fragments as morphisms in a category which has no parallel here.  In addition, morphisms between the fragments are defined via bisimulation.  Very roughly, the fragments are embellished game forms.  Accordingly, the paper's Section 4.7 shows that a standard Bayesian game can be modelled as a fragment (i.e.\ an ``open game'') plus a root node and utility functions (called a ``context'').}}

\vspace{.8mm}{\em Related Literature}.  Other papers have also developed categories for games or game forms.{\NonCat}  Most relevant is Bolt, Hedges, and Zahn 2019,\nocite{BoltHZ19} which defines a category admitting (finite-horizon) Bayesian game forms.{\BoltHZ}  SP, SF, and the present paper all differ from their paper by admitting both infinite-horizon trees and arbitrary information sets, by defining morphisms which are not isomorphisms, and by using very different categorical tools.  Further, the present paper differs by building a category for games, as opposed to game forms.

Other contributions develop categories which are less relevant to this paper because their information sets are less general.  These include Lapitsky 1999\nocite{Lapi99} and Jim\'{e}nez 2014\nocite{Jime14}, which define categories for simultaneous-move games, and Machover and Terrington 2014\nocite{MachTerr14} which defines a category for some cooperative games.  Also included are Abramsky, Jagadeesan, and Malacaria 2000\nocite{AbraJagaM00}, Hyland and Ong 2000\nocite{HylaOngL00}, and McCusker 2000\nocite{McCu00}, which develop categories for some specialized games in computer science.  Lastly included are Honsell, Lenisa, and Redamalla 2012\nocite{HonseLR12}, Abramsky and Winschel 2017\nocite{AbramW17}, Hedges 2018\nocite{Hedge18}, and Ghani, Kupke, Lambert, and Forsberg 2018\nocite{GhaniKLF18} and 2019\nocite{GhaniKLF19}, all of which define categories for various games with relatively trivial information sets.

\ssec{Organization}\label{9271}

Section \rf{9255} reviews the category \ct{Tree} for functioned trees, the category \ct{NCP} for node-and-choice preforms, and the category \ct{NCF} for node-and-choice forms (all from SP and SF).  In addition, Section~\rf{9255} defines subtrees, subpreforms, and subforms; and provides new results about how tree morphisms interact with plays.  Next, Section~\rf{9272} defines and develops the category \ct{NCG} for node-and-choice games.  Finally, Section~\rf{9087} studies some of this new category's subcategories.  Several proofs and many lemmas are relegated to the appendices: Appendix~\rf{9285} concerns \ct{Tree}, Appendix~\rf{9289} concerns \ct{NCP}, Appendix~\rf{9290} concerns \ct{NCF}, and Appendix~\rf{9311} concerns \ct{NCG}.

\section{Preliminaries}\label{9255} 
\markb{\sc \rf{9255}. Preliminaries}

\ssec{Explicitness for those learning category theory}\label{9256}

As discussed in Section~\rf{9735}, this paper tries to help game theorists learning category theory.  Hence the paper tries to avoid the notational and conceptual shortcuts that expert category theorists use to suppress routine details that are of no interest to them.  For example, the structure of a structured set is kept explicit, contrary to the habits of expert category theorists who routinely make such a structure implicit.  More generally, the papers try to avoid the specialized shortcuts that any mathematical discipline uses to suppress routine details.  For example, a function is kept distinct from its graph, contrary to the habits of expert set theorists who routinely identify a function with its graph (similar distinctions are made between a correspondence and its graph, and between a relation and its graph).  

Such matters are discussed more fully in SF Section 1.6.  In addition, this paper is yet more explicit than SP and SF in three minor regards.  [1] Let the superscript $\T$ (rather than $^{-1}$) mark the transpose of a function or correspondence (by default, transposes are correspondences).
[2] For a function $f{:}X→Y$, define $\bar{f}{:}\PP(X)→\PP(Y)$ by $\bar{f}(A) = ⎨f(x)|x∈A⎬$.\footnote{An alternative notation for $\bar{f}$ would have been $\PP f$, as in Mac Lane 1998, page~13.\nocite{Macla98}}  Similarly, for a correspondence $G{:}X⇉Y$, define $\bar{G}{:}\PP(X)→\PP(Y)$ by $\bar{G}(A) = ∪_{x∈A}G(x)$.  [3] For a function $f{:}X→Y$ and a set $A⋅⊆⋅X$, let $f\sj_A$ be the surjective restriction of $f$ to $A$.  For example, if $f{:}Ṛ→Ṛ$ is defined by $f(x) = 2x$, then $f\sj_{[0,1]}{:}[0,1]→[0,2]$.

At the same time, this paper is less explicit than SP and SF in one significant regard:  Here it is assumed that trees, preforms, forms, and games are implicitly accompanied by their components and derivatives.  See Table~\rf{8935} for an overview.

\newcommand{\dve}{\dra\,\,}
\renewcommand{\capp}{Trees, preforms, forms, and games are implicitly accompanied by their components and derivatives (\protect\rotatebox[origin=c]{180}{$\Lsh$}).  The symbol * means new to this paper.} 
\begin{table}[h]
{\small 
\begin{tabular}{cl} 
$(T,p)$ & (Functioned) Tree satisfying \rf{T1}--\rf{T2} \\ \hline
$T$ & set of nodes $t$ \\
$p$ & immediate-predecessor function \\
$t^o$ & $\dve$ root node \\
$X$ & $\dve$ set of decision nodes $t$ \\
$k$ & $\dve$ stage function \\
$≺$ & $\dve$ strict precedence relation\\
$≼$ & $\dve$ weak precedence relation\\
$\ZZ$ & $\dve$ collection of plays (i.e., maximal chains) $Z$ \\
$\ZZf$ & $\dve$ collection of finite plays $Z$ \\
$\ZZi$ & $\dve$ collection of infinite plays $Z$ \\
$P$ & $\dve$ strict-predecessor correspondence (Section~\rf{9257}.1)\,*\\[2mm]

$Π$ & Preform $(T,C,⊗)$ satisfying \rf{P1}--\rf{P3} \\ \hline
$C$ & set of choices $c$ \\
$⊗$ & node-and-choice operator \\
$F$ & $\dve$ feasibility correspondence \\
$\HH$ & $\dve$ collection of information sets $H$ \\
$q$ & $\dve$ previous-choice function \\
$\Ss$ & $\dve$ collection of grand strategies $S$ (Section~\rf{9440})\,*\\
$ζ$ & $\dve$ grand-strategy-to-play function (Section~\rf{9440})\,* \\[2mm]

$Φ$ & Form $(I,T,(C_i)_{i∈I},⊗)$ satisfying \rf{F1}--\rf{F3} \\ \hline
$I$ & set of players $i$ \\
$C_i$ & player $i$'s set of choices $c$ \\
$X_i$ & $\dve$ player $i$'s set of decision nodes $t$ \\
$\HH_i$ & $\dve$ player $i$'s collection of information sets $H$ \\
$\Ss_i$ & $\dve$ player $i$'s collection of strategies $S_i$ (Section~\rf{9441})\,* \\[2mm]

$Γ$ & Game $(I,T,(C_i)_{i∈I},⊗,(U_i)_{i∈I})$ satisfying \rf{G1}--\rf{G2}\,* \\ \hline
$U_i$ & player $i$'s utility function (Section~\rf{9442})\,* \\[2mm]
\end{tabular} }
\caption{\small \capp} \label{8935}
\end{table}

\pagebreak
\ssec{The category \ct{Tree}}\label{9257}

{\sc \rf{9257}.1.\ Basics}. Let $T$ be a set of {\em nodes} $t$.  A {\em (functioned) tree} (SP Section 2.1) is a pair $(T,p)$ such that there are $t^o⋅∈⋅T$ and $X⋅⊆⋅T$ satisfying\begin{gather}
\zz
p⋅\text{is a nonempty function from}⋅T⧷⎨t^o⎬⋅\text{onto}⋅X,⋅\text{and} \ttt{[T1]}{T1}\\
(∀t∈T⧷⎨t^o⎬)(∃m∈\tNa)⋅p^m(t) = t^o.\footnotemark 
\ttt{[T2]}{T2} 
\zz
\end{gather}\footnotetext{\label{8097}Let $\tNz = ⎨0,1,2,...⎬$ and $\tNa = ⎨1,2,...⎬$.  Further,  for any $m⋅∈⋅\tNz$ and any function $f$, let $f^m(x)$ be the result of applying $f$ to $x$, $m$ times.}Call $p$ the {\em (immediate) predecessor} function, call $t^o$ the {\em root} node, and call $X$ the set of {\em decision} nodes.  Further, define a tree's {\em stage} function $k{:}T→\tNz$ by [a] $k(t^o) = 0$ and [b] $(∀t∈T⧷⎨t^o⎬)$ $p^{k(t)}(t) = t^o$.  Define its {\em (strict) precedence relation} $≺$ on $T$ by $(∀t^1∈T,t^2∈T)$ $t^1⋅≺⋅t^2$ iff $(∃m∈\tNa)$ $t^1 = p^m(t^2)$.  Relatedly, define its {\em weak precedence relation} $≼$ on $T$ by $(∀t^1∈T,t^2∈T)$ $t^1⋅≼⋅t^2$ iff $(∃m∈\tNz)$ $t^1 = p^m(t^2)$.  Finally, let $\ZZ$ be the collection of maximal chains in $(T,≼)$.  Call $Z⋅∈⋅\ZZ$ a {\em play}.  $\ZZ$ can be split into the (possibly empty) collection $\ZZf$ of finite plays and the (possibly empty) collection $\ZZi$ of infinite plays.  

The above are developed in SP Sections 2.1--2.2.  In addition, define $P{:}T⇉T$ by $P(t) = ⎨t\fl∈T|t\fl≺t⎬$.  Call $P$ the {\em strict-predecessor correspondence}.\footnote{An alternative notation for $P(t)$ would have been $⇣_{≺}(t)$, where $⇣$ suggests order theory's down-set, as in Davey and Priestley 2002, page 20.\nocite{DaveyP02}}  As might be expected, Lemma~\rf{8327}(\rf{8452},\rf{9401},\rf{8956}) shows that $P(t^o) = ∅$, that $(∀t∈T⧷⎨t^o⎬)⋅p(t)⋅∈⋅P(t)$, and that $(∀t∈T)$ $P(t) = ⎨p^m(t)|k(t)≥m{>}0⎬$. 

All the above are summarized in the first section of Table~\rf{8935}.  (By assumption, each tree is implicitly accompanied by all its derivatives.) Figure~\rf{9718} depicts an example for those learning game theory.\footnote{\label{9721}Figure~\rf{9718}'s tree $(T,p)$ is defined by setting $T = ⎨\f0,\f1,\f2,\f3,\f4,\f5,\f6,\f7,\f8⎬$ and by letting $p$ be the surjective function with graph $⎨(\f1,\f0),(\f2,\f1),(\f3,\f0),(\f4,\f1),(\f5,\f3),(\f6,\f3),(\f7,\f4),(\f8,\f4)⎬$.  As depicted in the figure, $t^o\,{=}\,\f0$ and $X\,{=}\,⎨\f0,\f1,\f3,\f4⎬$.  Also, $k(\f0)\,{=}\,0$, $k(\f1)\,{=}\,k(\f3)\,{=}\,1$, $k(\f2)\,{=}\,k(\f4)\,{=}\,k(\f5)\,{=}\,k(\f6)\,{=}\,2$, and $k(\f7)\,{=}\,k(\f8)\,{=}\,3$.  Also, $≺$ is the transitive closure of the binary relation on $T$ whose graph is the transpose of $p\gr$.  For example, $\f0⋅≺⋅\f1$ and $\f1⋅≺⋅\f2$ by the transpose of $p\gr$, and $\f0⋅≺⋅\f2$ by transitivity.  Also, $t^A⋅≼⋅t^B$ iff $t^A⋅≺⋅t^B$ or $t^A = t^B$.  $\ZZ$ is the collection consisting of $⎨\f0,\f3,\f5⎬$, $⎨\f0,\f3,\f6⎬$, $⎨\f0,\f1,\f4,\f7⎬$, $⎨\f0,\f1,\f4,\f8⎬$, and $⎨\f0,\f1,\f2⎬$.  $\ZZf = \ZZ$ and $\ZZi = ∅$.  Finally, $P(\f0) = ∅$, $P(\f1) = P(\f3) = ⎨\f0⎬$, $P(\f2) = P(\f4) = ⎨\f0,\f1⎬$, $P(\f5) = P(\f6) = ⎨\f0,\f3⎬$, and $P(\f7) = P(\f8) = ⎨\f0,\f1,\f4⎬$.}

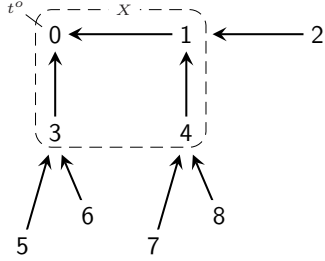
\renewcommand{\capp}{The tree $(T,p)$ under Figure~\rf{9259}(a)'s game.  $X$ is the set of decision nodes, and $t^o$ is the root node.  Each arrow [a] takes an argument of $p$ to its value and [b] ``points'' in the same direction as $≺$.  Arrowheads are later suppressed.  Details in note~\rf{9721}.}
\begin{figure}[h]
  \newcommand{\hgth}{95}  
  \begin{picture}(0,\hgth) 
  \put(-102,-12){\scalebox{.96}{ 
    \begin{pspicture}(4,1)(12,-5) 
      \end{pspicture}
    }} \end{picture}
  \caption{\small \capp} \label{9718} 
\end{figure} 

A {\em tree morphism} (SP Section 2.3) is a triple $θ = [(T,p),(T′,p′),τ]$ such that $(T,p)$ and $(T′,p′)$ are (functioned) trees,\begin{gather}
\zz
τ{:}T→T′⋅\text{and} \ttt{[t1]}{t1}\\
⎨⋅(τ(t\sh),τ(t))⋅|⋅(t\sh,t)∈p\gr⋅⎬⋅⊆⋅p\pgr. \ttt{[t2]}{t2}
\zz
\end{gather} 
The category \ct{Tree} (SP Section~2.4) has trees as its objects and tree morphisms as its arrows.  SP Section 2.5 shows that \ct{Tree} is isomorphic to the full subcategory of \ct{Grph} for converging arborescences.

\newcommand{\notesub}{\footnote{\label{9454}Since a subtree inclusion is a special kind of tree inclusion, a subtree is a special kind of included tree.  Although it might seem natural to call {\em any} included tree a ``subtree'', the present terminology accords with the extremely well-known concept of subgame-perfection (Selten 1975).}}

\myskip{\sc \rf{9257}.2.\ Subtrees}. Let a {\em tree inclusion} be a \ct{Tree} morphism that satisfies $T⋅⊆⋅T′$ and $τ = \inc_{T,T′}$, where $\inc_{T,T′}$ is the inclusion function\footnote{\label{9169}In general, suppose $A⋅⊆⋅B$.  Then $\inc_{A,B}$ is the function $f{:}A→B$ defined by $(∀a∈A)$ $f(a) = a$.} from $T$ into $T′$.  Further, let a {\em subtree inclusion} be a tree inclusion that also satisfies $T\,{=}\,⎨t′∈T′|t^o≼′t′⎬$.  Finally, call $(T,p)$ a {\em subtree} of $(T′,p′)$ iff $[(T,p),(T′,p′),\inc_{T,T′}]$ is a subtree inclusion.{\notesub}

\newcommand{\noteXX}{\footnote{\label{9453}The only difference is that subtrees here must have more than one node simply because trees here must have more than one node (SF Lemma~A.1(a)).}}

Lemma~\rf{9172} shows that one tree $(T,p)$ is a subtree of another tree $(T′,p′)$ iff $T\,{=}\,⎨t′∈T′|t^o≼′t′⎬$ and $p = p′\sj_{T⧷⎨t^o⎬}$.  Also, Lemma~\rf{9185} takes a tree $(T′,p′)$ and constructs the subtree starting from an arbitrary node in $X′$.  Both Lemma~\rf{9172}'s characterization and Lemma~\rf{9185}'s construction are consistent{\noteXX} with the standard concept of a subgame in Selten 1975 and Myerson 1991, page 184. \nocite{Myers91}

\myskip{\sc \rf{9257}.3.\ How \ct{Tree} morphisms interact with plays}. SP Sections 2.3 and 2.4 show the various ways in which \ct{Tree} morphisms and isomorphisms preserve the various components and derivatives of trees.  The SP results about plays are especially important to this paper because plays appear in the domains of utility functions.  For example, the following proposition shows that \ct{Tree} isomorphisms preserve plays in a strong sense.

\begin{prop}\label{9050} If $[(T,p),(T′,p′),τ]$ is an isomorphism, then $\dtau\sj_{\ZZ}$ is a bijection from $\ZZ$ onto $\ZZ′$. (SP Proposition 2.7(h,i).) \end{prop}

\ct{Tree} morphisms, as opposed to isomorphisms, preserve plays in a much weaker sense.  In particular, SP Proposition 2.4(g,h) shows that if $[(T,p),(T′,p′),τ]$ is a morphism, then $(∀Z∈\ZZ)(∃Z′∈\ZZ′)$ $\dtau(Z)⋅⊆⋅Z′$.  Stronger results seem to be precluded by examples like that of Figure~\rf{9020}.\footnote{\label{9021}The example $θ = [(T,p),(T′,p′),τ]$ is defined by setting $T = ⎨\f1,\f2,\f3,\f4⎬$, letting $p$ be with surjective function with graph $p\gr = ⎨(\f2,\f1),$ $(\f3,\f1),$ $(\f4,\f1)⎬$, setting $T′ = ⎨\f{10},\f{11},\f{12},\f{13},\f{14},\f{15},\f{16},\f{17}⎬$, letting $p′$ be the surjective function with graph $p\pgr = ⎨(\f{11},\f{10}),$ $(\f{17},\f{10}),$ $(\f{12},\f{11}),$ $(\f{13},\f{11}),$ $(\f{14},\f{11}),$ $(\f{15},\f{14}),$ $(\f{16},\f{14})⎬$, and defining $τ{:}T→T′$ by $τ\gr = ⎨(\f1,\f{11}),$ $(\f2,\f{12}),$ $(\f3,\f{13}),$ $(\f4,\f{14})⎬$.}  There $\ZZ = ⎨⎨\f1,\f2⎬,⎨\f1,\f3⎬,⎨\f1,\f4⎬⎬$ and $\ZZ′ = ⎨⎨\f{10},\f{17}⎬,$ $⎨\f{10},\f{11},\f{12}⎬,$ $\!⎨\f{10},\f{11},\f{13}⎬,$ $\!⎨\f{10},\f{11},\f{14},\f{15}⎬,$ $\!⎨\f{10},\f{11},\f{14},\f{16}⎬⎬$.  Since the example's $τ$ maps each source node $\f{n}$ to the target node $\f{10{+}n}$, there is no $Z⋅∈⋅\ZZ$ such that $\dtau(Z)⋅∈⋅\ZZ′$.  Thus the image of a source play is generally not a target play.

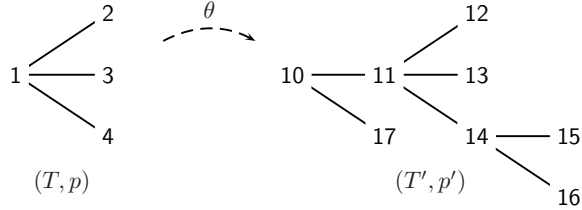
\renewcommand{\capp}{The source tree and target tree of the morphism $θ = [(T,p),(T′,p′),τ]$ defined in note~\rf{9021}.  The function $τ{:}T→T′$ maps each source node $\f{n}$ to the target node $\f{10{+}n}$.}
\begin{figure}[h]
  \newcommand{\hgth}{82}  
  \begin{picture}(0,\hgth) 
  \put(-125,-12){\scalebox{.9}{  
    \begin{pspicture}(-6,-5)(8,3) 
      \end{pspicture}
    }} 
    \end{picture}
  \caption{\small \capp} \label{9020}  
  \end{figure} 

Relatedly, Proposition~\rf{8940t} below describes only limited ways in which a morphism $θ$ preserves plays.  For part (a), say that a chain $S⋅⊆⋅T$ in a tree $(T,p)$ is {\em consecutive} iff $(∀s^1∈S,s^2∈S,t∈T)$ $s^1⋅≺⋅t⋅≺⋅s^2$ implies $t⋅∈⋅S$.  As might be expected, Lemma~\rf{8327}(\rf{8957}) shows every play is a consecutive chain.  In a similar vein, part (a) of the proposition below shows that the image of each source play $Z$ is a consecutive chain of target nodes.

Part (b) shows that the image of each source play $Z$ is preceded by $P′○τ(t^o)$, which is conveniently independent of $Z$.  For example, in Figure~\rf{9020}, $P′○τ(t^o) = P′○τ(\f1) = P′(\f{11}) = ⎨\f{10}⎬$.  

For part (c), consider an arbitrary tree $(T,p)$ and let the {\em end} of a finite play be its maximum (an infinite play does not have an end).  Then consider a morphism $θ = [(T,p),(T′,p′),τ]$ and define \begin{gather}
\zz
\ZZ^θ = \ZZi⋅∪⋅⎨\,Z∈\ZZf\,|\,(∄t^{\prime +}∈T′)\,τ(\max Z)\,≺′\,t^{\prime +}\,⎬.\notag
\zz
\end{gather} Call $\ZZ^θ$ the collection of $θ$'s {\em end-preserved} source plays.\footnote{The symbol $\ZZ^θ$ does not belong in Table~\rf{8935} because it is derived from a morphism $θ$ rather than from a tree $(T,p)$.} Thus an infinite source play is end-preserved by definition, and a finite source play is end-preserved iff its end's image is not succeeded by a target node (conveniently $τ(\max Z) = \text{max}\,\dtau(Z)$ by Lemma~\rf{8940}(\rf{8988})).  For example, in Figure~\rf{9020}, the end-preserved source plays are $⎨\f1,\f2⎬$ and $⎨\f1,\f3⎬$.  Finally, the proposition's part (\rf{8945t}) shows that the end-preserved source plays are identical to the source plays $Z$ that are mapped to target plays by the rule $Z \mapsto$ $P′○τ(t^o)∪\dtau(Z)$.  In this sense, the end-preserved source plays are the source plays that can be reasonably mapped to target plays.  For example, in Figure~\rf{9020}, the end-preserved plays $⎨\f1,\f2⎬$ and $⎨\f1,\f3⎬$ are mapped to the target plays \begin{gather}
\zz
P′○τ(t^o)∪\dtau(⎨\f1,\f2⎬) = ⎨\f{10}⎬∪⎨\f{11},\f{12}⎬ = ⎨\f{10},\f{11},\f{12}⎬⋅\text{and} \nt
P′○τ(t^o)∪\dtau(⎨\f1,\f3⎬) = ⎨\f{10}⎬∪⎨\f{11},\f{13}⎬ = ⎨\f{10},\f{11},\f{13}⎬. \notag
\zz
\end{gather} In contrast, $⎨\f1,\f4⎬$ is not end-preserved, and accordingly, $P′○τ(t^o)∪\dtau(⎨\f1,\f4⎬) = ⎨\f{10}⎬∪⎨\f{11},\f{14}⎬ = ⎨\f{10},\f{11},\f{14}⎬$ is not a target play.

\begin{prop}\label{8940t} Suppose $θ = [(T,p),(T′,p′),τ]$ is a morphism.  Then the following hold.\begin{tlist}
\vspace{1mm}
\yl{8941t} $(∀Z∈\ZZ)$ $\dtau(Z)$ is a consecutive chain.
\yl{8942t} $(∀Z∈\ZZ)$ $P′(\min\dtau(Z)) = P′○τ(t^o)$.
\yl{8945t} $\ZZ^θ = ⎨\,Z∈\ZZ\,|\,P′○τ(t^o)∪\dtau(Z)\,∈\,\ZZ′\,⎬$, where $\ZZ^θ$ is the collection of $θ$'s end-preserved source plays.
(Lemma~\rf{8940}(\rf{8941},\rf{8942},\rf{8945}).)
\end{tlist} \end{prop}

A morphism $θ$ is said to be {\em end-preserving} iff $\ZZ^θ = \ZZ$.  Proposition~\rf{9043} provides three broad classes of end-preserving morphisms.  For such a morphism, Proposition \rf{8940t}(\rf{8945t}) implies that {\em every} source play is sent to a target play by the map $\ZZ⋅∋⋅Z \mapsto P′○τ(t^o)∪\dtau(Z)⋅∈⋅\ZZ′$.  When the morphism is a subtree inclusion (as in the Proposition \rf{9043}(\rf{9046})), this map is injective because of the injectivity of the inclusion function $τ = \mathsf{inc}_{T,T′}$.  When the morphism is an isomorphism, the map simplifies to $\ZZ⋅∋⋅Z \mapsto \dtau(Z)⋅∈⋅\ZZ′$ by Proposition \rf{9043}(\rf{9045})[2], and further, the map is bijective by Proposition~\rf{9050}.

\begin{prop}\label{9043} Suppose $θ = [(T,p),(T′,p′),τ]$ is a morphism.  Then the following hold. \begin{tlist}
\yl{9044} If $(T,p)$ has only infinite plays, then $\ZZ^θ = \ZZ$. 
\yl{9046} If $θ$ is a subtree inclusion, then $\ZZ^θ = \ZZ$.
\yl{9045} If $θ$ is an isomorphism, then [1] $\ZZ^θ = \ZZ$ and [2] $P′○τ(t^o) = ∅$. (Proof~\rf{9043p}.)
\end{tlist} \end{prop}

\pagebreak
\ssec{The category \ct{NCP}}\label{9440}

\newcommand{\noteFtranspose}{\footnotetext{\label{8128}To be clear,  $F{:}T⇉C$ means that $F$ is a correspondence from $T$ to $C$, which means that $F$ is a triple $(T,C,F\gr)$ such that $F\gr⋅⊆⋅T×C$ (this accords with SF Section 2.1).  In accord with this paper's Section~\rf{9256}, $F\T(c) = ⎨t∈T|c∈F(t)⎬$ and $\overline{F\T}(C) = ∪_{c∈C}F\T(c)$.  The latter expression, by the definition of $F\T(c)$, equals  $∪_{c∈C}⎨t∈T|c∈F(t)⎬$, which equals $⎨t∈T|(∃c∈C)c∈F(t)⎬$, which equals $⎨t∈T|F(t)≠∅⎬$.}}

\newcommand{\notewellp}{\footnotetext{SP Lemma C.1(a) shows that \rf{P1} implies the well-definition and surjectivity of the function $p$.  Thus, if \rf{P1} holds, then \rf{P2} holds iff both $T⧷⎨t^o⎬⋅≠⋅∅$ and $(∀t∈T⧷⎨t^o⎬)(∃m≥1)$ $t^o = p^m(t)$.  (This follows from inspecting the definition of a functioned tree.)}}

Let $C$ be a set of {\em choices} $c$.  A triple $Π = (T,C,⊗)$ is a {\em (node-and-choice) preform} (SF Section 2.1)\footnote{The definition of a preform originally appeared in SP Section 3.1.  The definition there is slightly less explicit.} iff 
\begin{center}
\zz
\vspace{-3mm}$\mqi \text{\ttc{[P1]}{P1}} & \text{there is a correspondence}\,\footnotemark⋅F{:}T⇉C⋅\text{and a}⋅t^o∈T \\   
&\text{such that}⋅⊗⋅\text{is a bijection from}⋅F\gr⋅\text{onto}⋅T⧷⎨t^o⎬, \mqo$
\noteFtranspose\\[.6mm]
$\mqi \text{\ttc{[P2]}{P2}} & (T,p)⋅\text{is a (functioned) tree where}⋅p{:}T⧷⎨t^o⎬→\overline{F\T}(C)\,^{\text{\rf{8128}}} \\
  & \text{is defined}\,\footnotemark⋅\text{by}⋅p\gr = ⎨(t\sh,t)∈T^2|(∃c∈C)(t,c,t\sh)∈⊗\gr⎬,⋅\text{and} \mqo$
\notewellp\\[.6mm]
$\mqi \text{\ttc{[P3]}{P3}} & \HH⋅\text{partitions}⋅\overline{F\T}(C)\,^{\text{\rf{8128}}} \\
  & \text{where}⋅\HH⋅⊆⋅\PP(T)⋅\text{is defined by}⋅\HH = ⎨F\T(c)|c∈C⎬.\mqo$
\zz
\end{center}%
\vspace{1.4mm}Call $⊗$ the {\em node-and-choice} operator, and let $t⊗c$ denote its value at $(t,c)⋅∈⋅F\gr$.  Call $F$ the {\em feasibility} correspondence, call $F(t)$ the set of {\em feasible} choices at node $t$, call $t^o$ the {\em root} node, call $p$ the {\em immediate-predecessor} function, and call $\HH$ the collection of {\em information sets}.  For convenience, let $X$ equal $\overline{F\T}(C)$, and call $X$ the set of {\em decision nodes}.  Further, define the {\em previous-choice} function $q{:}T⧷⎨t^o⎬→C$ by $q\gr = ⎨(t\sh,c)∈T×C|(∃t∈T)(t,c,t\sh)∈⊗\gr⎬$.  Finally, note \rf{P2} defines the tree $(T,p)$, which in turn defines the preform's $k$, $≺$, $≼$, $\ZZ$, $\ZZf$, $\ZZi$, and $P$ by means of the previous subsection.\footnote{SP Lemma~C.1(b,c) implies that a preform's $t^o$ and $X = \overline{F\T}(C)$ coincide with the underlying tree's $t^o$ and $X$.  Hence the symbols $t^o$ and $X$ are unambiguous.  (Inconsequentially, SP uses the notation $F^{-1}(C)$ rather than either the notation $X$ or the notation $\overline{F\T}(C)$.)} 

The above are developed in SP Sections 3.1--3.2.  In addition, two further constructions will be useful.  First, define $\Ss = ⎨⋅S⊆C⋅|⋅(∀H∈\HH)\,|S∩\dF(H)|{=}1⋅⎬$.  Call each $S⋅∈⋅\Ss$ a {\em grand strategy}.  Thus a grand strategy names exactly one feasible choice at each information set.  Second, Lemma~\rf{8401} shows that, for each $S⋅∈⋅\Ss$, there is exactly one $Z⋅∈⋅\ZZ$ such that $(∀t∈Z⧷⎨t^o⎬)$ $q(t)⋅∈⋅S$.  Define $ζ{:}\Ss→\ZZ$ accordingly.  Call $ζ$ the {\em grand-strategy-to-play function}. 

All the above are summarized in Table~\rf{8935}'s first two sections.  (By assumption, each preform $Π$ is implicitly accompanied by all its components and derivatives.)  Figure~\rf{9719} continues Figure~\rf{9718}'s example for those learning game theory.\footnote{\label{9722}Figure~\rf{9719}'s preform $(T,C,⊗)$ is defined by defining $T$ as in note \rf{9721}, letting $C$ equal $⎨\fa,\fb,\fg,\fd,\fe,\ff⎬$, and letting $⊗$ be the surjective function with graph $⎨((\f0,\fa),\f1),$ $((\f0,\fb),\f3),$ $((\f1,\fg),\f2),$ $((\f1,\fd),\f4),$ $((\f3,\fe),\f5),$ $((\f3,\ff),\f6),$ $((\f4,\fe),\f7),$ $((\f4,\ff),\f8)⎬$ (pairs of the form $((t,c),t\sh)$ are usually written as triples of the form $(t,c,t\sh)$).  $F$ is the correspondence from $T$ to $C$ whose graph is $⎨(\f0,\fa),(\f0,\fb),(\f1,\fg),(\f1,\fd),(\f3,\fe),(\f3,\ff),(\f4,\fe),(\f4,\ff)⎬$.  Thus, as stated in Figure~\rf{9719}'s caption, $F\T(\fa) = F\T(\fb) = ⎨\f0⎬$, $F\T(\fg) = F\T(\fd) = ⎨\f1⎬$, and $F\T(\fe) = F\T(\ff) = ⎨\f3,\f4⎬$.  Further, $q$ is the surjective function with graph $⎨(\f1,\fa),(\f2,\fg),(\f3,\fb),(\f4,\fd),(\f5,\fe),(\f6,\ff),(\f7,\fe),(\f8,\ff)⎬$.  $\Ss$ is the eight-set collection consisting of $⎨\fa,\fg,\fe⎬$, $⎨\fa,\fg,\ff⎬$, $⎨\fa,\fd,\fe⎬$, $⎨\fa,\fd,\ff⎬$, $⎨\fb,\fg,\fe⎬$, $⎨\fb,\fg,\ff⎬$, $⎨\fb,\fd,\fe⎬$, and $⎨\fb,\fd,\ff⎬$.  Finally, $ζ{:}\Ss→\ZZ$ is defined by $ζ(⎨\fa,\fg,\fe⎬) = ζ(⎨\fa,\fg,\ff⎬) = ⎨\f0,\f1,\f2⎬$, by $ζ(⎨\fa,\fd,\fe⎬) = ⎨\f0,\f1,\f4,\f7⎬$, by $ζ(⎨\fa,\fd,\ff⎬) = ⎨\f0,\f1,\f4,\f8⎬$, by $ζ(⎨\fb,\fg,\fe⎬) = ζ(⎨\fb,\fd,\fe⎬) = ⎨\f0,\f3,\f5⎬$, and by $ζ(⎨\fb,\fg,\ff⎬) = ζ(⎨\fb,\fd,\ff⎬) = ⎨\f0,\f3,\f6⎬$.}

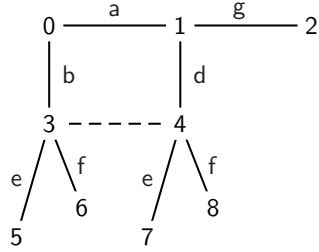
\renewcommand{\capp}{The preform $(T,C,⊗)$ under Figure~\rf{9259}(a)'s game.  The tree defined in \rf{P2} is in Figure~\rf{9718}.  The information sets defined in \rf{P3} are $F\T(\fa){=}F\T(\fb){=}⎨\f0⎬$, $F\T(\f{g}){=}F\T(\fd){=}⎨\f1⎬$, and $F\T(\f{e}){=}F\T(\f{f}){=}⎨\f3,\f4⎬$ (the last is depicted by the dashed line).  Details in note~\rf{9722}.}
\begin{figure}[h]
  \newcommand{\hgth}{92}  
  \begin{picture}(0,\hgth) 
  \put(-102,-12){\scalebox{.96}{ 
    \begin{pspicture}(4,1)(12,-5) 
      \end{pspicture}
    }} \end{picture}
  \caption{\small \capp} \label{9719} 
  \end{figure}  

A {\em preform morphism} (SP Section 3.3) is a quadruple $[Π,Π′,τ,δ]$ such that $Π$ and $Π′$ are preforms, \begin{gather}
\zz
τ{:}T→T′,⋅δ{:}C→C′,⋅\text{and} \ttt{[p1]}{p1}\\
⎨(τ(t),δ(c),τ(t\sh))|(t,c,t\sh)∈⊗\gr⎬⋅⊆⋅⊗\pgr. \ttt{[p2]}{p2}
\zz
\end{gather} The category \ct{NCP} (SP Section 3.4) has preforms as its objects and preform morphisms as its arrows.  SP Theorem~3.9 shows there is a ``forgetful'' functor $\TB$ from \ct{NCP} to \ct{Tree} (there $\TB$ is called ``$\FB$'').  See Figure~\rf{8307} below. 

Let a {\em preform inclusion} be an \ct{NCP} morphism that satisfies $T⋅⊆⋅T′$, $τ = \inc_{T,T′}$, $C⋅⊆⋅C′$, and $δ = \inc_{C,C′}$ (where note~\rf{9169} defines $\inc_{A,B}$ to be the inclusion function from $A$ to $B$).  Further, let a {\em subpreform inclusion} be a preform inclusion that also satisfies  $T\,{=}\,⎨t′∈T′|t^o≼′t′⎬$ and $\HH⋅⊆⋅\HH′$.  Finally call $Π$ a {\em subpreform} of $Π′$ iff $[Π,Π′,\inc_{T,T′},\inc_{C,C′}]$ is a subpreform inclusion.

Lemma~\rf{9173} shows that $Π$ is a subpreform of $Π′$ iff $T\,{=}\,⎨t′∈T′|t^o≼′t′⎬$, $C⋅⊆⋅C′$, $⊗ = ⊗\sj_{F\gr}$, and $\HH⋅⊆⋅\HH′$.  These four conditions are consistent with the standard definition of a subgame in Selten 1975 (see note~\rf{9453} for a trivial difference about the minimal number of nodes).  The condition $\HH⋅⊆⋅\HH′$ holds trivially if both $Π$ and $Π′$ have perfect-information.\footnote{Perfect-information appears later in Section~\rf{9278}.  It is the property of having singleton information sets.  In this event, $\HH = ⎨⎨t⎬|t∈X⎬$ and $\HH′ = ⎨⎨t′⎬|t′∈X′⎬$.  Thus $\HH⋅⊆⋅\HH′$ follows from $X⋅⊆⋅X′$, which follows from Lemma~\rf{9022}(\rf{9187}).}

\ssec{The category \ct{NCF}}\label{9441}

Let $I$ be a set of {\em players} $i$.  A {\em (node-and-choice) form} (SF Section 2.1) is a quadruple $Φ = (I,T,(C_i)_{i∈I},⊗)$ such that \begin{gather}
\zz
(T,C,⊗)⋅\text{is an (\ct{NCP}) preform where}⋅C = ∪_{i∈I}C_i, \ttt{[F1]}{F1}\\
(∀i∈I,j∈I⧷⎨i⎬)⋅C_i∩C_j = ∅,⋅\text{and} \ttt{[F2]}{F2}\\
(∀t∈T)(∃i∈I)⋅F(t)⋅⊆⋅C_i. \ttt{[F3]}{F3}
\zz
\end{gather} Each $C_i$ is the set of choices that are assigned to player $i$.  Further, define $(X_i)_{i∈I}$ at each $i$ by $X_i = ∪_{c∈C_i}F\T(c)$.  Each $X_i$ is the set of decision nodes that are assigned to player $i$.  Finally, define $(\HH_i)_{i∈I}$ at each $i$ by $\HH_i = ⎨F\T(c)|c∈C_i⎬$.  Each $\HH_i$ is the collection of information sets that are assigned to player $i$.  Note \rf{F1} defines the form's preform $Π = (T,C,⊗)$, which in turn determines the form's $F$, $t^o$, $p$, $q$, $\HH$, $X$, $k$, $≺$, $≼$, $P$, $\ZZ$, $\ZZf$, $\ZZi$, $\Ss$, and $ζ$ by means of the previous subsection. 

The above are developed in SF Section 2.1.  In addition, for each player $i$, define $\Ss_i = ⎨⋅S_i⊆C_i⋅|⋅(∀H∈\HH_i)\,|S_i∩\dF(H)|{=}1⋅⎬$.  Call each $S_i⋅∈⋅\Ss_i$ a (pure) {\em strategy} for player $i$.\footnote{Remark [2] after SF Proposition~2.1 shows how \ct{NCF} allows vacuous players, that is, players $i$ for which $C_i=∅$.  If $i$ is vacuous, $\Ss_i = ⎨∅⎬$.  In other words, the only strategy of a vacuous player is $∅$.}  Thus a strategy for player $i$ names exactly one feasible choice at each of the player's information sets.  The following proposition shows that there is a straightforward bijection between grand strategies $S⋅∈⋅\Ss$ and strategy profiles $(S_i)_{i∈I}⋅∈⋅∏_{i∈I}S_i$.  Grand strategies economize on notation by not referring to players.

\begin{prop}\label{9072} Suppose $Φ$ is a form.  Then $\Ss⋅∋⋅S \mapsto (S∩C_i)_{i∈I}⋅∈⋅∏_{i∈I}\Ss_i$ is a bijection.  Its inverse is $\Ss⋅∋⋅∪_{i∈I}S_i$ $\mapsfrom$ $(S_i)_{i∈I}⋅∈⋅∏_{i∈I}\Ss_i$.  (Proof~\rf{9072p}.) \end{prop}

All the above are summarized in Table~\rf{8935}'s first three sections.  (By assumption, each form $Φ$ is implicitly accompanied by all its components and derivatives.)  Figure~\rf{9720} continues Figure~\rf{9719}'s example for those learning game theory.\footnote{\label{9723}Figure~\rf{9720}'s form $(I,T,(C_i)_{i∈I},⊗)$ is defined by defining $T$ and $⊗$ as in note \rf{9722}, by defining $I = ⎨\py1,\py2,\py3⎬$, and by defining $C_{\py1} = ⎨\fa,\fb⎬$, $C_{\py2} = ⎨\fg,\fd⎬$, and $C_{\py3} = ⎨\fe,\ff⎬$.  Note $X_{\py1} = ⎨\f0⎬$, $X_{\py2} = ⎨\f1⎬$, and $X_{\py3} = ⎨\f3,\f4⎬$.  Also, $\HH_{\py1} = ⎨⎨\f0⎬⎬$, $\HH_{\py2} = ⎨⎨\f1⎬⎬$, and $\HH_{\py3} = ⎨⎨\f3,\f4⎬⎬$.  Finally, $\Ss_{\py1} = ⎨⎨\fa⎬,⎨\fb⎬⎬$, $\Ss_{\py2} = ⎨⎨\fg⎬,⎨\fd⎬⎬$, and $\Ss_{\py3} = ⎨⎨\fe⎬,⎨\ff⎬⎬$.}

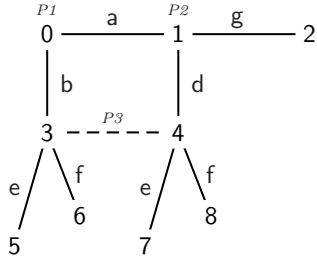
\renewcommand{\capp}{The form $(I,T,(C_i)_{i∈I},⊗)$ under Figure~\rf{9259}(a)'s game.  The form assigns choices, and hence decision nodes and information sets, to players.  Details in note~\rf{9723}.  (Story in note~\rf{9263}.)}
 \begin{figure}[h]
  \newcommand{\hgth}{95}  
  \begin{picture}(0,\hgth) 
  \put(-102,-12){\scalebox{.96}{ 
    \begin{pspicture}(4,1)(12,-5)
      \end{pspicture}
    }} \end{picture}
  \caption{\small \capp} \label{9720} 
  \end{figure} 

A {\em form morphism} (SF Section 2.2)\footnote{The definition here is equivalent to the definition of an \ct{NCF} morphism in SF Section 2.2: \rf{f1}'s last two statements and \rf{f2} are together identical to SF [FM1]; \rf{f1}'s first statement is identical to SF [FM2]; and \rf{f3} is identical to SF [FM3].} is a quintuple $[Φ,Φ′,ι,τ,δ]$ such that $Φ$ and $Φ′$ are forms,\begin{gather}
\zz
ι{:}I→I′,⋅τ{:}T→T′,⋅δ{:}C→C′, \ttt{[f1]}{f1}\\
⎨\,(τ(t),δ(c),τ(t\sh))\,|\,(t,c,t\sh)∈⊗\gr\,⎬⋅⊆⋅⊗^{\prime\mathsf{gr}},⋅\text{and} \ttt{[f2]}{f2}\\
(∀i∈I)⋅\dd{δ}(C_i)⋅⊆⋅C′_{ι(i)}. \ttt{[f3]}{f3}
\zz
\end{gather} The category \ct{NCF} (SF Section 2.3) has forms as its objects and form morphisms as its arrows.  SF Theorem 2.7 shows there is a ``forgetful'' functor $\PB$ from \ct{NCF} to \ct{NCP}. See Figure~\rf{8307} below.  

Let a {\em form inclusion} be an \ct{NCF} morphism that satisfies $I\,⊆\,I′$, $ι\,{=}\,\inc_{I,I′}$, $T⋅⊆⋅T′$, $τ\,{=}\,\inc_{T,T′}$, $C\,⊆\,C′$, and $δ\,{=}\,\inc_{C,C′}$.  Further, let a {\em subform inclusion} be a form inclusion that also satisfies $T\,{=}\,⎨t′∈T′|t^o≼′t′⎬$ and $\HH\,⊆\,\HH′$.  Finally call $Φ$ a {\em subform} of $Φ′$ iff $[Φ,Φ′,\inc_{I,I′},\inc_{T,T′},\inc_{C,C′}]$ is a subform inclusion.  Lemma~\rf{9174} shows that $Φ$ is a subform of $Φ′$ iff it satisfies five conditions which are consistent with the standard concept of a subgame in Selten 1975 (see note~\rf{9453} for a trivial difference).

\section{The Category of Node-and-Choice Games}\label{9272}
\markb{\sc \rf{9272}. The Category of Node-and-Choice Games}

\ssec{Objects}\label{9442}

A {\em (node-and-choice) game} is a quintuple $Γ = (I,T,(C_i)_{i∈I},⊗,(U_i)_{i∈I})$ such that \begin{gather}
\zz
(I,T,(C_i)_{i∈I},⊗)⋅\text{is an (\ct{NCF}) form and} \ttt{[G1]}{G1}\\
(∀i∈I)⋅U_i⋅\text{is a surjective real-valued function from}⋅\ZZ. \ttt{[G2]}{G2} 
\zz
\end{gather} Condition \rf{G1} defines the game's form $Φ = (I,T,(C_i)_{i∈I},⊗)$, which in turn defines the game's $C$, $Π$, $F$, $t^o$, $p$, $q$, $X$, $(X_i)_{i∈I}$, $\HH$, $(\HH_i)_{i∈I}$, $k$, $≺$, $≼$, $\ZZ$, $\ZZf$, $\ZZi$, $P$, $\Ss$, $(\Ss_i)_{i∈I}$, and $ζ$ by means of Section~2.  All these components and derivatives are listed in Table~\rf{8935}, and by assumption, each game $Γ$ is implicitly accompanied by all of them.  Thus the symbol $\ZZ$ appearing in \rf{G2} is the game's collection of plays.     

At the same time, the symbol $U_i$ appearing in \rf{G2} is new.  Call $U_i$ the {\em utility function} of player $i$.  In accord with the explicit notation of Section~2.1, $\dU_i(\ZZ)$ is the range of $U_i$.   The surjectivity in \rf{G2} requires that the codomain of $U_i$ is $\dU_i(\ZZ)$.  Call $\dU_i(\ZZ)$ the set of player-$i$ {\em utilities} $u_i$.  For those learning game theory, the utility functions for Figure~\rf{9259}(a)'s game are defined in a footnote.\footnote{\label{9724}Figure~\rf{9259}(a)'s game $(I,T,(C_i)_{i∈I},⊗,(U_i)_{i∈I})$ is defined by defining $(I,T,(C_i)_{i∈I},⊗)$ as in note \rf{9723}, and by defining $(U_i)_{i∈I}$ to coincide with Figure~\rf{9259}(a)'s utility vectors.  Thus $U_{\py1}{:}\ZZ→Ṛ$ is defined by $U_{\py1}(⎨\f0,\f3,\f5⎬) = 1$, by $U_{\py1}(⎨\f0,\f3,\f6⎬) = U_{\py1}(⎨\f0,\f1,\f4,\f7⎬) = U_{\py1}(⎨\f0,\f1,\f2⎬) = 0$, and by $U_{\py1}(⎨\f0,\f1,\f4,\f8⎬) = -1$.  Similarly, $U_{\py2}{:}\ZZ→Ṛ$ is defined by  $U_{\py2}(⎨\f0,\f1,\f4,\f7⎬) = U_{\py2}(⎨\f0,\f1,\f4,\f8⎬) = 1$, and by $U_{\py2}(⎨\f0,\f3,\f5⎬) = U_{\py2}(⎨\f0,\f3,\f6⎬) = U_{\py2}(⎨\f0,\f1,\f2⎬) = 0$.  Finally, $U_{\py3}{:}\ZZ→Ṛ$ is defined by  $U_{\py3}(⎨\f0,\f3,\f6⎬) = U_{\py3}(⎨\f0,\f1,\f4,\f7⎬) = U_{\py3}(⎨\f0,\f1,\f2⎬) = 1$, and by $U_{\py3}(⎨\f0,\f3,\f5⎬) = U_{\py3}(⎨\f0,\f1,\f4,\f8⎬ = 0$.}

\ssec{Arrows}\label{9443}

A {\em game morphism} is a sextuple $γ = [Γ,Γ′,ι,τ,δ,\vb]$ such that $Γ$ and $Γ′$ are games, \begin{gather}
\zz
[Φ,Φ′,ι,τ,δ]⋅\text{is an (\ct{NCF}) form morphism}, \ttt{[g1]}{g1}\\
\vb = (⋅β_i : \dU_i(\ZZ^θ)→\dU′_{ι(i)}(\ZZ′)⋅)_{i∈I}, \ttt{[g2]}{g2}\\
(∀i∈I)⋅β_i⋅\text{is weakly increasing, and} \ttt{[g3]}{g3}\\
(∀i∈I,Z∈\ZZ^θ)⋅β_i(U_i(Z)) = U′_{ι(i)}(P′○τ(t^o)∪\dd{τ}(Z)), \ttt{[g4]}{g4}
\zz
\end{gather} where $θ = [(T,p),(T′,p′),τ]$,\footnote{Equivalently, by \rf{g1} and Figure~\rf{8307}'s functors, $θ = \mathsf{T_1}○\PA([Φ,Φ′,ι,τ,δ])$.} and $\ZZ^θ$ is $θ$'s collection of end-preserved source plays, as defined in Section~\rf{9257}.3.

Essentially, conditions \rf{g2}--\rf{g4} require that each $β_i$ preserves the ordinal content of the source player-$i$ utility function $U_i$ over as many plays $Z$ as possible.  The essential background is Proposition~\rf{8940t}(\rf{8945t}), which shows that the members of $\ZZ^θ$ (that is, the end-preserved source plays) coincide with the source plays $Z$ that are mapped to target plays by the rule $Z \mapsto P′○τ(t^o)∪\dtau(Z)$.  Condition \rf{g2} states that the domain of each $β_i$ consists of the player-$i$ source utilities that are generated by the end-preserved source plays.  These utilities are mapped by $β_i$ to a player-$ι(i)$ target utilities.  Next, condition \rf{g3} requires that each $β_i$ is weakly increasing in the sense that $(∀i∈I,u_i^1∈\dU_i(\ZZ^θ),u_i^2∈\dU_i(\ZZ^θ))$\begin{gather}
\zz
u_i^1⋅≥⋅u_i^2⋅⋅{⇒}⋅⋅β_i(u_i^1)⋅≥⋅β_i(u_i^2). \notag
\zz
\end{gather} Finally, conditions \rf{g2}--\rf{g4} imply that $(∀i∈I,Z^1∈\ZZ^θ,Z^2∈\ZZ^θ)$ \begin{gather}
\zz
U_i(Z^1)\,≥\,U_i(Z^2)⋅⋅{⇒}⋅⋅U′_{ι(i)}(P′○τ(t^o)∪\dtau(Z^1))\,≥\,U′_{ι(i)}(P′○τ(t^o)∪\dtau(Z^2)).\notag
\zz
\end{gather} 

In contrast, \rf{g2}--\rf{g4} say nothing about the source utility of a (non-end-preserved) source play in $\ZZ⧷\ZZ^θ$.  Arguably, such a source utility should say nothing about target utilities.  For example, recall Figure~\rf{9020}.  There, $⎨\f1,\f4⎬$ is in $\ZZ⧷\ZZ^θ$.  Arguably, the source utility assigned to the source play $⎨\f1,\f4⎬$ should imply nothing about the target utility assigned to the target play $⎨\f{10},\f{11},\f{14},\f{15}⎬$ because the connection between $⎨\f1,\f4⎬$ and $⎨\f{10},\f{11},\f{14},\f{15}⎬$ is so nebulous.

Finally, \rf{g2} and \rf{g4} can often be simplified. Specifically, assume \rf{g1}.  Then $\ZZ^θ$ simplifies to $\ZZ$ if either (i) there are no finite source plays, or (ii) $θ$ is a subtree inclusion, or (iii) $θ$ is an isomorphism (all via Proposition~\rf{9043}).  Further, $P′○τ(t^o)∪\dtau(Z)$ simplifies to $\dtau(Z)$ if $θ$ is an isomorphism (via Proposition~\rf{9043}(c)[2]).

\ssec{The category itself}\label{8637}

This paragraph, the next two paragraphs, and Theorem~\rf{8231} define the category \ct{NCG}, which is called the {\em category of node-and-choice games}.  Let an object be a game $Γ$.  Let an arrow be a game morphism $γ = [Γ,Γ′,ι,τ,δ,\vb]$.  Let source, target, and identity be \begin{gather}
\zz
[Γ,Γ′,ι,τ,δ,\vb]^\src = Γ, \nt
[Γ,Γ′,ι,τ,δ,\vb]^\trg = Γ′,⋅\text{and} \nt
\id_Γ = [Γ,Γ,\id_I,\id_T,\id_C,(\id_{\dU_i(\ZZ)})_{i∈I}], \notag 
\zz
\end{gather} where each member of $⎨\id_I,\id_T,\id_C⎬\,∪\,⎨\id_{\dU_i(\ZZ)}|i∈I⎬$ is an identity in \ct{Set}. 

For composition, consider $γ = [Γ,Γ′,ι,τ,δ,\vb]$ and $γ′ = [Γ′,Γ″,ι′,τ′,δ′,\vb′]$ along with $θ = [(T,p),(T′,p′),τ]$ and \rule[-.9ex]{0ex}{3.2ex} $θ′ = [(T′,p′),(T″,p″),τ′]$.  Firstly, for each player $i$, define $β′_{ι(i)}{*}β_i$ : $\overline{β\T_i}(\dU′_{ι(i)}(\ZZ^{\prime\,θ′}))$ $→$ $\dU″_{ι′○ι(i)}(\ZZ″)$ by\begin{gather}
\zz
β′_{ι(i)}{*}β_i(u_i) = β′_{ι(i)}(β_i(u_i)). \notag
\zz
\end{gather} Straightforwardly, the codomain equals the codomain of $β′_{ι(i)}$ by \rf{g2} for $γ′$ at $i′ = ι(i)$.  Less clearly, the domain is the set of $u_i$ for which the righthand side is well-defined.  To spell this out, note that $\dU′_{ι(i)}(\ZZ^{\prime\,θ′})$ is the domain of $β′_{ι(i)}$ by \rf{g2} for $γ′$ at $i′ = ι(i)$.  Thus $\overline{β\T_i}(\dU′_{ι(i)}(\ZZ^{\prime\,θ′}))$ consists of the arguments of $β_i$ which $β_i$ maps to an argument of $β′_{ι(i)}$.  This complexity arises because [a] $\dU′_{ι(i)}(\ZZ′)$ is the codomain of $β_i$ by \rf{g2}, and [b] this may be larger than $\dU′_{ι(i)}(\ZZ^{\prime\,θ′})$ when $\ZZ′$ is larger than $\ZZ^{\prime\,θ′}$. 

Secondly and lastly, define\begin{gather}
\zz
γ′○γ = [Γ,Γ″,ι′○ι,τ′○τ,δ′○δ,(β′_{ι(i)}{*}β_i|_{\dU_i(\ZZ^{θ′○θ})})_{i∈I}],\notag
\zz
\end{gather} where $\ZZ^{θ′○θ}$ is $θ′○θ$'s collection of end-preserved plays, as defined in Section \rf{9257}.3.  Intuitively, $\ZZ^{θ′○θ}$ consists of the plays from $(T,p)$ that are ``twice end-preserved''.  If $\ZZ^{\prime\,θ′} = \ZZ′$ (that is, if the second morphism is end-preserving), then $\ZZ^{θ′○θ} = \ZZ^θ$ (Lemma~\rf{9313}(\rf{9352})), and further, for each player $i$ \begin{gather}
\zz
β′_{ι(i)}{*}β_i|_{\dU_i(\ZZ^{θ′○θ})} = β′_{ι(i)}○β_i \notag
\zz
\end{gather} (Lemma~\rf{9350}(b)).  This important special case arises when $(T′,p′)$ has only infinite plays, or when $θ′$ is a subtree inclusion, or when $θ′$ is an isomorphism (all by Proposition~\rf{9043} applied to $θ′$). 

\begin{nthm}\label{8231} \ct{NCG} is a category. (Proof~\rf{8231p}.) \end{nthm}

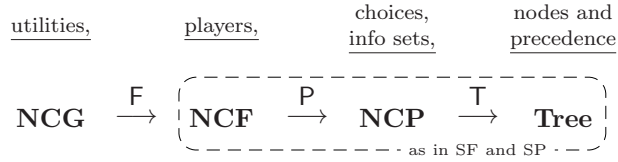
\renewcommand{\capp}{The ``forgetful'' functors of Theorem~\rf{8647}, SF Theorem~2.7, and SP Theorem 3.9 (there $\mathsf{T}$ is denoted ``$\FB$'').}
\begin{figure}[h]
  \newcommand{\hgth}{59}  
  \begin{picture}(0,\hgth) 
  \put(-133,-5){\scalebox{1}{  
    \begin{pspicture}(-3,-1)(5,4) 
      \end{pspicture}
    }} 
    \end{picture}
  \caption{\small \capp} \label{8307} 
  \end{figure} 

As already seen, the definition of a game incorporates an \ct{NCF} form, and the definition of a game morphism incorporates an \ct{NCF} morphism.  Correspondingly, Theorem~\rf{8647} shows there is a ``forgetful'' functor $\FB$ from \ct{NCG} to \ct{NCF}.  Similarly, SF Theorem~2.7 shows there is a functor $\mathsf{P}$ from \ct{NCF} to \ct{NCP}, and SP Theorem~3.9 shows there is a functor $\TB$ from \ct{NCP} to \ct{Tree} (there $\TB$ is denoted ``$\FB$'').

All three functors are shown in Figure~\rf{8307}.  Three of the figure's column headers end in commas.  This conveys the notion that each column subsumes any column(s) to its right.  In particular, the category (\ct{NCG}) in the first column concerns not only utilities, but also players, choices, information sets, nodes, and precedence.  Similarly, the category (\ct{NCF}) in the second column concerns not only players, but also choices, information sets, nodes, and precedence.  Finally, the category (\ct{NCP}) in the third column concerns not only choices and information sets, but also nodes and precedence.

\begin{nthm}\label{8647} Define $\FB$ from \ct{NCG} to \ct{NCF} by\begin{gather}
\zz
\FO⋅{:}⋅(I,T,(C_i)_{i∈I},⊗,(U_i)_{i∈I}) \mapsto (I,T,(C_i)_{i∈I},⊗)⋅\text{and} \nt
\FA⋅{:}⋅[Γ,Γ′,ι,τ,δ,\vb] \mapsto [\mathsf{F_0}(Γ),\mathsf{F_0}(Γ′),ι,τ,δ]. \notag
\zz
\end{gather} Then $\FB$ is a well-defined functor.  (Proof~\rf{8647p}.) \end{nthm}

\ssec{Subgames}\label{9444}

Let a {\em game inclusion} be a morphism $[Γ,Γ′,ι,τ,δ,\vb]$ which satisfies \begin{gather}
\zz
I\,⊆\,I′,⋅ι\,{=}\,\inc_{I,I′},⋅T\,⊆\,T′,⋅τ\,{=}\,\inc_{T,T′},⋅C\,⊆\,C′,⋅δ\,{=}\,\inc_{C,C′},⋅\text{and} \nt 
(∀i∈I)⋅\text{both}⋅\dU_i(\ZZ^θ)\,⊆\,\dU′_i(\ZZ′)⋅\text{and}⋅β_i = \inc_{\dU_i(\ZZ^θ),\dU′_i(\ZZ′)}, \notag
\zz
\end{gather} where $θ = [(T,p),(T′,p′),τ]$.  Further, let a {\em subgame inclusion} be a game inclusion which also satisfies  $T\,{=}\,⎨t′∈T′|t^o≼′t′⎬$ and $\HH\,⊆\,\HH′$.  In this important special case, $θ$ is a subtree inclusion [easily, by Lemmas \rf{9221}(a$⇒$b), \rf{9174}(a$⇒$b), and \rf{9173}(a$⇒$b)], and thus $\ZZ^θ = \ZZ$ [by Proposition~\rf{9043}(\rf{9046})].  Thus, without loss of generality, a subgame inclusion has the form \begin{gather}
\zz
[Γ,Γ′,\inc_{I,I′},\inc_{T,T′},\inc_{C,C′},(\inc_{\dU_i(\ZZ),\dU′_i(\ZZ′)})_{i∈I}]. \notag
\zz
\end{gather} Call $Γ$ a {\em subgame} of $Γ′$ iff $[Γ,Γ′,\inc_{I,I′},\inc_{T,T′},$ $\inc_{C,C′},(\inc_{\dU_i(\ZZ),\dU′_i(\ZZ′)})_{i∈I}]$ is a subgame inclusion.

Proposition~\rf{9171} characterizes a subgame without reference to morphisms.  The characterization's six conditions are consistent with the standard concept of a subgame in Selten 1975 (see note~\rf{9453} for a trivial difference about the minimal number of nodes).

\begin{prop}\label{9171} Suppose $Γ$ and $Γ′$ are \ct{NCG} games.  Then $Γ$ is a subgame of $Γ′$ iff [1] $I\,⊆\,I′$, [2] $T\,{=}\,⎨t′∈T′|t^o≼′t′⎬$, [3] $(∀i∈I)\,C_i\,⊆\,C′_i$, [4] $⊗\,{=}\,⊗′\sj_{F\gr}$, [5]~$\HH\,⊆\,\HH′$, and [6] $(∀i∈I,Z∈\ZZ)$ $U_i(Z) = U′_i(P′(t^o)∪Z)$. (Lemma~\rf{9221}(a$⟺$c).) \end{prop}

\ssec{Isomorphisms}\label{9445}

\begin{nthm}\label{8248} Suppose $γ = [Γ,Γ′,ι,τ,δ,\vb]$ is an \ct{NCG} morphism.  Then the following hold.
\par(a) $γ$ is an isomorphism iff every member of $⎨ι,τ,δ⎬∪⎨β_i|i∈I⎬$ is a bijection.
\par(b) If $γ$ is an isomorphism, then $γ^{-1} = [Γ′,Γ,ι^{-1},τ^{-1},δ^{-1},(β^{-1}_{ι^{-1}(i′)})_{i′∈I′}]$.\linebreak (Proof~\rf{8248p}.) \end{nthm}

The following corollary expresses Theorem~\rf{8248}(a)'s characterization in alternative ways.  In parts (\rf{9128}) and (\rf{9129}), $β_i$ is {\em strictly increasing} iff $(∀u_i^1∈\dU_i(\ZZ^θ),u_i^2∈\dU_i(\ZZ^θ))$ $u_i^1 > u_i^2$ implies $β_i(u_i^1) > β_i(u_i^2)$.

\begin{ncrly}\label{8636} Suppose $γ = [Γ,Γ′,ι,τ,δ,\vb]$ is an \ct{NCG} morphism.  Then the following are equivalent. \begin{tlist}
\yl{9126} $γ$ is an isomorphism.
\yl{9128} $ι$, $τ$, and $δ$ are bijections and $(∀i∈I)$ $β_i$ is strictly increasing.
\yl{9127} $[Φ,Φ′,ι,τ,δ]$ is an \ct{NCF} isomorphism and $(∀i∈I)$ $β_i$ is a bijection.
\yl{9129} $[Φ,Φ′,ι,τ,δ]$ is an \ct{NCF} isomorphism and $(∀i∈I)$ $β_i$ is strictly increasing. \end{tlist} \end{ncrly}

\begin{pf} {\em (\rf{9126})$⟺$(\rf{9128})}. By Theorem~\rf{8248}(a), it suffices to show that each $β_i$ is bijective iff it is strictly increasing.  This equivalence holds because each $β_i$ is weakly increasing by \rf{g3}.  {\em (\rf{9126})$⟺$(\rf{9127})}.  Theorem~\rf{8248}(a) and SF Theorem 2.4(a).  {\em (\rf{9128})$⟺$(\rf{9129})}. SF Theorem 2.4(a). \end{pf}

Proposition~\rf{9334} lists several properties of game isomorphisms.  Parts (\rf{9429})--(\rf{9428}) are new.  Meanwhile, parts (\rf{9434}) and (\rf{9425}) are derived, via Figure~\rf{8307}'s functors, from the properties of tree isomorphisms in Propositions \rf{9050} and \rf{9043}(\rf{9045}).  Further, nineteen additional properties of game isomorphisms can be similarly derived, via Figure~\rf{8307}'s functor $\FB$, from the nineteen properties of form isomorphisms in SF Proposition 2.6.

\begin{prop}\label{9334} Suppose $[Γ,Γ′,ι,τ,δ,\vb]$ is an \ct{NCG} isomorphism.  Then the following hold. \begin{tlist}
\yl{9434} $\dtau\sj_{\ZZ}$ is a bijection from $\ZZ$ onto $\ZZ′$.
\yl{9425} [1] $\ZZ^θ = \ZZ$ where $θ = [(T,p),(T′,p′),τ]$, and [2] $P′○τ(t^o) = ∅$.
\yl{9429} $(∀H∈\HH)$ $δ\sj_{\dF(H)}$ is a bijection from $\dF(H)$ onto $\dF′(\dtau(H))$.
\yl{9433}\vspace{-.5mm} $(∀i∈I)$ $\dd{δ}\sj_{\Ss_i}$ is a bijection from $\Ss_i$ onto $\Ss′_{ι(i)}$.
\yl{9430}\vspace{-.5mm} $\dd{δ}\sj_{\Ss}$ is a bijection from $\Ss$ onto $\Ss′$.
\yl{9432} $\dtau\sj_{\ZZ}○ζ = ζ′○\dd{δ}\sj_{\Ss}$.
\yl{9427} $(∀i∈I)$ $β_i$ is a strictly increasing bijection from $\dU_i(\ZZ)$ onto $\dU′_{ι(i)}(\ZZ′)$.
\yl{9428}\vspace{-.5mm} $(∀i∈I)$ $β_i○U_i = U′_{ι(i)}○\dtau\sj_{\ZZ}$. (Proof~\rf{9334p}.) \end{tlist} \end{prop}

\newcommand{\ba}{^*}
\newcommand{\bb}{^{**}}
\newcommand{\bc}{^{***}}
\ssec{Nash-equilibria}\label{9446}

\newcommand{\notealt}{\footnote{\label{9756}It must be shown that, for all $S⋅∈⋅\Ss$, \rf{NE} holds iff \rf{NE*} holds.  Toward that end, take $S⋅∈⋅\Ss$.  The forward direction is obvious.  For the reverse direction, suppose \rf{NE*} and take $i⋅∈⋅I$.  Because of \rf{NE*}, it suffices to show that \rf{NE}'s inequality holds at $S^+_i = S∩C_i$.  This inequality holds with equality because its right-hand side at $S^+_i = S∩C_i$ is $U_i○ζ((S⧷C_i)∪(S∩C_i)) = U_i○ζ(S)$.}} %

Consider an \ct{NCG} game $Γ$.  A {\em Nash-equilibrium} is an $S⋅∈⋅\Ss$ such that\begin{gather}
\zz
(∀i∈I,S^+_i∈\Ss_i)⋅U_i○ζ(S)⋅≥⋅U_i○ζ((S⧷C_i)∪S^+_i). \ttt{[NE]}{NE}
\zz
\end{gather}To explore this definition, take $S⋅∈⋅\Ss$ and $i⋅∈⋅I$.  Then Proposition~\rf{9072} implies that $S^+_i⋅∈⋅\Ss_i$ iff $(S⧷C_i)∩S^+_i⋅∈⋅\Ss$.  Thus a Nash-equilibrium is a grand strategy $S$ such that no player can benefit by altering their component of $S$.  

Alternatively, it can be shown{\notealt} that a Nash-equilibrium is an $S⋅∈⋅\Ss$ such that\begin{gather}
\zz
(∀i∈I,S^+_i∈\Ss_i⧷⎨S∩C_i⎬)⋅U_i○ζ(S)⋅≥⋅U_i○ζ((S⧷C_i)∪S^+_i). \ttt{[NE*]}{NE*} \notag
\zz
\end{gather} To explore this characterization, take $S⋅∈⋅\Ss$ and $i⋅∈⋅I$.  Proposition~\rf{9072} implies that one of player-$i$'s strategies is the strategy $S∩C_i$ that is included in $S$ itself.  Straightforwardly, the inequality holds with equality at $S^+_i = S∩C_i$.  Thus it suffices to consider the other strategies, that is, the $S^+_i⋅∈⋅\Ss_i⧷⎨S∩C_i⎬$.  These other strategies can be called player $i$'s ``deviations'' from $S∩C_i$.  Thus a Nash-equilibrium can be understood as a grand strategy from which no player has a beneficial deviation.  For those learning game theory, $⎨\fb,\fd,\ff⎬$ is a Nash-equilibrium in Figure~\rf{9259}(a)'s game.\footnote{\label{9725}Recall that note~\rf{9724} defined Figure~\rf{9259}(a)'s game $(I,T,(C_i)_{i∈I},⊗,(U_i)_{i∈I})$, that note~\rf{9723} derived the player strategy collections $\Ss_{\py1} = ⎨⎨\fa⎬,⎨\fb⎬⎬$, $\Ss_{\py2} = ⎨⎨\fg⎬,⎨\fd⎬⎬$, and $\Ss_{\py3} = ⎨⎨\fe⎬,⎨\ff⎬⎬$, and that note~\rf{9722} derived the grand-strategy collection $\Ss$.  Note that $⎨\fb,\fd,\ff⎬⋅∈⋅\Ss$.  To see that $S = ⎨\fb,\fd,\ff⎬$ satisfies \rf{NE*}, begin by observing that \begin{gather}
\zz
\mqi \text{[1]} & U_{\py1}○ζ(⎨\fb,\fd,\ff⎬) = U_{\py1}(⎨\f0,\f3,\f6⎬) = 0⋅⋅⋅≥⋅⋅{-}1 = U_{\py1}(⎨\f0,\f1,\f4,\f8⎬) = U_{\py1}○ζ(⎨\fa,\fd,\ff⎬),⋅⋅  \mqo \nt
\mqi \text{[2]} & U_{\py2}○ζ(⎨\fb,\fd,\ff⎬) = U_{\py2}(⎨\f0,\f3,\f6⎬) = 0⋅⋅⋅≥⋅⋅⋅0 = U_{\py2}(⎨\f0,\f3,\f6⎬) = U_{\py2}○ζ(⎨\fb,\fg,\ff⎬),⋅\text{and} \mqo \nt
\mqi \text{[3]} & U_{\py3}○ζ(⎨\fb,\fd,\ff⎬) = U_{\py3}(⎨\f0,\f3,\f6⎬) = 1⋅⋅⋅≥⋅⋅⋅0 = U_{\py3}(⎨\f0,\f3,\f5⎬) = U_{\py3}○ζ(⎨\fb,\fd,\fe⎬).⋅⋅⋅⋅⋅⋅ \mqo \notag
\zz
\end{gather} Then, for $i = \py1$, $S∩C_{\py1} = ⎨\fb,\fd,\ff⎬∩⎨\fa,\fb⎬ = ⎨\fb⎬$, so $\Ss_{\py1}⧷⎨S∩C_{\py1}⎬ = ⎨⎨\fa⎬,⎨\fb⎬⎬⧷⎨⎨\fb⎬⎬ = ⎨⎨\fa⎬⎬$.  \rf{NE*} holds at $S^+_{\py1} = ⎨\fa⎬$ by [1].  Similarly, for $i = \py2$, $S∩C_{\py2} = ⎨\fb,\fd,\ff⎬∩⎨\fg,\fd⎬ = ⎨\fd⎬$, so $\Ss_{\py2}⧷⎨S∩C_{\py2}⎬ = ⎨⎨\fg⎬,⎨\fd⎬⎬⧷⎨⎨\fd⎬⎬ = ⎨\fg⎬$.  \rf{NE*} holds at $S^+_{\py2} = ⎨\fg⎬$ by [2].  Finally, for $i = \py3$, $S∩C_{\py3} = ⎨\fb,\fd,\ff⎬∩⎨\fe,\ff⎬ = ⎨\ff⎬$, so $\Ss_{\py3}⧷⎨S∩C_{\py3}⎬ = ⎨⎨\fe⎬,⎨\ff⎬⎬⧷⎨⎨\ff⎬⎬ = ⎨\fe⎬$.  \rf{NE*} holds at $S^+_{\py3} = ⎨\fe⎬$ by [3].  (Incidentally, the only other Nash-equilibrium is $⎨\fb,\fg,\ff⎬$.)} 

\begin{prop}\label{8261} Suppose $[Γ,Γ′,ι,τ,δ,\vb]$ is an \ct{NCG} isomorphism.  Then $(∀S∈\Ss)$ $S$ is a Nash-equilibrium in $Γ$ iff $\bar{δ}(S)$ is a Nash-equilibrium in $Γ′$. \linebreak (Proof \rf{8261p}.) \end{prop}

\section{Some Full Subcategories}\label{9087} \markb{\sc \rf{9087}. Some Full Subcategories} 

\ssec{Isomorphic Inclusion}\label{9277}

Isomorphic
\xin{%%%%%%%%%%%%%%%%%%%%%%%%%%%%%%%%%%%%%%%%%%%%%%%%%%%%%%%%
\begin{picture}(0,0)
\put(35,56){\color{white} \rule{40ex}{3ex}}
\put(68,60){\sc \SMALL 4. Some Full Subcategories}
\end{picture}}%%%%%%%%%%%%%%%%%%%%%%%%%%%%%%%%%%%%%%%%%%%%%%%
inclusion (SF Sections 3.2 and 3.3)\footnote{In SF, isomorphic inclusion is ``isomorphic enclosure''.  Further, [a] \ie{⊆} is ``\ud{→}'', [b] \ie{\simeq} is ``\ud{⟷}'', and [c] \ie{⊋} is ``$\sencloses$''.} is a means of comparing full subcategories.  It is recapitulated here with slight modifications.  In general, consider two full subcategories \ct{A} and \ct{B} of some overarching category \ct{Z}.  Then \ct{A} is {\em isomorphically included} in \ct{B} (in symbols, \ct{A} \ie{⊆} \ct{B}) iff every object of \ct{A} is isomorphic (in \ct{Z}) to an object of \ct{B}.  Conveniently, isomorphic inclusions can be composed in the sense that \ct{A} \ie{⊆} \ct{B} and \ct{B} \ie{⊆} \ct{C} imply \ct{A} \ie{⊆} \ct{C}.

In addition, let \ct{A} \ie{\simeq} \ct{B} mean that both \ct{A} \ie{⊆} \ct{B} and \ct{A} \ie{⊇} \ct{B} hold.  Call \ie{\simeq} {\em isomorphic equivalence}.  Isomorphic equivalence implies the standard categorical concept of equivalence in Mac Lane 1998, page 18.  Clearly isomorphic equivalences can be composed.  Further, the following proposition shows that isomorphic inclusions and equivalences can be restricted by isomorphically invariant properties.\footnote{A subscript on a category name refers to an isomorphically invariant property.  One example is the ``\,\ct{_r}\,'' appearing in the following propositions.  Examples in the next subsection are ``\,\ct{_{\ga}}\,'' and ``\,\ct{_p}\,''.}  

\begin{prop}\label{9109} Suppose that \ct{A} and \ct{B} are full subcategories of \ct{Z}, and that $r$ is an invariant property defined for the objects of \ct{Z}.  Let \ct{A_r} be the full subcategory of \ct{A} for objects satisfying $r$, and let \ct{B_r} be the full subcategory of \ct{B} for objects satisfying $r$.  Then (a) \ct{A} \ie{⊆} \ct{B} implies \ct{A_r} \ie{⊆} \ct{B_r}.  Further (b) \ct{A} \ie{\simeq} \ct{B} implies \ct{A_r} \ie{\simeq} \ct{B_r}.  ((a) is SF Proposition 3.6.  (b) holds by two applications of (a).) \end{prop}

Finally, let \ct{A} \ie{⊊} \ct{B} mean that \ct{A} \ie{⊆} \ct{B} but not \ct{A} \ie{⊇} \ct{B}.  Call \ie{⊊} {\em strict isomorphic inclusion}.  Because isomorphic inclusions can be composed, \ct{A} \ie{⊊} \ct{B} and \ct{B} \ie{⊆} \ct{C} together imply \ct{A} \ie{⊊} \ct{C}.\footnote{\label{9088}The proof is straightforward.  Suppose \ct{A} \ie{⊊} \ct{B} and \ct{B} \ie{⊆} \ct{C}.  Then \ct{A} \ie{⊆} \ct{C} holds by composition.  To show \ct{A} \ie{⊇} \ct{C} is false, suppose it were true.  Then \ct{B} \ie{⊆} \ct{C} and composition would imply \ct{A} \ie{⊇} \ct{B}, which was assumed to be false.} Similarly, \ct{B} \ie{⊆} \ct{C} and \ct{C} \ie{⊊} \ct{D} together imply \ct{B}\,\,\ie{⊊}\,\,\ct{D}.  Further, the following shows that strict isomorphic inclusions can be readily constructed.  In this result, a property defined on the objects of \ct{Z} is said to be {\em nontrivial} iff there is an object of \ct{Z} which violates it.

\begin{prop}\label{9110} Suppose that \ct{Z} is a category, and that $r$ is a nontrivial invariant property defined on the objects of \ct{Z}.  Let \ct{Z_r} be the full subcategory of \ct{Z} for objects satisfying $r$.  Then \ct{Z_r} \ie{⊊} \ct{Z}. (Corollary of SF Proposition 3.4.)\end{prop}

\ssec{No-absentmindedness and perfect-information}\label{9278}

An \ct{NCF} form $Φ$ is said to have {\em no-absentmindedness} (Piccione and Rubinstein 1997;\nocite{PiRu97a} SF Section 2.4) iff $(∄H∈\HH,$ $t^A∈H,$ $t^B∈H)$ $t^A\,≺\,t^B$.  An \ct{NCG} game $Γ$ is said to have {\em no-absentmindedness} iff its form has no-absentmindedness.  Finally, let \ct{NCG_\ga} be the full \ct{NCG} subcategory for games with no-absentmindedness.

\begin{ncrly}\label{8268} (a) No-absentmindedness is isomorphically invariant in \ct{NCG}. (b) \ct{NCG_\ga} \ie{⊊} \ct{NCG}. \end{ncrly} 

\begin{pf} (a) follows from SF Proposition 2.8(b) and Figure~\rf{8307}'s ``forgetful'' functor $\FB$.  (b) holds by part (a), by Proposition~\rf{9110} at \ct{Z} =\ct{NCG}, and by the existence of a game which violates no-absentmindedness.\footnote{\label{9108}SF note 13 defines a form $Φ$ which violates no-absentmindedness.  Note $\ZZ$ is the collection consisting of $⎨⎨⎬,(\fb)⎬$, $⎨⎨⎬,(\fa),(\fa,\fb)⎬$, and $⎨⎨⎬,(\fa),(\fa,\fa)⎬$.  $Φ$ can be extended to a game by letting $U_1$ be any surjective real-valued function over $\ZZ$.} \end{pf}

An \ct{NCF} form $Φ$ is said to have {\em perfect-information} (Myerson 1991, page 185; SF Section 2.4) iff $(∀H∈\HH)$ $|H| = 1$.  An \ct{NCG} game $Γ$ is said to have {\em perfect-information} iff its form has perfect-information.  Finally, let \ct{NCG_p} be the full \ct{NCG} subcategory for games with perfect-information.

\begin{ncrly}\label{8269} (a) Perfect-information is isomorphically invariant in \ct{NCG}.  (b) \ct{NCG_p} \ie{⊊} \ct{NCG_\ga}. \end{ncrly} 

\begin{pf} (a) follows from SF Proposition 2.9(b) and Figure~\rf{8307}'s ``forgetful'' functor $\FB$.  For (b), note [1] \ct{NCG_{\ga p}} \ie{⊊} \ct{NCG_\ga} by part (a), by Proposition~\rf{9110} at \ct{Z} = \ct{NCG_\ga}, and by the existence of a no-absentminded game which violates perfect-information.\footnote{\label{9107}SF note 15 defines a form $Φ$ which satisfies no-absentmindedness but not perfect-information.  Note $\ZZ$ is the collection consisting of $⎨⎨⎬,(\fa),(\fa,\fc)⎬$, $⎨⎨⎬,(\fa),(\fa,\fd)⎬$, $⎨⎨⎬,(\fb),(\fb,\fc)⎬$, and $⎨⎨⎬,(\fb),(\fb,\fd)⎬$.  $Φ$ can be extended to a game by letting $U_1$ and $U_2$ be any surjective real-valued functions over~$\ZZ$.}  Also note [2] \ct{NCG_{\ga p}} = \ct{NCG_p} because perfect-information is stronger than no-absentmindedness (SF note 14).  [1] and [2] imply \ct{NCG_p} \ie{⊊} \ct{NCG_\ga}. \end{pf}

Corollaries~\rf{8268}(b) and \rf{8269}(b) contribute to game theory.  The first formalizes the notion that no-absentmindedness is a ``substantial'' property, and the second formalizes the notion that perfect-information is ``substantially'' stronger than no-absentmindedness.  In contrast, Corollary~\rf{8271} will show that the property of using choice-sequences is ``insubstantial'', and similarly, Corollary~\rf{8310} will show a strong sense in which the property of using choice-sets is ``insubstantial''.

\ssec{Choice-sequence games}\label{9279}

An \ct{NCF} form $Φ$ is said to {\em use choice-sequences} (SF Section~3.1) iff \begin{gather}
\zz
\mqi\text{[Csq1]} & T⋅\text{is a collection of finite sequences which contains}⋅⎨⎬⋅\text{and} \mqo\nt
\mqi\text{[Csq2]} & (∀\,(t,c,t\sh)\,∈\,⊗\gr)⋅t±(c) = t\sh,\mqo\notag
\zz
\end{gather} where $t±(c)$ is the concatenation of the sequence $t$ with the one-element sequence $(c)$.  Equivalently, the form is said to be a {\em choice-sequence} \ct{NCF} form. 

An \ct{NCG} game $Γ$ is said to {\em use choice-sequences} iff its form uses choice-{\linebreak}sequences.  Let \ct{CsqG} be the full \ct{NCG} subcategory for choice-sequence games.  Theorem~\rf{8270} shows that using choice-sequences is not restrictive in the sense that \ct{NCG} \ie{⊆} \ct{CsqG}. 

\begin{nthm}\label{8270} \ct{NCG} \ie{⊆} \ct{CsqG}.  In particular, suppose $Γ$ is an \ct{NCG} game.  Define $\zT = ⎨\,(q(p^{k(t)-ℓ}(t)))^{k(t)}_{ℓ=1}\,|\,t∈T\,⎬$, define $\ztau{:}T→\check{T}$ by $\check{τ}(t) = (q(p^{k(t)-ℓ}(t)))^{k(t)}_{ℓ=1}$, and define $\check{⊗}$ by surjectivity and \begin{gather}
\zz
\check{⊗}\gr = ⎨\,(\check{τ}(t),c,\check{τ}(t\sh))\,|\,(t,c,t\sh)∈⊗\gr\,⎬.\notag
\zz
\end{gather} Then (a) $(I,\zT,(C_i)_{i∈I},\check{⊗})$ is a form and $\bar{\ztau}\sj_{\ZZ}{:}\ZZ→\zZZ$ is a bijection.  Further, at each $i\,∈\,I$, define  $\check{U}_i = U_i○(\bar{\ztau}\sj_{\ZZ})^{-1}$.  Then (b) $\check{Γ} = (I,\check{T},(C_i)_{i∈I},\check{⊗},(\check{U}_i)_{i∈I})$ is a \ct{CsqG} game, and (c) $[Γ,\check{Γ},\id_I,\check{τ},\id_C,$ $\!(\id_{\dU_i(\ZZ)})_{i∈I}]$ is an \ct{NCG} isomorphism. \end{nthm}

\begin{pf} Let \ct{CsqF} be the full \ct{NCF} subcategory for choice-sequence forms.  SF Theorem 3.2(b) implies that \lic{9397} $\check{Φ} = (I,\zT,(C_i)_{i∈I},\check{⊗})$ is a \ct{CsqF} form and that \li{9398} $[Φ,\check{Φ},\id_I,\ztau,\id_C]$ is an \ct{NCF} isomorphism (SF uses $\bar{⋅}$ rather than $\check{⋅}$ to suggest ``choice-sequence'').  \rf{9397} and \rf{9398} imply the assumptions of Lemma~\rf{9391} when the lemma's $Γ$ is $Γ$, and the lemma's $[Φ,Φ′,ι,τ,δ]$ is  $[Φ,\check{Φ},\id_I,\ztau,\id_C]$.  Therefore, \rf{9397} and the lemma's (a) imply (a), \rf{9397} and the lemma's (b) imply (b), and the lemma's (c) implies (c).  \end{pf}

\begin{ncrly}\label{8271} \ct{NCG} \ie{\simeq} \ct{CsqG}. \end{ncrly}

\begin{pf} The forward direction follows from Theorem~\rf{8270}.  The reverse direction is immediate since, by definition, each \ct{CsqG} game is an \ct{NCG} game. \end{pf}

In the following corollary, \ct{CsqG_\ga} is the full subcategory for choice-sequence games with no-absentmindedness.

\begin{ncrly}\label{8272} \ct{NCG_\ga} \ie{\simeq} \ct{CsqG_\ga}. \end{ncrly}

\begin{pf} Corollary~\rf{8268}(a) shows no-absentmindedness is invariant.  Thus Corollary~\rf{8271} and Proposition~\rf{9109}(b) imply the result. \end{pf}

\renewcommand{\capp}{Two isomorphic equivalences and two strict isomorphic inclusions among \ct{NCG} subcategories (C$=$Corollary).}
\begin{figure}[h]
  \newcommand{\hgth}{65}  
  \begin{picture}(0,\hgth) 
  \put(-72,-5){\scalebox{1}{  
    \begin{pspicture}(-6,-1)(2,5) 
    \end{pspicture}
    }} \end{picture}
  \caption{\small \capp} \label{8297} 
  \end{figure} 

Figure~\rf{8297} illustrates Corollaries \rf{8268}(b), \rf{8271}, and \rf{8272}.  The righthand inclusion  \ct{CsqG_\ga} \ie{⊊} \ct{CsqG} was obtained by composing the other three inclusions (details in note~\rf{9088}).  Similarly, \ct{NCG_\ga} \ie{⊊} \ct{CsqG} could be obtained by composing the lefthand and top inclusions, and \ct{CsqG_\ga} \ie{⊊} \ct{NCG} could be obtained by composing the bottom and lefthand inclusions.  The latter two could have been shown diagonally in Figure~\rf{8297}.

\ssec{Choice-set games}\label{9280}

An \ct{NCF} form $Φ$ is said to {\em use choice-sets} (SF Section 4.1) iff \begin{gather}
\zz
\mqi\text{[Cset1]} & T⋅\text{is a collection of finite sets which contains}⋅⎨⎬⋅\text{and} \mqo\nt
\mqi\text{[Cset2]} & (∀\,(t,c,t\sh)\,∈\,⊗\gr)⋅t∪⎨c⎬ = t\sh.\mqo\notag
\zz
\end{gather} Equivalently, the form is said to be a {\em choice-set} \ct{NCF} form.

An \ct{NCG} game $Γ$ is said to {\em use choice-sets} iff its form uses choice-sets.  Let \ct{CsetG} be the full \ct{NCG} subcategory for choice-set games.  Theorem~\rf{8298} shows that \ct{CsqG_\ga} \ie{⊆} \ct{CsetG}, where \ct{CsetG_{\ga}} is the full \ct{NCG} subcategory for choice-sequence games with no-absentmindedness.  Thus using choice-sets is not restrictive in a qualified sense which excludes absentminded games.  In the theorem statement, $R$ is the function mapping a finite sequence $\zt = (\zt_1,\zt_2,...\zt_{\check{k}(\zt)})$ to its range $⎨\zt_1,\zt_2,...\zt_{\check{k}(\zt)}⎬$.

\begin{nthm}\label{8298} \ct{CsqG_\ga} \ie{⊆} \ct{CsetG}.  In particular, suppose $\check{Γ}$ is a \ct{CsqG_\ga} game.  Define $T = ⎨R(\zt)|\zt∈\zT⎬$, define $τ{:}\zT→T$ by $τ(\zt) = R(\zt)$, and define $⊗$ by surjectivity and \begin{gather}
\zz
⊗\gr = ⎨\,(τ(\zt),\zc,τ(\zt\sh))\,|\,(\zt,\zc,\zt\sh)∈\check{⊗}\gr\,⎬. \notag
\zz
\end{gather} Then (a) $(\zI,T,(\zC_{\zi})_{\zi∈\zI},⊗)$ is a form and $\dtau\sj_{\zZZ}{:}\zZZ→\ZZ$ is a bijection.  Further, at each $\zi⋅∈\zI$, define $U_{\zi} = \zU_{\zi}○(\dtau\sj_{\zZZ})^{-1}$.  Then (b) $Γ = (\zI,T,(\zC_{\zi})_{\zi∈\zI},⊗,(U_{\zi})_{\zi∈\zI})$ is a \ct{CsetG} game, and (c) $[\check{Γ},Γ,\id_{\zI},τ,\id_{\zC},$ $\!(\id_{\bar{\zU}_{\zi}(\check{\ZZ})})_{\zi∈\zI}]$ is an \ct{NCG} isomorphism. \end{nthm}

\begin{pf} Let \ct{CsetF} be the full \ct{NCF} subcategory for choice-set forms.  SF Theorem 4.2(b) implies that \lic{9399} $Φ = (\zI,T,(\zC)_{\zi∈\zI},⊗)$ is a \ct{CsetF} form and that \li{9400} $[\check{Φ},Φ,\id_{\zI},τ,\id_{\zC}]$ is an \ct{NCF} isomorphism (SF Theorem 4.2 uses $\bar{⋅}$ rather than $\check{⋅}$ to suggest ``choice-sequence'', and uses $R|_{\dT}$ rather than $τ$).  \rf{9399} and \rf{9400} imply the assumptions of Lemma~\rf{9391} when the lemma's $Γ$ is $\check{Γ}$, and the lemma's $[Φ,Φ′,ι,τ,δ]$ is $[\check{Φ},Φ,\id_{\zI},τ,\id_{\zC}]$.  Therefore, \rf{9399} and the lemma's (a) implies (a), \rf{9399} and the lemma's (b) imply (b), and the lemma's (c) implies (c). \end{pf}

\begin{ncrly}\label{8310} \ct{CsqG_\ga} \ie{\simeq} \ct{CsetG}. \end{ncrly}

\begin{pf} The forward direction follows from Theorem~\rf{8298}.  The reverse direction holds because [1]  \ct{CsetG} \ie{⊆} \ct{NCG_\ga} since each choice-set game is no-absentminded by SF Proposition~4.1(g), and [2] \ct{NCG_\ga} \ie{\simeq} \ct{CsqG_\ga} by Corollary~\rf{8272}.\footnote{Categorical proofs that compose isomorphic inclusions can save pages of work.  This observation is made carefully in the last paragraph of SF Section 4.2, in connection with the proof of SF Corollary 4.3(b).  That proof closely resembles the proof of Corollary~\rf{8310} here.  Other categorical arguments which compose isomorphic inclusions become routine in the paragraphs discussing Figures \rf{8297} and \rf{8313}.} \end{pf}

In the following corollary, \ct{NCG_p} is the subcategory for games with perfect-information, \ct{CsqG_p} is the subcategory for choice-sequence games with perfect-information, and \ct{CsetG_p} is the subcategory for choice-set games with perfect-information.

\begin{ncrly}\label{8311} \ct{NCG_p} \ie{\simeq} \ct{CsqG_p} \ie{\simeq} \ct{CsetG_p}. \end{ncrly}

\begin{pf} Corollaries \rf{8272} and \rf{8310} imply \ct{NCG_\ga} \ie{\simeq} \ct{CsqG_\ga} \ie{\simeq} \ct{CsetG}.  Thus Proposition \rf{9109}(b) and Corollary \rf{8269}(a) imply \lic{9405} \ct{NCG_{\ga p}} \ie{\simeq} \ct{CsqG_{\ga p}} \ie{\simeq} \ct{CsetG_p}, where \ct{NCG_{\ga p}} is the full subcategory for games with both no-absentmindedness and perfect-information, and \ct{CsqG_{\ga p}} is the full subcategory for choice-sequence games with both no-absentmindedness and perfect-information.  Since no-absentminded\-ness is weaker than perfect-information (SF note 14), \il{9406} \ct{NCG_{\ga p}} = \ct{NCG_p} and{\linebreak} \il{9407} \ct{CsqG_{\ga p}} = \ct{CsqG_p}.  The result follows from \rf{9405}, \rf{9406}, and \rf{9407}. \end{pf}

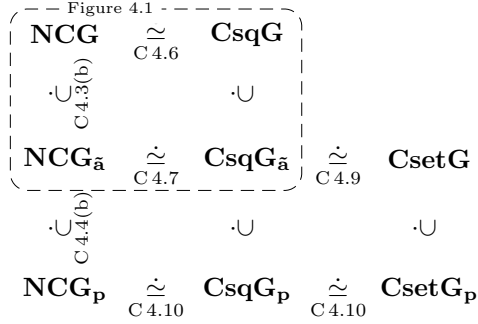
\renewcommand{\capp}{Some isomorphic equivalences and strict isomorphic inclusions among subcategories of \ct{NCG} (C$=$Corollary).}
\begin{figure}[h]
  \newcommand{\hgth}{118}  
  \begin{picture}(0,\hgth) 
  \put(-106,-4){\scalebox{1}{  
    \begin{pspicture}(-6,-5.1)(6,5) 
    \end{pspicture}
    }} \end{picture}
  \caption{\small \capp} \label{8313} 
  \end{figure}

Figure~\rf{8313} illustrates Corollaries \rf{8268}(b), \rf{8269}(b), \rf{8271}, \rf{8272}, \rf{8310}, and \rf{8311}.  By composing various combinations of these corollaries, one can compare any of the figure's eight subcategories with any of the other seven subcategories.  The figure shows three of these many results, namely, the strict inclusions \ct{CsqG_p} \ie{⊊} \ct{CsqG_\ga} \ie{⊊} \ct{CsqG} and the strict inclusion \ct{CsetG_p} \ie{⊊} \ct{CsetG}.

These many comparisons promise to be useful.  For a small example, Streufert 2019 (Section 7) observes that choice-set forms are less general than choice-sequence forms to the extent that the latter can accommodate absentmindedness while the former cannot.  That observation about forms is extended to games by the comparisons \ct{CsetG} \ie{\simeq} \ct{CsqG_{\ga}} \ie{⊊} \ct{CsqG} in the top right quadrant of Figure~\rf{8313}.

\appendix 

\section{For \ct{Tree}}\label{9285} \markb{\sc A. For \ct{Tree}}

\ssec{Basics}\label{9286}

\begin{lemma}\label{8327} Let $(T,p)$ be a (functioned) tree. Then the following hold.\begin{tlist}
\yl{9192} $(∀t∈T)⋅t^o⋅≼⋅t$.
\yl{8980} $(∀t∈T⧷⎨t^o⎬)⋅t^o⋅≺⋅t$.
\yl{8989} $(∀t∈T)⋅t\,∈\,X$ iff $(∃t^*∈T)\,t\,≺\,t^*$.
\yl{8470} $(∀t∈T,t^*∈T,n≥0)⋅t\,{=}\,p^n(t^*)⋅\text{implies}⋅k(t)\,{+}\,n\,{=}\,k(t^*)$.
\yl{8450} $(∀t∈T,m|k(t)≥m≥0,n|k(t)≥n≥0)⋅m\,{=}\,n⋅\text{iff}⋅p^m(t)\,{=}\,p^n(t)$.
\vspace{1.5mm}
\yl{8956} $(∀t∈T)$ $P(t) = ⎨p^m(t)|k(t)≥m{>}0⎬$.
\yl{9401} $(∀t∈T⧷⎨t^o⎬)⋅p(t)⋅∈⋅P(t)$.
\yl{8451} $(∀t∈T)⋅|P(t)| = k(t)$.
\yl{8456} $(∀t∈T)⋅|P(t)∪⎨t⎬| = k(t){+}1$.
\yl{8452} $(∀t∈T)⋅t\,{=}\,t^o$ iff $P(t) = ∅$.
\yl{8453} $(∀S⊆T)⋅∪_{t∈S}P(t) = ⎨p^m(t)|t∈S,k(t)≥m{>}0⎬$.
\vspace{1.5mm}
\yl{8992} $(∀t∈T)$ $P(t)$ is a chain.
\yl{8993} $(∀t∈T)$ $P(t)∪⎨t⎬$ is a chain.
\yl{8947} Suppose $S⊆T$ is a nonempty chain.  Then min\,$S$ exists.
\yl{8948} Suppose $S⊆T$ is a nonempty chain.  Then $S$ is finite iff $\max S$ exists.
\yl{8994} Suppose $S⊆T$ is a finite nonempty chain.  Then $S$ is consecutive iff $S = ⎨\,t∈T\,|\,\min S\,≼\,t\,≼\,\max S\,⎬$. 
\vspace{1.5mm}
\yl{8455} $(∀Z∈\ZZ)⋅Z⋅≠⋅∅$ and $t^o = \min Z$. 
\yl{8459} $(∀Z∈\ZZf)⋅Z = P(\max Z)∪⎨\max Z⎬$.
\yl{8454} $(∀Z∈\ZZf)⋅|Z| = k(\max Z) + 1$.
\yl{8957} $(∀Z∈\ZZ)$ $Z$ is consecutive.
\end{tlist}\end{lemma}
\pagebreak
\begin{pf} {\em(\rf{9192})}.  This follows immediately from \rf{T2} and the definition of $≼$.

{\em(\rf{8980})}.  This follows immediately from \rf{T2} and the definition of $≺$.

{\em(\rf{8989})}.  Take $t⋅∈⋅T$.  For the forward direction, suppose $t⋅∈⋅X$.  Then by \rf{T1} there is $t^*⋅∈⋅T⧷⎨t^o⎬$ such that $t = p(t^*)$.  Thus the definition of $≺$ implies $t⋅≺⋅t^*$.  Conversely, suppose there is $t^*$ such that $t⋅≺⋅t^*$.  Then the definition of $≺$ implies there is $m⋅≥⋅1$ such that $t = p^m(t^*)$.  Hence \rf{T1} implies $t⋅∈⋅X$.

{\em(\rf{8470})}.  Suppose such $t$, $t^*$, and $n$ satisfy \lic{8471} $t = p^n(t^*)$.  By the definition of $k(t)$, \il{8472} $t^o = p^{k(t)}(t)$.  By substitution, \rf{8471} and \rf{8472} imply $t^o = p^{k(t)+n}(t^*)$.  Thus the definition of $k(t^*)$ implies $k(t^*) = k(t){+}n$.

{\em(\rf{8450})}.  Consider such $t$, $m$, and $n$.  The forward direction is immediate.  For the contrapositive of the reverse direction, assume $m⋅≠⋅n$.  Without loss of generality, assume $m > n$.  Then $p^m(t) = p^{m-n}(p^n(t))$.  Thus the definition of $≺$ implies $p^m(t)⋅≺⋅p^n(t)$.  Thus $p^m(t)⋅≠⋅p^n(t)$ (by SP Lemma~A.1(b)).

\myskip

{\em(\rf{8956})}.  Take $t⋅∈⋅T$.  By definition, $P(t) = ⎨t\fl∈T|t\fl≺t⎬$.  Thus the result is equivalent to the statement that $(∀t\fl∈T)$ $t\fl⋅≺⋅t$ iff $(∃m|k(t)≥m{>}0)$ $t\fl = p^m(t)$.  The reverse direction follows from the definition of $≺$.  For the forward direction, suppose $t\fl⋅≺⋅t$.  Then the definition of $≺$ implies there is $m⋅≥⋅1$ such that \lic{8954} $t\fl = p^m(t)$.  Hence it remains to show that $k(t)⋅≥⋅m$.  Toward that end, suppose \il{8955} $m > k(t)$.  Then \rf{8954} implies $p^m(t)$ exists, which by \rf{8955} implies $p^{k(t)+1}(t)$ exists, which by the definition of $k(t)$ implies $p(t^o)$ exists, which contradicts claim [ii] in the second paragraph of SP Section 2.1.

{\em(\rf{9401})}.  Take $t⋅∈⋅T⧷⎨t^o⎬$.  Since $t⋅≠⋅t^o$, $k(t)⋅≠⋅0$.  Thus $p(t)⋅∈⋅P(t)$ by part (\rf{8956}).

{\em(\rf{8451})}.  This follows from part (\rf{8450}) and (\rf{8956}). 

{\em(\rf{8456})}.  This follows from part (\rf{8451}) and the fact that $t⋅∉⋅P(t)$ by inspection.

{\em(\rf{8452})}.  Take $t⋅∈⋅T$.  In steps, $t = t^o$ by the definition of $k$ is equivalent to $k(t) = 0$, which by part (\rf{8451}) is equivalent to $P(t) = ∅$.

{\em(\rf{8453})}.  Take $S⋅⊆⋅T$.  In steps, $∪_{t∈S}P(t)$ by (\rf{8956}) is equal to $∪_{t∈S}⎨p^m(t)|k(t)≥m{>}0⎬$, which by manipulation is equal to $⎨p^m(t)|t∈S,k(t)≥m{>}0⎬$.

\myskip

{\em(\rf{8992})}.  Take $t⋅∈⋅T$, $t^1⋅∈⋅P(t)$, and $t^2⋅∈⋅P(t)$.  Since part (\rf{8956}) implies $P(t) = ⎨p^m(t)|$ $\!k(t)≥m{>}0⎬$, there are $m^1$ and $m^2$ such that \lic{8996} $t^1 = p^{m^1}(t)$ and \il{8997} $t^2 = p^{m^2}(t)$.  Without loss of generality, assume \il{8995} $m^1⋅≥⋅m^2$.  Then $p^{m^1}(t) = p^{m^1-m^2}○p^{m^2}(t)$.  Thus, \rf{8996} and \rf{8997} imply $t^1 = p^{m^1-m^2}(t^2)$.  Thus, \rf{8995} and the definition of $≼$ imply $t^1⋅≼⋅t^2$.

{\em(\rf{8993})}.  Take $t⋅∈⋅T$.  Since $P(t)$ is a chain by part (\rf{8992}), it suffices to show that $(∀t\fl∈P(t))$ $t\fl⋅≺⋅t$.  This follows immediately from the definition of $P$.

{\em(\rf{8947})}.  Take $t^1⋅∈⋅S$.  By part~(\rf{8451}), $P(t^1)$ is finite.  Thus [1] $t^* =$ min\,$P(t^1)∩S$ exists.  It suffices to show that $(∀t^2∈S)⋅t^*⋅≼⋅t^2$.  Toward that end, take $t^2⋅∈⋅S$.  Then since $S$ is a chain and $t^1$ is also in $S$, either [A] $t^2⋅≺⋅t^1$ or [B] $t^1⋅≼⋅t^2$.  Suppose [A].  Then $t^2⋅∈⋅P(t^1)∩S$, which by [1] implies $t^*⋅≼⋅t^2$.  Suppose [B].  Then [1] implies $t^*⋅≼⋅t^1$.  This and [B] imply $t^*⋅≼⋅t^2$. 

{\em(\rf{8948})}.  The forward direction is immediate.  For the reverse direction, suppose $t^{**} =$ max\,$S$.  Then $S⋅⊆⋅P(t^{**})∪⎨t^{**}⎬$.  This superset is finite by part~(\rf{8456}). 

{\em(\rf{8994})}.  Since $S$ is finite and nonempty, parts (\rf{8947}) and (\rf{8948}) imply $\min S$ and $\max S$ exist.  For the forward direction of the equivalence, suppose \lic{8998} $S$ is consecutive.  Then the $⊆$ half of the equality follows from the definitions of min and max.  For the $⊇$ half of the equality, take any $t$ such that $\min S⋅≼⋅t⋅≼⋅\max S$.  On the one hand, if $t$ equals $\min S$ or $\max S$, the definitions of min and max imply $t⋅∈⋅S$.  On the other hand, if $\min S⋅≺⋅t⋅≺⋅\max S$, then \rf{8998} implies $t⋅∈⋅S$.  For the reverse direction, suppose \il{8999} $S = ⎨\,t∈T\,|\,\min S\,≼\,t\,≼\,\max S\,⎬$.  Then suppose \il{9000} $s^1⋅∈⋅S$, \il{9001} $s^2⋅∈⋅S$, and $t⋅∈⋅T$ satisfy $s^1⋅≺⋅t⋅≺⋅s^2$.  This, \rf{9000}, and \rf{9001} imply $\min S⋅≺⋅t⋅≺⋅\max S$. Hence \rf{8999} implies $t⋅∈⋅S$.

\myskip

{\em(\rf{8455})}.  Take $Z⋅∈⋅\ZZ$.  Since $Z⋅⊆⋅T$, part (\rf{9192}) implies \lic{8985} $(∀t∈Z)⋅t^o⋅≼⋅t$.  Thus since $Z$ is a chain, $⎨t^o⎬∪Z$ is a chain.  Thus since $Z$ is a maximal chain, \li{9283} $t^o⋅∈⋅Z$.  \rf{9283} implies $Z⋅≠⋅∅$.  Further, \rf{9283} and \rf{8985} imply $t^o = \text{min}\,Z$.

{\em(\rf{8459})}.  Take $Z⋅∈⋅\ZZf$.  To set up, note part (\rf{8455}) implies $Z⋅≠⋅∅$.  Thus the finiteness of $Z$ and part (\rf{8948}) imply max\,$Z$ exists.  

To show $⊆$, take \lic{9408} $t^*⋅∈⋅Z$ and suppose \li{9409} $t^*⋅∉⋅P(\max Z)∪⎨\max Z⎬$.  \rf{9409} implies \li{9410} $t^* \not\preccurlyeq \max Z$.  Yet, because $Z$ is a chain and $\max Z⋅∈⋅Z$, \rf{9408} and \rf{9410} implies $t^*⋅≻⋅\max Z$.  This contradicts \rf{9408}.  

To show $⊇$, consider $⋅P(\max Z)∪⎨\max Z⎬$.  Since $\max Z⋅∈⋅Z$, it suffices to show $P(\max Z)⋅⊆⋅Z$.  Toward that end, take $t^*⋅∈⋅P(\max Z)$.  Then part (\rf{8956}) implies there is $m$ such that $k(\max Z)⋅≥⋅m > 0$ and $t^* = p^m(\max Z)$.  Thus SP Lemma~A.1(h) at $t = \max Z$ implies $t^*⋅∈⋅Z$.
 
{\em(\rf{8454})}.  Take $Z⋅∈⋅\ZZf$.  Part~(\rf{8459}) implies that $|Z|$ equals $|P(\text{max}\,Z)∪⎨\text{max}\,Z⎬|$, which by part (\rf{8456}) equals $k(\text{max}\,Z){+}1$.

{\em(\rf{8957})}.  Take $Z⋅∈⋅\ZZ$.  Then take $t^1⋅∈⋅Z$, $t^2⋅∈⋅Z$, and $t⋅∈⋅T$ such that $t^1⋅≺⋅t⋅≺⋅t^2$.  The definition of $≺$ and $t⋅≺⋅t^2$ implies there is $m⋅≥⋅1$ such that $t = p^m(t^2)$.  Thus SP Lemma A.1(h) implies $t⋅∈⋅Z$.  (The existence of $t^1$ is superfluous.)  \end{pf}

\begin{lemma}\label{8449} Suppose $(T,p)$ is a (functioned) tree.  Define \begin{gather}
\zz
\WWt = ∪_{\dv≥1}⎨⋅(t^v)^{\dv}_{v=1}∈T^{\dv}⋅|⋅t^o{=}p(t^1),⋅(∀v|2≤v≤\dv)\,t^{v-1}{=}p(t^v)⋅⎬. \notag
\zz
\end{gather} Also define $w{:}T⧷⎨t^o⎬→\WWt$ by $w(t) = (p^{k(t)-v}(t))^{k(t)}_{v=1}$.  Then (a) $w$ is a well-defined bijection, and (b) its inverse is \begin{gather}
\zz
T⧷⎨t^o⎬⋅∋⋅t^{\dv}⋅\mapsfrom⋅(t^v)^{\dv}_{v=1}⋅∈⋅\WWt.\notag
\zz
\end{gather} \end{lemma}

\begin{pf} Call the purported inverse $f$.  It suffices to show that [1] $(∀t∈T⧷⎨t^o⎬)$ $f○w(t) = t$ and [2] $(∀(t^v)^{\dv}_{v=1}∈\WWt)$ $w○f((t^v)^{\dv}_{v=1}) = (t^v)^{\dv}_{v=1}$.

For [1], take $t⋅∈⋅T⧷⎨t^o⎬$.  First, it must be shown that $w(t)⋅∈⋅\WWt$.  By the definition of $w$, it suffices to show $t^o = p(p^{k(t)-1})$ and $(∀v|2≤v≤\dv)$ $p^{k(t)-(v-1)}(t) = p(p^{k(t)-v}(t))$.  The former holds by the definition of $k$, the latter holds by manipulation.  Second, it must be shown that $f○w(t) = t$.  In steps, $f○w(t)$ by the definition of $w$ is equal to $f((p^{k(t)-v}(t))^{k(t)}_{v=1})$, which by the definition of $f$ is equal to $p^{k(t)-v}(t)|_{v=k(t)}$, which is equal to $t$.

For [2], take $(t^v)^{\dv}_{v=1}⋅∈⋅\WWt$.  Note membership in $\WWt$ implies \ilc{9411} $\dv⋅≥⋅1$, \il{8473} $t^o = p^{\dv}(t^{\dv})$ and \il{8474} $(∀v|1≤v≤\dv)$ $t^v = p^{\dv-v}(t^{\dv})$.  Also note \rf{8473} and the definition of $k$ imply \il{8475} $k(t^{\dv}) = \dv$.  Also note the definition of $f$ implies \il{9412} $f((t^v)^{\dv}_{v=1}) = t^{\dv}$.  

First, it must be shown that $f((t^v)^{\dv}_{v=1})⋅∈⋅T⧷⎨t^o⎬$.  \rf{8475} and \rf{9411} imply $k(t^{\dv})⋅≥⋅1$.  Thus the definition of $k$ implies $t^{\dv}⋅∈⋅T⧷⎨t^o⎬$.  Thus \rf{9412} implies $f((t^v)^{\dv}_{v=1})⋅∈⋅T⧷⎨t^o⎬$.  Second, it must be shown that $w○f((t^v)^{\dv}_{v=1}) = (t^v)^{\dv}_{v=1}$.  In steps, $w○f((t^v)^{\dv}_{v=1})$ by \rf{9412} equals $w(t^{\dv})$, which by the definition of $w$ equals $(p^{k(t^{\dv})-v}(t^{\dv}))^{k(t^{\dv})}_{v=1}$, which by \rf{8475} equals $(p^{\dv-v}(t^{\dv}))^{\dv}_{v=1}$, which by \rf{8474} equals $(t^v)^{\dv}_{v=1}$. \end{pf}

\pagebreak
\begin{lemma}\label{8421} Suppose $(T,p)$ is a (functioned) tree.  Define \begin{gather}
\zz
\WWf = ∪_{\dv≥1}⎨⋅(t^v)^{\dv}_{v=1}∈T^{\dv}⋅|⋅t^o{=}p(t^1),⋅(∀v|2≤v≤\dv)\,t^{v-1}{=}p(t^v),⋅t^{\dv}∉X⋅⎬.\notag
\zz
\end{gather}  Also define $\Ef{:}\ZZf→\WWf$ by $\Ef(Z) = (t^v)^{|Z|-1}_{v=1}$, where each $t^v$ is the unique $t⋅∈⋅Z$ such that $k(t) = v$.  Then (a) $\Ef$ is a well-defined bijection, and (b) its inverse is \begin{gather}
\zz
\ZZf⋅∋⋅⎨t^o⎬∪⎨t^v|1≤v≤\dv⎬⋅\mapsfrom⋅(t^v)^{\dv}_{v=1}⋅∈⋅\WWf⋅. \notag
\zz
\end{gather} \end{lemma}

\begin{pf} SP Proposition 2.2(a) implies that the following is a well-defined inverse pair: \begin{gather}
\zz
\ZZf⋅∋⋅Z⋅⋅\mapsto⋅⋅\text{max}\,Z⋅∈⋅T⧷X⋅⋅\text{and} \nt 
\ZZf⋅∋⋅⎨p^m(t)|k(t)≥m≥0⎬⋅⋅\mapsfrom⋅⋅t⋅∈⋅T⧷X. \notag 
\zz
\end{gather} Also, note that $T⧷X⋅⊆⋅T⧷⎨t^o⎬$ because $t^o⋅∈⋅X$ by claim [iv] in the paragraph following SP equation (1).  Thus Lemma~\rf{8449}'s function $w$ can be restricted to $T⧷X$.  Hence, by inspection, Lemma~\rf{8449} implies that the following is a well-defined inverse pair: \begin{gather}
\zz
T⧷X⋅∋⋅t⋅⋅\mapsto⋅⋅(p^{k(t)-v}(t))^{k(t)}_{v=1}⋅∈⋅\WWf⋅⋅\text{and} \nt 
T⧷X⋅∋⋅t^{\dv}⋅⋅\mapsfrom⋅⋅(t^v)^{\dv}_{v=1}⋅∈⋅\WWf. \notag 
\zz
\end{gather} Composing these two inverse pairs implies that the following is a well-defined inverse pair: \begin{gather}
\zz
\ZZf⋅∋⋅Z⋅⋅\mapsto⋅⋅(p^{k(\text{max}\,Z)-v}(\text{max}\,Z))^{k(\text{max}\,Z)}_{v=1}⋅∈⋅\WWf⋅⋅\text{and} \label{8465} \\
\ZZf⋅∋⋅⎨p^m(t^{\dv})|k(t^{\dv})≥m≥0⎬⋅⋅\mapsfrom⋅⋅(t^v)^{\dv}_{v=1}⋅∈⋅\WWf. \label{8466}
\zz
\end{gather} 

For the lemma's part (a), it suffices to prove that (\rf{8465}) coincides with $\Ef$.  By inspection, the domain and codomain of (\rf{8465}) coincide with the domain and codomain of $\Ef$.  Thus it suffices to show that\begin{gather}
\zz
(∀Z∈\ZZf)⋅(p^{k(\text{max}\,Z)-v}(\text{max}\,Z))^{k(\text{max}\,Z)}_{v=1} = (t^v)^{|Z|-1}_{v=1},\notag
\zz
\end{gather} where each $t^v$ is defined as the unique stage-$v$ element of $Z$ (the well-definition of $(t^v)^{|Z|-1}_{v=1}$ is not yet assured).  Toward that end, take $Z⋅∈⋅\ZZf$.  Since $k(\text{max}\,Z) = |Z|{-}1$ by Lemma~\rf{8327}(\rf{8454}), it suffices to show that\begin{gather}
\zz
(∀v|1≤v≤|Z|{-}1)⋅p^{|Z|-1-v}(\text{max}\,Z)⋅\text{is the unique stage-$v$ element of $Z$}.\notag
\zz
\end{gather} Toward that end, take $v$ such that $1⋅≤⋅v⋅≤⋅|Z|{-}1$.  Since $Z$ is a chain by definition, and since distinct nodes in a chain have distinct stages by SP Lemma~A.1(a), it suffices to show that $k(p^{|Z|-1-v}(\text{max}\,Z))= v$.  In steps, $k(p^{|Z|-1-v}(\text{max}\,Z))$ by Lemma~\rf{8327}(\rf{8470}) equals $k(\text{max}\,Z) - (|Z|{-}1{-}v)$, which by Lemma~\rf{8327}(\rf{8454}) equals $(|Z|{-}1) - (|Z|{-}1{-}v)$, which equals $v$.

For the lemma's part (b), it suffices to prove that (\rf{8466}) coincides with the lemma's purported inverse.  By inspection, the domains and codomains coincide.  Thus it suffices to show that\begin{gather}
\zz
(∀(t^v)^{\dv}_{v=1}∈\WWf)⋅⎨p^m(t^{\dv})|k(t^{\dv})≥m≥0⎬ = ⎨t^o⎬∪⎨t^v|1≤v≤\dv⎬.\notag
\zz
\end{gather} Toward that end, take $(t^v)^{\dv}_{v=1}⋅∈⋅\WWf$.  Membership in $\WWf$ implies \ilc{9413} $\dv⋅≥⋅1$, \il{8467} $t^o = p^{\dv}(v^{\dv})$ and \il{8468} $(∀v|1≤v≤\dv)$ $t^v = p^{\dv-v}(t^{\dv})$.  Also, \rf{8467} and the definition of $k$ imply \il{8469} $k(t^{\dv}) = \dv$.  It will be argued that\pagebreak\begin{align}
\zz
&⋅⎨p^m(t^{\dv})|k(t^{\dv})≥m≥0⎬ \nt
=&⋅⎨p^m(t^{\dv})|\dv≥m≥0⎬ \nt
=&⋅⎨p^{\dv}(t^{\dv})⎬⋅∪⋅⎨p^m(t^{\dv})|\dv{-}1≥m≥0⎬ \nt
=&⋅⎨p^{\dv}(t^{\dv})⎬⋅∪⋅⎨p^{\dv-v}(t^{\dv})|\dv{-}1≥\dv{-}v≥0⎬ \nt
=&⋅⎨p^{\dv}(t^{\dv})⎬⋅∪⋅⎨p^{\dv-v}(t^{\dv})|1≤v≤\dv⎬ \nt
=&⋅⎨t^o⎬⋅∪⋅⎨t^v|1≤v≤\dv⎬⋅. \notag
\zz
\end{align} The first equality holds by \rf{8469}, the second holds by \rf{9413}, and the third holds by changing the variable $m$ to $\dv{-}v$.  The fourth holds because $\dv{-}1≥\dv{-}v≥0$ is equivalent to ${-}1≥{-}v≥{-}\dv$, which is equivalent to $1≤v≤\dv$.  The fifth holds by \rf{8467} and \rf{8468}. \end{pf}

\ssec{Subtrees}\label{9447}

\begin{lemma}\label{9022} Let $(T,p)$ be a subtree of $(T′,p′)$. Then the following hold. \begin{tlist}
\yl{9186} $T⧷⎨t^o⎬⋅⊆⋅T′⧷⎨t\po⎬$.
\yl{9188} $p = p′\sj_{T⧷⎨t^o⎬}$.
\yl{9187} $X = X′∩T$.
\yl{9455} $t^o⋅∈⋅X′$.
\yl{9024} $T⧷X⋅⊆⋅T′⧷X′$.
\end{tlist}\end{lemma}

\begin{pf} {\em (\rf{9186})}.  Take \lic{9461} $t⋅∈⋅T⧷⎨t^o⎬$.  By tree inclusion, $T⋅⊆⋅T′$.  Thus, \li{9460} $t^o⋅∈⋅T′$, and also, \rf{9461} implies \il{9457} $t⋅∈⋅T′$.  Further, \rf{9461} and the subtree condition $T\,{=}\,⎨t′∈T′|$ $\!t^o≼′t′⎬$ imply \il{9458} $t^o⋅≺′⋅t$.  Note \rf{9460} and Lemma~\rf{8327}(\rf{9192}) imply \il{9459} $t\po⋅≼′⋅t^o$.  \rf{9458} and \rf{9459} imply $t\po⋅≺′⋅t$.  This and \rf{9457} imply $t⋅∈⋅T′⧷⎨t\po⎬$.

{\em (\rf{9188})}.  \rf{T1} for $(T,p)$ implies that $p$ has domain $T⧷⎨t^o⎬$ and is surjective.  Thus it suffices to show that $p\gr⋅⊆⋅p^{\prime\,\mathsf{gr}}$.  This follows from \rf{t2} for the tree inclusion $[(T,p),(T′,p′),\inc_{T,T′}]$. 

{\em (\rf{9187})}.  For the $⊆$ direction, it suffices to show $X⋅⊆⋅X′$.  It is argued, in steps, that $X$ by \rf{T1} for $(T,p)$ is equal to the range of $p$, which by part (\rf{9188}) is included in the range of $p′$, which by \rf{T2} for $(T′,p′)$ is equal to $X′$.  For the $⊇$ direction, take $t⋅∈⋅X′∩T$.  Since $t⋅∈⋅X′$, there exists $t\ps⋅∈⋅T′$ such that \lic{9198} $t = p′(t\ps)$.  Since $t⋅∈⋅T$, the subtree condition $T = ⎨t′∈T′|t^o≼′t′⎬$ implies \il{9199} $t^o⋅≼′⋅t$.  Since \rf{9198} implies $t⋅≺′⋅t\ps$, \rf{9199} implies $t^o⋅≺′⋅t\ps$.  Thus the subtree condition $T = ⎨t′∈T′|t^o≼′t′⎬$ implies $t\ps⋅∈⋅T⧷⎨t^o⎬$.  This, \rf{9198}, and part (\rf{9188}) imply $t = p(t\ps)$.  This and \rf{T1} for $(T,p)$ imply $t⋅∈⋅X$.  

{\em (\rf{9455})}.  Remark [iv] in the paragraph after SP equation (1) implies $t^o⋅∈⋅X$.  Hence part (\rf{9187}) implies $t^o⋅∈⋅X′$.  

{\em (\rf{9024})}.  It is argued, in steps, that $T⧷X$ by (\rf{9187}) equals $T⧷(X′∩T)$, which by inspection equals $T⧷X′$, which by $T\,⊆\,T′$ is a subset of $T′⧷X′$. \end{pf}

\begin{lemma}\label{9172} Suppose $(T,p)$ and $(T′,p′)$ are trees.  Then $(T,p)$ is a subtree of $(T′,p′)$ iff $T\,{=}\,⎨t′∈T′|t^o≼′t′⎬$ and $p = p′\sj_{T⧷⎨t^o⎬}$. \end{lemma}

\begin{pf} Suppose $(T,p)$ is a subtree of $(T′,p′)$.  The first condition holds by the definition of subtree inclusion.  The second holds by Lemma~\rf{9022}(\rf{9188}). 

Conversely, suppose \lic{9177} $T\,{=}\,⎨t′∈T′|t^o≼t′⎬$ and \il{9178} $p = p′\sj_{T⧷⎨t^o⎬}$.  By the definition of subtree, it suffices to show that $[(T,p),(T′,p′),\inc_{T,T′}]$ is a morphism satisfying $T\,{=}\,⎨t′∈T′|t^o≼t′⎬$.  Thus, by \rf{9177}, it remains to show that the triple $[(T,p),(T′,p′),\inc_{T,T′}]$ is a morphism.  It will be argued that the triple satisfies \rf{t1} and \rf{t2}.  For \rf{t1}, note \rf{9177} implies $T⋅⊆⋅T′$ and thus $\inc_{T,T′}{:}T→T′$ is well-defined.  For \rf{t2}, it suffices to show that $p\gr⋅⊆⋅p\pgr$.  This holds by \rf{9178}. \end{pf}

\begin{lemma}\label{9185} Suppose $(T′,p′)$ is a tree and $t^\star⋅∈⋅X′$.  Define $T = ⎨t′∈T′|t^\star≼′t′⎬$ and $p = p′\sj_{T⧷⎨t^\star⎬}$.  Then $(T,p)$ is a well-defined subtree of $(T′,p′)$, and $t^o = t^\star$.  \end{lemma}

\begin{pf} The lemma follows from Claims \rf{9196} and \rf{9197}.

\begin{cllist}\yl{9027} $T⧷⎨t^\star⎬⋅≠⋅∅$.  Since $t^\star⋅∈⋅X′$, Lemma~\rf{8327}(\rf{8989}) implies there is \lic{9028} $t′⋅∈⋅T′$ such that \li{9029} $t^\star⋅≺⋅t′$.  \rf{9028}, \rf{9029}, and the definition of $T$ imply $t′⋅∈⋅T$.  \rf{9029} implies $t′⋅≠⋅t^\star$.

\yl{9033} $T⧷⎨t^\star⎬⋅⊆⋅T′⧷⎨t\po⎬$.  Take $t⋅∈⋅T⧷⎨t^\star⎬$.  Thus the definition of $T$ implies \lic{9195} $t⋅∈⋅T′$ and \il{9191} $t^\star⋅≺′⋅t$.   Meanwhile, the assumption $t^\star⋅∈⋅X′$ implies $t^\star⋅∈⋅T′$, and thus Lemma~\rf{8327}(\rf{9192}) implies \il{9193} $t\po⋅≼′⋅t^\star$.  \rf{9191} and \rf{9193} imply $t\po⋅≺′⋅t$.  This and \rf{9195} imply $t⋅∈⋅T′⧷⎨t\po⎬$.

\yl{9196} {\em (a) $(T,p)$ is a well-defined tree, and (b) $t^o = t^\star$.}  Note that $p$ is surjective by construction.  Thus it suffices to show \begin{gather}
\zz
\mqi \text{[T1$^\star$]} & p⋅\text{is a nonempty function with domain}⋅T⧷⎨t^\star⎬,⋅\text{and} \mqo \nt
\mqi \text{[T2$^\star$]} & (∀t∈T⧷⎨t^\star⎬)(∃m≥1)⋅p^m(t) = t^\star. \mqo\notag
\zz
\end{gather} First consider [T1$^\star$].  \rf{T1} for $(T′,p′)$ states that $p′$ is a function from $T′⧷⎨t\po⎬$.  This and Claim~\rf{9033} imply that $p = p′\sj_{T⧷⎨t^\star⎬}$ is a well-defined function from $T⧷⎨t^\star⎬$.  Further, Claim~\rf{9027} implies that this function is nonempty.

Second consider [T2$^\star$].  Take $t⋅∈⋅T⧷⎨t^\star⎬$.  This and the definition of $T$ imply $t^\star⋅≺′⋅t$, which by the definition of $≺′$ implies there is $m⋅≥⋅1$ such that \lic{9040} $t^\star = (p′)^m(t)$.   Note that $(∀n|m{>}n≥0)$ $t^\star = (p′)^{m-n}○(p′)^n(t)$.  This and the definition of $≺′$ imply $(∀n|m{>}n≥0)$ $t^\star⋅≺⋅(p′)^n(t)$.  This and the definition of $T$ imply $(∀n|m{>}n≥0)$ $(p′)^n(t)⋅∈⋅T⧷⎨t^\star⎬$.  This and the definition of $p$ implies \il{9042} $(p′)^m(t) = p^m(t)$.  \rf{9040} and \rf{9042} imply $t^\star = p^m(t)$.

\yl{9197} $(T,p)$ {\em is a subtree of} $(T′,p′)$.  This follows from the reverse direction of Lemma~\rf{9172}.  In particular, the lemma's assumptions hold because $(T,p)$ is a tree by Claim~\rf{9196}(a) and because $(T′,p′)$ is a tree by assumption.  The lemma's two conditions are identical to the definitions of $T$ and $p$. \end{cllist} \unskipcl \end{pf}

\ssec{How \ct{Tree} morphisms interact with plays}\label{9287}

\begin{lemma}\label{8940} Suppose $θ = [(T,p),(T′,p′),τ]$ is a tree morphism with its $\ZZ^θ$.  Then the following hold.\begin{tlist}
\vspace{1.5mm}
\yl{8959} If $S⊆T$ is a consecutive chain, then $\dtau(S)$ is a consecutive chain.
\yl{8982} If $S⊆T$ is a nonempty chain, then $τ(\min S) = \min\dtau(S)$.
\yl{8983} If $S⊆T$ is a finite nonempty chain, then $τ(\max S) = \max\dtau(S)$.
\vspace{1.5mm}
\yl{8941} $(∀Z∈\ZZ)$ $\dtau(Z)$ is a consecutive chain.
\yl{8950} $(∀Z∈\ZZ)$ $\min\dtau(Z) = τ(t^o)$.
\yl{8942} $(∀Z∈\ZZ)$ $P′(\min\dtau(Z)) = P′○τ(t^o)$.
\yl{8988} $(∀Z∈\ZZf)$ $\max\dtau(Z) = τ(\max Z)$.
\yl{9312} $(∀Z∈\ZZf)$ $Z⋅∈⋅\ZZ^θ$ iff $τ(\max Z)⋅∉⋅X′$.
\yl{8943} $\ZZ^θ = \ZZi⋅∪⋅⎨\,Z∈\ZZf\,|\,τ(\max Z)\,∉\,X′\,⎬$.
\yl{8944} $\ZZ^θ = \ZZ$\, iff \,$\dtau(T⧷X)⋅⊆⋅T′⧷X′$. 
\vspace{1.5mm}
\yl{9010} $(∀Z∈\ZZi)⋅P′○τ(t^o)∪\dtau(Z)\,∈\,\ZZi′$.
\yl{8984} $(∀Z∈\ZZf)$ $P′○τ(t^o)∪\dtau(Z) = P′(\max\dtau(Z))∪⎨\max\dtau(Z)⎬$.
\yl{9335} $(∀Z∈\ZZf)⋅P′○τ(t^o)∪\dtau(Z)$ is finite.
\yl{8951} $(∀Z∈\ZZf)⋅Z\,∈\,\ZZ^θ$ iff $P′○τ(t^o)∪\dtau(Z)\,∈\,\ZZ′$.
\yl{8945} $\ZZ^θ = ⎨\,Z∈\ZZ\,|\,P′○τ(t^o)∪\dtau(Z)\,∈\,\ZZ′\,⎬$. 
\end{tlist} \end{lemma}

\begin{pf} {\em (\rf{8959})}.  Suppose $S$ is a consecutive chain.  SP Proposition~2.4(f) implies that $\dtau(S)$ is a chain.  Thus it remains to show that $\dtau(S)$ is consecutive.  Toward that end, take \ilc{8960} $t\pA⋅∈⋅\dtau(S)$, \il{8961} $t\pB⋅∈⋅\dtau(S)$, and $t′⋅∈⋅T′$ such that \il{8965} $t\pA⋅≺⋅t′⋅≺⋅t\pB$.  \rf{8960} implies there is \il{8966} $t^A⋅∈⋅S$ such that \il{8971} $t\pA = τ(t^A)$.  \rf{8961} implies there is \il{8967} $t^B⋅∈⋅S$ such that \il{8970} $t\pB = τ(t^B)$.  Finally, \rf{8965} implies there are \il{8963} $m⋅≥⋅1$ and \il{8964} $n⋅≥⋅1$ such that \il{8968} $t\pA = (p′)^m(t′)$ and \il{8969} $t′ = (p′)^n(t\pB)$. 

By substitution, \rf{8968} and \rf{8969} imply \lic{8972} $t\pA = (p′)^{m+n}(t\pB)$.  Thus Lemma~\rf{8327}(\rf{8470}) implies \li{8973} $m + n = k(t^B) - k(t^A)$.  By substitution, \rf{8969} and \rf{8970} imply \li{8974} $t′ = (p′)^n(τ(t^B))$.  Since $n⋅≤⋅k(t^B)$ by \rf{8973}, SP Proposition 2.4(b) implies $(p′)^n(τ(t^B)) = τ(p^n(t^B))$.  This and \rf{8974} imply \li{8975} $t′ = τ(p^n(t^B))$.

By substitution, \rf{8972}, \rf{8971}, and \rf{8970} imply \li{8976} $τ(t^A) = (p′)^{m+n}(τ(t^B))$.  Since $m+n⋅≤⋅k(t^B)$ by \rf{8973}, SP Proposition~2.4(b) implies $(p′)^{m+n}(τ(t^B)) = τ(p^{m+n}(t^B))$.  This and \rf{8976} imply \il{8977} $τ(t^A) = τ(p^{m+n}(t^B))$.  Because $τ|_S$ is injective by SP Proposition~2.4(f), \rf{8977} implies \il{8978} $t^A = p^{m+n}(t^B)$.  

Finally, \rf{8978}, \rf{8963}, and \rf{8964} imply $t^A⋅≺⋅p^n(t^B)⋅≺⋅t^B$.  This, \rf{8966}, \rf{8967}, and the assumed consecutiveness of $S$ imply that $p^n(t^B)⋅∈⋅S$.  This and \rf{8975} imply $t′⋅∈⋅\dtau(S)$.

\lstep{(\rf{8982})}. Suppose $S$ is a nonempty chain.  Lemma~\rf{8327}(\rf{8947}) implies $\min S$ exists.  Similarly, SP Proposition 2.4(f) implies $\dtau(S)$ is a chain, which by Lemma~\rf{8327}(\rf{8947}) implies $\min \dtau(S)$ exists.  Thus it remains to show the relationship between these two minima.  

To begin, $(∀t∈S)⋅\min S⋅≼⋅t$, which by SP Proposition 2.4(e) implies $(∀t∈S)$ $τ(\min S)⋅≼′⋅τ(t)$, which implies \lic{8986} $τ(\min S)⋅≼′⋅\min\dtau(S)$.  Conversely, $\min S⋅∈⋅S$, which implies $τ(\min S)⋅∈⋅\dtau(S)$, which implies \il{8987} $τ(\min S)⋅≽′⋅\min\dtau(S)$.  \rf{8986} and \rf{8987} imply $τ(\min S) = \min\dtau(S)$

\lstep{(\rf{8983})}. Suppose $S$ is a finite nonempty chain.  Lemma~\rf{8327}(\rf{8948}) implies $\max S$ exists.  Inspection shows $\dtau(S)$ is finite and nonempty, and SP Proposition 2.4(f) shows $\dtau(S)$ is a chain.  Thus Lemma~\rf{8327}(\rf{8948}) implies $\max \dtau(S)$ exists.  It remains to show the relationship between these two maxima.  This is done as in the second paragraph of the proof of part (\rf{8982}).

\myskip \lstep{(\rf{8941})}.  This follows from part (\rf{8959}) and Lemma~\rf{8327}(\rf{8957}).   

\lstep{(\rf{8950})}.  Take $Z⋅∈⋅\ZZ$.  In steps, $\min\dtau(S)$ by part (\rf{8982}) is equal to $τ(\min Z)$, which by Lemma~\rf{8327}(\rf{8455}) is equal to $τ(t^o)$.

\lstep{(\rf{8942})}.  This follows immediately from part (\rf{8950}).

\lstep{(\rf{8988})}.  Take $Z⋅∈⋅\ZZf$.  Lemma~\rf{8327}(\rf{8455}) implies $Z$ is nonempty.  Thus part (\rf{8983}) implies $τ(\max Z) = \max\dtau(Z)$.

\lstep{(\rf{9312})}.  Take $Z⋅∈⋅\ZZf$.  In steps, $Z⋅∈⋅\ZZ^θ$ by the definition of $\ZZ^θ$ is equivalent to $(∄t′∈T′)$ \!$τ(\max Z)≺′t′$, which by Lemma~\rf{8327}(\rf{8989}) is equivalent to $τ(\max Z)\,∉\,X′$.

\lstep{(\rf{8943})}.  By definition, $\ZZ^θ$ includes $\ZZi$.  Thus part (\rf{8943}) follows from (\rf{9312}).

\lstep{(\rf{8944})}.  In steps, $\ZZ^θ = \ZZ$ by part (\rf{8943}) is equivalent to $(∄Z∈\ZZf)$ $τ(\max Z)⋅∈⋅X′$, which by SP Proposition 2.2(a) is equivalent to $(∄t∈T⧷X)$ $τ(t)⋅∈⋅X′$, which by \rf{t1} is equivalent to $τ(T⧷X)⋅⊆⋅T′⧷X′$.

\myskip \lstep{(\rf{9010})}.  Take \ilc{9019} $Z⋅∈⋅\ZZi$.  SP Proposition 2.4(f) implies $\dtau(Z)⋅⊆⋅T′$ is an infinite chain.  Thus SP Lemma A.1(f) implies $\dtau(Z)∪⎨(p′)^m(t′)|t′∈\dtau(Z),k′(t′)≥m{>}0⎬$ $∈$ $\ZZi′$.  Hence it suffices to show that \begin{gather}
\zz
\dtau(Z)\,∪\,⎨(p′)^m(t′)|t′∈\dtau(Z),k′(t′)≥m{>}0⎬ = P′○τ(t^o)\,∪\,\dtau(Z). \notag
\zz
\end{gather} 

For the $⊇$ direction, it suffices to show that $P′○τ(t^o)$ is included in the left-hand side.  Note part (\rf{8950}) implies \il{9011} $τ(t^o)⋅∈⋅\dtau(Z)$.  It is then argued that $P′○τ(t^o)$ by Lemma~\rf{8327}(\rf{8956}) is equal to $⎨(p′)^m(τ(t^o))|k′(τ(t^o))≥t′{>}0⎬$, which by \rf{9011} is a subset of $⎨(p′)^m(t′)|t′∈\dtau(Z),k′(t′)≥m{>}0⎬$.

For the $⊆$ direction, it suffices to show that $⎨(p′)^m(t′)|t′∈\dtau(Z),k′(t′)≥m{>}0⎬$ is included in the right-hand side.  Toward that end, take \il{9012} $t′⋅∈⋅\dtau(Z)$ and $m$ such that \il{9013} $k′(t′)⋅≥⋅m > 0$.  \rf{9012} implies there is \il{9014} $t⋅∈⋅Z$ such that \il{9015} $t′ = τ(t)$.  Note \il{9016} $k′(t′) = k′○τ(t^o) + k(t)$, because $k′(t′)$ by \rf{9015} equals $k′○τ(t)$ which by SP Proposition 2.4(c) equals $k′○τ(t^o) + k(t)$.  \rf{9013} and \rf{9016} imply either [1] $k′(t′)⋅≥⋅m > k(t)$ or [2] $k(t)⋅≥⋅m > 0$.

First assume [1].  Note \il{9017} $(p′)^m(t′) = (p′)^{m-k(t)}(τ(t^o))$, because $(p′)^m(t′)$ by \rf{9015} equals $(p′)^m(τ(t))$, which by [1] equals $(p′)^{m-k(t)}((p′)^{k(t)}(τ(t)))$, which by SP Proposition 2.4(b) equals $(p′)^{m-k(t)}(τ(p^{k(t)}(t)))$, which by the definition of $k$ equals $(p′)^{m-k(t)}(τ(t^o))$.  \rf{9017} and [1] imply $(p′)^m(t′)⋅∈⋅P′○τ(t^o)$.

Second assume [2].  Note \il{9018} $(p′)^m(t′) = τ(p^m(t))$, because $(p′)^m(t′)$ by \rf{9015} equals $(p′)^m(τ(t))$, which by SP Proposition 2.4(b) and [2] equals $τ(p^m(t))$.  SP Lemma A.1(h), \rf{9019}, \rf{9014}, and [2] imply $p^m(t)⋅∈⋅Z$.  Thus \rf{9018} implies $(p′)^m(t′)⋅∈⋅\dtau(Z)$.

\lstep{(\rf{8984})}.  Take $Z⋅∈⋅\ZZf$.  Lemma~\rf{8327}(\rf{8455}) shows $Z$ is nonempty.  Thus part (\rf{8941}) implies that $\dtau(Z)$ is a consecutive finite nonempty chain.  Thus Lemma~\rf{8327}(\rf{8994}) implies \ilc{9003} $\dtau(Z) = ⎨\,t′∈T′\,|\,\min\dtau(Z)\,≼\,t′\,≼\,\max\dtau(Z)\,⎬$.  Meanwhile, part (\rf{8942}) implies \il{8990} $P′○τ(t^o) = ⎨\,t′∈T′\,|\,t′\,≺′\,\min\dtau(Z)\,⎬$.  \rf{9003} and \rf{8990} imply $P′○τ(t^o)∪\dtau(Z)$ $=${\linebreak} $⎨\,t′∈T′\,|\,t′\,≼′\,\max\dtau(Z)\,⎬$.

\lstep{(\rf{9335})}. This follows from part (\rf{8984}) and Lemma~\rf{8327}(\rf{8456}).

\lstep{(\rf{8951})}.  Take $Z⋅∈⋅\ZZf$.  Consider the following statements.\begin{gather}
\zz
\mqi \text{[1]} & Z⋅∈⋅\ZZ^θ. \mqo \nt
\mqi \text{[2]} & (∄t^{\prime+}∈T′)\,τ(\max Z)\,≺′\,t^{\prime+}. \mqo \nt
\mqi \text{[3]} & (∄t^{\prime+}∈T′)\,\max\dtau(Z)\,≺′\,t^{\prime+}. \mqo \nt
\mqi \text{[4]} & P′(\max\dtau(Z))∪⎨\max\dtau(Z)⎬⋅∈⋅\ZZ′. \mqo \nt
\mqi \text{[5]} & P′○τ(t^o)∪\dtau(Z)⋅∈⋅\ZZ′. \mqo \notag
\zz
\end{gather} [1]$⟺$[2] by the definition of $\ZZ^θ$.  [2]$⟺$[3] by part (\rf{8988}).  To show the contrapositive of [3]$⇒$[4], suppose [4] is violated.  Then since $P′(\max\dtau(Z))∪⎨\max\dtau(Z)⎬$ is a chain by Lemma~\rf{8327}(\rf{8993}), this chain fails to be maximal.  Thus [3] is violated.  To show the contrapositive of [3]$⇐$[4], suppose [3] is violated.  Then, by inspection, $P′(\max\dtau(Z))∪⎨\max\dtau(Z)⎬$ cannot be a maximal chain.  Thus [4] is violated.  Finally [4]$⟺$[5] by part (\rf{8984}).  

\lstep{(\rf{8945})}.  For $⊆$, suppose $Z⋅∈⋅Z^θ$.  On the one hand, if $Z⋅∈⋅\ZZi$, part (\rf{9010}) implies $P′○τ(t^o)∪\dtau(Z)⋅∈⋅\ZZ′$.  On the other hand, if $Z⋅∈⋅\ZZf$, the forward direction of part (\rf{8951}) implies $P′○τ(t^o)∪\dtau(Z)⋅∈⋅\ZZ′$.  For $⊇$, suppose $Z⋅∈⋅\ZZ$ is such that $P′○τ(t^o)∪\dtau(Z)⋅∈⋅\ZZ′$.  On the one hand, if $Z⋅∈⋅\ZZi$, the definition of $\ZZ^θ$ implies $Z⋅∈⋅\ZZ^θ$.  On the other hand, if $Z⋅∈⋅\ZZf$, the reverse direction of part (\rf{8951}) implies $Z⋅∈⋅Z^θ$. \end{pf}

\begin{npf}[for Proposition~\rf{9043}]\label{9043p} {\em (\rf{9044})}.  Suppose $(T,p)$ has only infinite plays, that is, suppose \lic{9414} $\ZZ = \ZZi$.  The definition of $\ZZ^θ$ implies \il{9415} $\ZZi⋅⊆⋅\ZZ^θ⋅⊆⋅\ZZ$.  \rf{9414} and \rf{9415} imply $\ZZ^θ = \ZZ$. 

{\em (\rf{9046})}.  Suppose $θ = [(T,p),(T′,p′),\inc_{T,T′}]$ is a subtree inclusion.  Then Lemma \rf{9022}(\rf{9024}) implies $T⧷X⋅⊆⋅T′⧷X′$.  Thus $\inc_{T,T′}(T⧷X)⋅⊆⋅T′⧷X′$.  Thus the reverse direction of Lemma~\rf{8940}(\rf{8944}) implies $\ZZ^θ = \ZZ$.

{\em (\rf{9045})}.  Suppose $θ$ is an isomorphism.  

[1]. SP Theorem~2.6 implies $τ$ is a bijection from $T$ onto $T′$, and SP Proposition~2.7(a) implies $τ\sj_X$ is a bijection from $X$ onto $X′$.  Thus $τ\sj_{T⧷X}$ is a bijection from $T⧷X$ onto $T′⧷X′$, which implies a fortiori that $\dtau(T⧷X)⋅⊆⋅T′⧷X′$.  Thus the reverse direction of Lemma~\rf{8940}(\rf{8944}) implies $\ZZ^θ = \ZZ$.

[2]. SP Proposition 2.7(c) implies $τ(t^o) = t\po$.  Thus Lemma~\rf{8327}(\rf{8452}) implies $P′○τ(t^o) = ∅$. \end{npf}

\begin{lemma}\label{8331} Let $[(T,p),(T',p′),τ]$ and $[(T′,p′),(T″,p″),τ′]$ be morphisms.\linebreak  Then $k″○τ′○τ(t^o) = k″○τ′(t\po) + k′○τ(t^o)$. \end{lemma}

\begin{pf} By the definition of $k″$, it suffices to argue \begin{align}
\zz
t^{\prime\prime o} =&⋅[p″]^{k″○τ′(t\po)}(⋅τ′(t\po)⋅) \nt
=&⋅[p″]^{k″○τ′(t\po)}(⋅τ′(⋅[p′]^{k′○τ(t^o)}(⋅τ(t^o)⋅)⋅)⋅) \nt
=&⋅[p″]^{k″○τ′(t\po)}(⋅[p″]^{k′○τ(t^o)}(⋅τ′○τ(t^o)⋅)⋅) \nt
=&⋅[p″]^{k″○τ′(t\po)+k′○τ(t^o)}(⋅τ′○τ(t^o)⋅). \notag
\zz
\end{align} The first equality holds by the definition of $k″$, the second by the definition of $k′$, the third by SP Proposition 2.4(b), and the fourth by rearrangement. \end{pf}

\hyphenation{mor-phism mor-phisms}

\begin{lemma}\label{8335} Suppose $[(T,p),(T',p′),τ]$ and $[(T′,p′),(T″,p″),τ′]$ are morphisms. Then $(∀Z∈\ZZ)$ $P″○τ′○τ(t^o)⋅∪⋅\dtau′○\dtau(Z)$ $=$ $P″○τ′(t\po)⋅∪⋅\dtau′(P′○τ(t^o)∪\dtau(Z))$. \end{lemma}

\begin{pf} Take $Z⋅∈⋅\ZZ$.  First, it is argued that \begin{align}
\zz
&⋅P″○τ′(t\po) \tag{i} \\
=&⋅⎨⋅[p″]^{m″}(⋅τ′(t\po)⋅)⋅|⋅k″○τ′(t\po)≥m″{>}0⋅⎬ \nt
=&⋅⎨⋅[p″]^{m″}(⋅τ′(⋅[p′]^{k′○τ(t^o)}(⋅τ(t^o)⋅)⋅)⋅)⋅|⋅k″○τ′(t\po)≥m″{>}0⋅⎬ \nt 
=&⋅⎨⋅[p″]^{m″}(⋅[p″]^{k′○τ(t^o)}(⋅τ′○τ(t^o)⋅)⋅)⋅|⋅k″○τ′(t\po)≥m″{>}0⋅⎬ \nt 
=&⋅⎨⋅[p″]^{m″+k′○τ(t^o)}(⋅τ′○τ(t^o)⋅)⋅|⋅k″○τ′(t\po)≥m″{>}0⋅⎬ \nt 
=&⋅⎨⋅[p″]^{m^*}(⋅τ′○τ(t^o)⋅)⋅|⋅k″○τ′(t\po){+}k′○τ(t^o)≥m^*{>}k′○τ(t^o)⋅⎬ \nt 
=&⋅⎨⋅[p″]^{m^*}(⋅τ′○τ(t^o)⋅)⋅|⋅k″○τ′○τ(t^o)≥m^*{>}k′○τ(t^o)⋅⎬, \notag
\zz
\end{align} where the first equality holds by Lemma~\rf{8327}(\rf{8956}), the second by the definition of $k′$, the third by SP Proposition~2.4(b), the fourth by rearrangement, the fifth by replacing $m″$ with $m^*{-}k′○τ(t^o)$, and the sixth by Lemma~\rf{8331}.  Second, it is argued that \begin{align}
\zz
&⋅\dtau′(P′○τ(t^o)) \tag{ii} \\
=&⋅\dtau′(⎨⋅[p′]^{m′}(⋅τ(t^o)⋅)⋅|⋅k′○τ(t^o)≥m′{>}0⋅⎬) \nt
=&⋅⎨⋅τ′(⋅[p′]^{m′}(⋅τ(t^o)⋅)⋅)⋅|⋅k′○τ(t^o)≥m′{>}0⋅⎬ \nt
=&⋅⎨⋅[p″]^{m′}(⋅τ′○τ(t^o)⋅)⋅|⋅k′○τ(t^o)≥m′{>}0⋅⎬, \notag
\zz
\end{align} where the first equality holds by Lemma~\rf{8327}(\rf{8956}), the second by rearrangement, and the third by SP Proposition~2.4(b).  Finally, I argue \begin{align}
\zz
&⋅P″○τ′(t\po)⋅∪⋅\dtau′(P′○τ(t^o)∪\dtau(Z)) \nt
=&⋅P″○τ′(t\po)⋅∪⋅\dtau′(P′○τ(t^o))⋅∪⋅\dtau′○\dtau(Z) \nt
=&⋅⎨⋅[p″]^{m^*}(⋅τ′○τ(t^o)⋅)⋅|⋅k″○τ′○τ(t^o)≥m^*{>}k′○τ(t^o)⋅⎬ \nt
 &⋅⋅⋅∪⋅⎨⋅[p″]^{m′}(⋅τ′○τ(t^o)⋅)⋅|⋅k′○τ(t^o)≥m′{>}0⋅⎬⋅∪⋅\dtau′○\dtau(Z) \nt
=&⋅⎨⋅[p″]^{m″}○τ′○τ(t^o)⋅|⋅k″○τ′○τ(t^o)≥m″{>}0⋅⎬⋅∪⋅\dtau′○\dtau(Z) \nt
=&⋅P″○τ′○τ(t^o)⋅∪⋅\dtau′○\dtau(Z), \notag
\zz
\end{align} where the first equality holds by rearrangement, the second by (i) and (ii), the third by rearrangement, and the fourth by Lemma~\rf{8327}(\rf{8956}).  \end{pf}

\begin{lemma}\label{9313} Suppose $θ = [(T,p),(T′,p′),τ]$ and $θ′ = [(T′,p′),(T″,p″),τ′]$ are morphisms.  Then the following hold. \begin{tlist}
\yl{9314} $\ZZ^{θ′○θ}⋅⊆⋅\ZZ^θ$.
\yl{9352} If $\ZZ^{\prime\,θ′} = \ZZ′$, then $\ZZ^{θ′○θ} = \ZZ^θ$.
\yl{9315} $⎨P′○τ(t^o)∪\dtau(Z)|Z∈\ZZ^{θ′○θ}⎬⋅⊆⋅\ZZ^{\prime\,θ′}$.
\end{tlist} \end{lemma}

\begin{pf} {\em (\rf{9314})}. Take \lic{9316} $Z⋅∈⋅\ZZ^{θ′○θ}$.  On the one hand, suppose $Z⋅∈⋅\ZZi$.  Then $Z⋅∈⋅\ZZ^θ$ by the definition of $\ZZ^θ$.  On the other hand, suppose $Z⋅∈⋅\ZZf$.  Then \rf{9316} and the forward direction of Lemma~\rf{8940}(\rf{9312}) imply \li{9317} $τ′○τ(\max Z)⋅∉⋅X″$.  Meanwhile, the second half of SP Proposition~2.3(9a) implies $(∀t′∈T′)⋅t′∈X′⋅⇒⋅τ′(t′)∈X″$.  The contrapositive is \li{9318} $(∀t′∈T′)⋅τ′(t′)∉X″⋅⇒⋅t′∉X′$.  \rf{9317} and \rf{9318} at $t′{=}τ(\max Z)$ imply $τ(\max Z)∉X′$.  Hence the reverse direction of Lemma~\rf{8940}(\rf{9312}) implies $Z⋅∈⋅\ZZ^θ$.

{\em (\rf{9352})}. Suppose \lic{9351} $\ZZ^{\prime\,θ′} = \ZZ′$.  The $⊆$ direction holds by part (\rf{9314}) (independently of \rf{9351}).  For the $⊇$ direction, take \li{9353} $Z⋅∈⋅\ZZ^θ$.  On the one hand, suppose $Z⋅∈⋅\ZZi$.  Then $Z⋅∈⋅\ZZ^{θ′○θ}$ by the definition of $\ZZ^{θ′○θ}$.  On the other hand, suppose $Z⋅∈⋅\ZZf$.  Then \rf{9353} and the forward direction of Lemma~\rf{8940}(\rf{9312}) imply \il{9354} $τ(\max Z)⋅∈⋅T′⧷X′$.  Meanwhile, \rf{9351} and Lemma~\rf{8940}(\rf{8944}) imply \il{9355} $\dtau′(T′⧷X′)⋅⊆⋅T″⧷X″$.  \rf{9354} and \rf{9355} imply $τ′○τ(\max Z)⋅∈⋅T″⧷X″$.  Thus the reverse direction of Lemma~\rf{8940}(\rf{9312}) implies $Z⋅∈⋅\ZZ^{θ′○θ}$.

{\em (\rf{9315})}. Take \lic{9319} $Z⋅∈⋅\ZZ^{θ′○θ}$.  

On the one hand, suppose [A] $Z⋅∈⋅\ZZi$.  Then Lemma~\rf{8940}(\rf{9010}) implies\linebreak $P′○τ(t^o)∪Z⋅∈⋅\ZZi′$.  Thus the definition of $\ZZ^{\prime\,θ′}$ implies $P′○τ(t^o)∪Z⋅∈⋅\ZZ^{\prime\,θ′}$.  (This argument follows from [A] independently of \rf{9319}.)  

On the other hand, suppose [B] $Z⋅∈⋅\ZZf$.  \rf{9319} and part (\rf{9314}) imply $Z⋅∈⋅\ZZ^θ$.  Thus Lemma~\rf{8940}(\rf{8945}) implies $P′○τ(t^o)∪\dtau(Z)⋅∈⋅\ZZ′$.  Thus [B] and Lemma~\rf{8940}(\rf{9335}) imply \li{9322} $P′○τ(t^o)∪\dtau(Z)⋅∈⋅\ZZf′$.  

Further, \rf{9319}, [B], and the forward direction of Lemma~\rf{8940}(\rf{9312}) imply\linebreak \il{9324} $τ′○τ(\max Z)⋅∉⋅X″$.  Also, \begin{center}
\zz
\il{9323}⋅$\max P′○τ(t^o)∪\dtau(Z) = \max \dtau(Z) = τ(\max Z)$, 
\zz
\end{center} where the first equality holds by [B] and Lemma~\rf{8940}(\rf{8984}) and the second holds by [B] and Lemma~\rf{8940}(\rf{8988}).  Thus\begin{center}
\zz
\il{9325}⋅$τ′(\max P′○τ(t^o)∪\dtau(Z)) = τ′○τ(\max Z)⋅∉⋅X″$, 
\zz
\end{center} where the first equality holds by applying $τ′$ to both sides of \rf{9323}, and where the second holds by \rf{9324}.  

Finally, the reverse direction of Lemma~\rf{8940}(\rf{9312}) implies that \begin{center}
\zz
$(∀Z′∈\ZZf′)$ $τ′(\max Z′)\,∉\,X″⋅⇒⋅Z′\,∈\,\ZZ^{\prime\,θ′}$.  
\zz
\end{center} Consider this statement at $Z′ = P′○τ(t^o)∪\dtau(Z)$.  This, \rf{9322}, and \rf{9325} imply\linebreak $P′○τ(t^o)∪\dtau(Z)⋅∈⋅\ZZ^{\prime\,θ′}$.  \end{pf}

\section{For \ct{NCP}}\label{9289}\markb{\sc \rf{9289}. For \ct{NCP}}

\begin{lemma}\label{8401} Suppose $Π$ is an \ct{NCP} preform.  Then $(∀S∈\Ss)$ $(∃!Z∈\ZZ)$\linebreak $(∀t∈Z⧷⎨t^o⎬)$ $q(t)\,∈\,S$. \end{lemma}

\begin{pf} Take $S⋅∈⋅\Ss$. \begin{cllist}

\yl{8405} {\em Define $ε{:}X→T$ at each $t⋅∈⋅X$ by $ε(t) = t⊗c$, where $c$ is the unique element of $S∩F(t)$.  Then (a) $ε$ is well-defined.  Also (b) $(∀t∈X)⋅p(ε(t)) = t$.  Also (c) $(∀t∈X)⋅q(ε(t))⋅∈⋅S$.}  

{\em(a)}. Take $t⋅∈⋅X$.  \rf{P3} implies there is $H⋅∈⋅\HH$ such that $t⋅∈⋅H$.  Hence SP Proposition~3.2(16a) implies $\dF(H) = F(t)$.  Also, the definition of $\Ss$ implies $|S∩\dF(H)| = 1$.  The last two sentences imply $|S∩F(t)| = 1$.  Thus $c$ is well-defined as the set's unique element.  Finally, note $c⋅∈⋅F(t)$ implies $(t,c)⋅∈⋅F\gr$.  Thus $t⊗c$ is well-defined by \rf{P1}.  {\em(b) and (c)}.  Take $t⋅∈⋅X$.  By definition, $ε(t) = t⊗c$, where $c$ is the unique element of $S∩F(t)$. Hence $p(ε(t)) = p(t⊗c) = t$ and $q(ε(t)) = q(t⊗c) = c⋅∈⋅S$, where in both cases, the second equality holds by SP Proposition 3.1(b).

\yl{8406} {\em Recursively define $\rV⋅⊆⋅\tNa$ and $(\rt^v)_{v∈\rV}$ as follows.  [1] Let $\rt^1 = ε(t^o)$.  If $\rt^1⋅∉⋅X$, let $\rV = ⎨1⎬$ and terminate.  Else, [2] let $\rt^2 = ε(\rt^1)$.  If $\rt^2⋅∉⋅X$, let $\rV = ⎨1,2⎬$ and terminate.  Else, [3] let $\rt^3 = ε(\rt^2)$.  If $\rt^3⋅∉⋅X$, let $\rV = ⎨1,2,3⎬$ and terminate.  ...  If the recursion does not terminate (that is, if $(∀v∈\tNa)$ $\rt^v⋅∈⋅X$), let $\rV = \tNa$.  Then (a) $\rV$ and $(\rt^v)_{v∈\rV}$ are well-defined.  Also (b) $t^o = p(\rt^1)$ and $(∀v∈\rV⧷⎨1⎬)⋅\rt^{v-1} = p(\rt^v)$.  Also (c) $(∀v∈\rV)$ $q(\rt^v)⋅∈⋅S$.}  

{\em(a)}. By claim [iv] in the paragraph after SP equation (1), $t^o⋅∈⋅X$.  Thus Claim~\rf{8405}(a) implies $\rt^1$ is well-defined.  For each $v⋅≥⋅2$, step [$v{-}1$] assures $\rt^{v-1}⋅∈⋅X$ and thus Claim~\rf{8405}(a) implies $\rt^v$ is well-defined.  Hence by inspection, $\rV$ and $(\rt^v)_{v∈\rV}$ are well-defined. {\em(b)}.  At $v = 1$, $p(\rt^1)$ by definition equals $p(ε(t^o))$, which by Claim~\rf{8405}(b) equals $t^o$.  At $v⋅∈⋅\rV⧷⎨1⎬$, $p(\rt^v)$ by definition equals $p(ε(\rt^{v-1}))$, which by Claim~\rf{8405}(b) equals $\rt^{v-1}$.  {\em(c)}.   At $v = 1$, $q(\rt^1)$ by definition equals $q(ε(t^o))$, which by Claim~\rf{8405}(c) belongs to $S$.  At $v⋅∈⋅\rV⧷⎨1⎬$, $q(\rt^v)$ by definition equals $q(ε(\rt^{v-1}))$, which by Claim~\rf{8405}(c) belongs to $S$.

\pagebreak\yl{8422} {\em Define $\rZ = ⎨t^o⎬∪⎨\rt^v|v∈\rV⎬$.  Then (a) $\rZ⋅∈⋅\ZZ$ and (b) $(∀t∈\rZ⧷⎨t^o⎬)$ $q(t)⋅∈⋅S$.} 

{\em(a)}. On the one hand, suppose $\rV$ is finite and set $\dv =$ max\,$\rV$.  Then the definitions in Claim~\rf{8406} imply $\rV = ⎨1,2,...\,\dv⎬$ and $\rt^{\dv}⋅∉⋅X$.  Thus Claim~\rf{8406}(b) implies $(\rt^v)_{v∈\rV}⋅∈⋅\WWf$, where $\WWf$ is defined in Lemma~\rf{8421}.  Thus the reverse direction of Lemma~\rf{8421} implies $⎨t^o⎬∪⎨\rt^v|v∈\rV⎬⋅∈⋅\ZZf$.  Thus the definition of $\rZ$ implies $\rZ⋅∈⋅\ZZ$.  On the other hand, suppose $\rV$ is infinite.  Then the definitions in Claim~\rf{8406} imply $\rV = \tNa$.  Thus Claim~\rf{8406}(b) implies $(\rt^v)_{v∈\rV}⋅∈⋅\YY$, where $\YY$ is defined in SP Proposition 2.2(b).  Thus the reverse direction of SP Proposition 2.2(b) implies $⎨t^o⎬∪⎨\rt^v|v∈\rV⎬⋅∈⋅\ZZi$.  Thus the definition of $\rZ$ implies $\rZ⋅∈⋅\ZZ$.  {\em(b)}. Take $t⋅∈⋅\rZ⧷⎨t^o⎬$.  The definition of $\rZ$ implies there is $v⋅∈⋅\rV$ such that $t = \rt^v$.  Thus Claim~\rf{8406}(c) implies $q(t)⋅∈⋅S$.

\yl{8434} {\em Suppose $Z⋅∈⋅\ZZf$ satisfies $(∀t∈Z⧷⎨t^o⎬)⋅q(t)⋅∈⋅S$.  Then (a) $|Z|{-}1⋅≤$ max\,$\rV$ and (b) $(∀v|1≤v≤|Z|{-}1)$ $\Ef(Z)^v = \rt^v$, where $\Ef$ is defined in Lemma~\rf{8421}}. 

Note that $(∀v|1≤v≤|Z|{-}1)$ $\Ef(Z)^v⋅∈⋅Z⧷⎨t^o⎬$ by the definition of $\Ef$.  Thus the claim's assumption implies that \ilc{8429} $(∀v|1≤v≤|Z|{-}1)$ $q(\Ef(Z)^v)⋅∈⋅S$. Also, note that Lemma~\rf{8421} implies $\Ef(Z)⋅∈⋅\WWf$.  Thus \il{8426} $t^o = p(\Ef(Z)^1)$ and \il{8430} $(∀v|2≤v≤|Z|{-}1)$ $\Ef(Z)^{v-1} = p(\Ef(Z)^v)$. 

It suffices to show that $(∀v|1≤v≤|Z|{-}1)$ [1] $v⋅≤$ max\,$\rV$ and [2] $\Ef(Z)^v = \rt^v$.  This will be shown by induction.

For the initial step, suppose $v = 1$.  [1] holds easily by the definition of $\rV$ in Claim~\rf{8406}.  Further, \rf{8426} implies there is \il{8427} $c⋅∈⋅F(t^o)$ such that \il{8428} $t^o⊗c = \Ef(Z)^1$.  \rf{8428} implies $c⋅=⋅q(\Ef(Z)^1)$ and thus \rf{8429} implies \il{8431} $c⋅∈⋅S$.  \rf{8427}--\rf{8431} and the definition of $ε$ in Claim~\rf{8405} imply $\Ef(Z)^1 = ε(t^o)$.  Thus the definition of $\rt^1$ in Claim~\rf{8406} implies $\Ef(Z)^1 = \rt^1$.  Thus [2] holds.

For the inductive step, suppose $v$ satisfies $2⋅≤⋅v < |Z|{-}1$.  Then \rf{8430} and the inductive hypothesis imply \il{8445} $\rt^{v-1} = p(\Ef(Z)^v)$.  \rf{8445} implies $\rt^{v-1}⋅∈⋅X$ and thus [1] holds by the definition of $\rV$ in Claim~\rf{8406}.  Further, \rf{8445} implies there is \il{8432} $c⋅∈⋅F(\rt^{v-1})$ such that \il{8433} $\rt^{v-1}⊗c = \Ef(Z)^v$.  \rf{8433} implies $c⋅=⋅q(\Ef(Z)^v)$ and thus \rf{8429} implies \il{9291} $c⋅∈⋅S$.  \rf{8432}--\rf{9291} and the definition of $ε$ in Claim~\rf{8405} imply $\Ef(Z)^v = ε(\rt^{v-1})$.  Thus the definition of $\rt^v$ in Claim~\rf{8406} implies $\Ef(Z)^v = \rt^v$.  Thus [2] holds. 

\yl{8424} {\em Suppose $Z⋅∈⋅\ZZf$ satisfies $(∀t∈Z⧷⎨t^o⎬)⋅q(t)⋅∈⋅S$.  Then (a) $|Z|{-}1 =$ max\,$\rV$ and (b) $\Ef(Z) = (\rt^v)_{v∈\rV}$.}

Because of Claim~\rf{8434}, it suffices to show that $|Z|{-}1 =$ max\,$\rV$.  Lemma~\rf{8421} implies $\Ef(Z)⋅∈⋅\WWf$.  Thus the definitions of $\Ef$ and $\WWf$ imply $(\Ef(Z)^v)^{|Z|-1}_{v=1}$ $∈$  $∪_{\dv≥1}⎨⋅(t^v)^{\dv}_{v=1}∈T^{\dv}⋅|⋅t^{\dv}∉X⋅⎬$.   Hence $\Ef(Z)^{|Z|-1}⋅∉⋅X$.  Thus Claim~\rf{8434}(b) at $v = |Z|{-}1$ implies $\rt^{|Z|-1}⋅∉⋅X$.  Thus the recursive definition of $(\rt^v)_{v∈\rV}$ in Claim~\rf{8406} terminates at stage [\,$|Z|{-}1$\,].  Thus max\,$\rV$ $= |Z|{-}1$.   

\yl{9292} {\em Suppose $Z⋅∈⋅\ZZf$ satisfies $(∀t∈Z⧷⎨t^o⎬)⋅q(t)⋅∈⋅S$.  Then $Z = \rZ$.}

Claim~\rf{8424} implies, a fortiori, that the range of $\Ef(Z)$ equals the range of $(\rt^v)_{v∈\rV}$.  In other words, \ilc{9293} $⎨\Ef(Z)^v|1≤v≤|Z|{-}1⎬ = ⎨\rt^v|v∈\rV⎬$.  It is then argued, in steps, that $Z$ by Lemma~\rf{8421} is equal to $⎨t^o⎬∪⎨\Ef(Z)^v|1≤v≤|Z|{-}1⎬$, which by [a] is equal to $⎨t^o⎬∪⎨\rt^v|v∈\rV⎬$, which by the definition of $\rZ$ in Claim~\rf{8422} is equal to $\rZ$. 

\yl{8435} {\em Suppose $Z⋅∈⋅\ZZi$ satisfies $(∀t∈Z⧷⎨t^o⎬)⋅q(t)⋅∈⋅S$.  Then (a) $\rV⋅⊇⋅\tNa$ and (b) $(∀v∈\tNa)$ $E(Z)^v = \rt^v$, where $E$ is defined in SP Proposition 2.2(b)}. 

Note that $(∀v≥1)$ $E(Z)^v⋅∈⋅Z⧷⎨t^o⎬$ by the definition of $E$.  Thus the claim's assumption implies that \ilc{8436} $(∀v≥1)$ $q(E(Z)^v)⋅∈⋅S$. Also, note that SP Proposition~2.2(b) implies $E(Z)⋅∈⋅\YY$, where $\YY$ is defined in SP Proposition~2.2(b).  Thus \il{8437} $t^o = p(E(Z)^1)$ and \il{8438} $(∀v≥2)$ $E(Z)^{v-1} = p(E(Z)^v)$. 

It suffices to show that $(∀v≥1)$ [1] $v⋅∈⋅\rV$ and [2] $E(Z)^v = \rt^v$.  This will be shown by induction.

For the initial step, suppose $v = 1$.  [1] holds easily by the definition of $\rV$ in Claim~\rf{8406}.  Further, \rf{8437} implies there is \il{8439} $c⋅∈⋅F(t^o)$ such that \il{8440} $t^o⊗c = E(Z)^1$.  \rf{8440} implies $c⋅=⋅q(E(Z)^1)$ and thus \rf{8436} implies \il{8441} $c⋅∈⋅S$.  \rf{8439}--\rf{8441} and the definition of $ε$ in Claim~\rf{8405} imply $E(Z)^1 = ε(t^o)$.  Thus the definition of $\rt^1$ in Claim~\rf{8406} implies $E(Z)^1 = \rt^1$.  Thus [2] holds.

For the inductive step, suppose $v⋅≥⋅2$.  Then \rf{8438} and the inductive hypothesis imply \il{8476} $\rt^{v-1} = p(E(Z)^v)$.  \rf{8476} implies $\rt^{v-1}⋅∈⋅X$, so [1] holds by the definition of $\rV$ in Claim~\rf{8406}.  Further, \rf{8476} implies there is \il{8442} $c⋅∈⋅F(\rt^{v-1})$ such that \il{8443} $\rt^{v-1}⊗c = E(Z)^v$.  \rf{8443} implies $c⋅=⋅q(E(Z)^v)$ and thus \rf{8436} implies \il{8444} $c⋅∈⋅S$.  \rf{8442}--\rf{8444} and the definition of $ε$ in Claim~\rf{8405} imply $E(Z)^v = ε(\rt^{v-1})$.  Thus the definition of $\rt^v$ in Claim~\rf{8406} implies $E(Z)^v = \rt^v$.  Thus [2] holds.

\yl{8425} {\em Suppose $Z⋅∈⋅\ZZi$ satisfies $(∀t∈Z⧷⎨t^o⎬)⋅q(t)⋅∈⋅S$.  Then (a) $\rV = \tNa$ and (b) $E(Z) = (\rt^v)_{v∈\rV}$.}  

Because of Claim~\rf{8435}(b), it suffices to show that $\rV = \tNa$.  The definition of $\rV$ in Claim~\rf{8406} implies $\rV⋅⊆⋅\tNa$.  Claim~\rf{8435}(a) shows the converse.

\yl{8423} {\em Suppose $Z⋅∈⋅\ZZi$ satisfies $(∀t∈Z⧷⎨t^o⎬)⋅q(t)⋅∈⋅S$.  Then $Z = \rZ$.}

Claim~\rf{8425} implies, a fortiori, that the range of $E(Z)$ equals the range of $(\rt^v)_{v∈\rV}$.  In other words, \ilc{9294} $⎨E(Z)^v|v≥1⎬ = ⎨\rt^v|v∈\rV⎬$.  It is then argued, in steps, that $Z$ by SP Proposition~2.2(b) is equal to $⎨t^o⎬∪⎨E(Z)^v|v≥1⎬$, which by \rf{9294} is equal to $⎨t^o⎬∪⎨\rt^v|v∈\rV⎬$, which by the definition of $\rZ$ in Claim~\rf{8422} is equal to $\rZ$.  \end{cllist}

{\em Conclusion}.  Claim~\rf{8422} shows that $\rZ$ is a $Z\,∈\,\ZZ$ that satisfies $(∀t∈Z⧷⎨t^o⎬)⋅q(t)⋅∈⋅S$.  Claims \rf{9292} and \rf{8423} show that $\rZ$ is the only $Z\,∈\,\ZZ$ that does so. \end{pf}

\begin{lemma}\label{9173} Suppose $Π$ and $Π′$ are \ct{NCP} preforms.  Then the following are equivalent.\begin{tlist}
\yl{9244} $[Π,Π′,\inc_{T,T′},\inc_{C,C′}]$ is a subpreform inclusion.
\yl{9246} $[(T,p),(T′,p′),\inc_{T,T′}]$ is a subtree inclusion, $C\,⊆\,C′$\hspace{-.4mm}, $⊗\,{=}\,⊗′\sj_{F\gr}$\hspace{.2mm}, and $\HH\,⊆\,\HH′$.
\yl{9245} $T\,{=}\,⎨t′∈T′|t^o≼′t′⎬$, $C\,⊆\,C′$, $⊗ = ⊗′\sj_{F\gr}$\,, and $\HH\,⊆\,\HH′$.
\end{tlist} \end{lemma}

\begin{pf} {\em (\rf{9244})$⇒$(\rf{9246})}.  Suppose (\rf{9244}).  Then, by definition, \lic{9305} $[Π,Π′,\inc_{T,T′},\inc_{C,C′}]$ is an inclusion, \li{9303} $T\,{=}\,⎨t′∈T′|t^o≼′t′⎬$, and \li{9304} $\HH\,⊆\,\HH′$.  \rf{9305} and the ``forgetful'' functor of SP Theorem 3.9 imply $[(T,p),(T′,p′),\inc_{T,T′}]$ is an inclusion, and thus \rf{9303} implies it is a subtree inclusion.  \rf{9305} implies $C⋅⊆⋅C′$.  In light of \rf{9304} and the previous two sentences, it remains to show that $⊗ = ⊗′\sj_{F\gr}$\,.  \rf{P1} for $Π$ implies $⊗$ has domain $F\gr$ and is surjective.  Thus it suffices to show that $⊗\gr⋅⊆⋅⊗\pgr$.  This follows from \rf{p2} for $[Π,Π′,\inc_{T,T′},\inc_{C,C′}]$.  

{\em (\rf{9246})$⇒$(\rf{9245})}. Suppose (\rf{9246}).  Then, the definition of subtree inclusion implies $T\,{=}$ $⎨t′∈T′|t^o≼′t′⎬$, and the remaining conditions in (\rf{9246}) are identical to the remaining conditions in (\rf{9245}).

{\em (\rf{9244})$⇐$(\rf{9245})}. Suppose (\rf{9245}).  Then \lic{9202} $T\,{=}\,⎨t′∈T′|t^o≼′t′⎬$, \li{9203} $C\,⊆\,C′$, \li{9204} $⊗ = ⊗′\sj_{F\gr}$\,, and \li{9205} $\HH\,⊆\,\HH′$.  \rf{9202} implies \li{9206} $T⋅⊆⋅T′$.  By the definition of a subpreform inclusion, it suffices to show that [i] $T⋅⊆⋅T′$, that [ii] $C⋅⊆⋅C′$, and that [iii] $[Π,Π′,\inc_{T,T′},\inc_{C,C′}]$ is a morphism satisfying $T\,{=}\,⎨t′∈T′|t^o≼′t′⎬$ and $\HH\,⊆\,\HH′$.  By inspection, [i] is implied by \rf{9206}, [ii] is implied by \rf{9203}, and the two conditions in [iii] are implied by \rf{9202} and \rf{9205}.  Thus it remains to show that the quadruple $[Π,Π′,\inc_{T,T′},\inc_{C,C′}]$ is a morphism.  It will be argued that the quadruple satisfies \rf{p1} and \rf{p2}.  For \rf{p1}, \rf{9206} and \rf{9203} imply that $\inc_{T,T′}{:}T→T′$ and $\inc_{C,C'}{:}C→C′$ are well-defined.  For \rf{p2}, it suffices to show that $⊗\gr⋅⊆⋅⊗\pgr$.  This holds by \rf{9204}.  \end{pf}

\begin{lemma}\label{8598} Suppose $[Π,Π′,τ,δ]$ is an \ct{NCP} isomorphism.  Then the following hold. \begin{tlist} 
\yl{8488} $(∀H∈\HH,c∈C)$ $c∈\dF(H)$ $⟺$ $δ(c)∈\dF′(\dd{τ}(H))$.
\yl{8597} $(∀H∈\HH)⋅\dd{δ}○\dF(H) = \dF′(\dd{τ}(H))$.
\yl{8592} $(∀H∈\HH)⋅δ|_{\dF(H)}$ is a bijection from $\dF(H)$ onto $\dF′(\dd{τ}(H))$.
\yl{9297} $\dd{δ}\sj_\Ss$ is a bijection from $\Ss$ onto $\Ss′$.
\yl{8480} $(\dd{τ}\sj_{\ZZ})○ζ = ζ′○(\dd{δ}\sj_{\Ss})$.
\end{tlist}\end{lemma}

\begin{pf} Note SP Theorem 3.7 implies $τ{:}T→T′$ and $δ{:}C→C′$ are bijections.

\lstep{(\rf{8488})}. Take $H⋅∈⋅\HH$ and $c⋅∈⋅C$.  It is argued, in steps, that $c⋅∈\dd{F}(H)$ by inspection is equivalent to $(∃t∈H)⋅c⋅∈⋅F(t)$, which by SP Proposition 3.8(c) is equivalent to $(∃t∈H)⋅δ(c)⋅∈⋅F′(τ(t))$, which by inspection is equivalent to $(∃t∈T)$ $t⋅∈⋅H$ and $δ(c)⋅∈⋅F′(τ(t))$, which by the bijectivity of $τ$ is equivalent to $(∃t∈T)$ $τ(t)⋅∈⋅\dd{τ}(H)$ and $δ(c)⋅∈⋅F′(τ(t))$, which by the bijectivity of $τ$ is equivalent to $(∃t′∈T′)$ $t′⋅∈⋅\dd{τ}(H)$ and $δ(c)⋅∈⋅F′(t′)$, which by inspection is equivalent to $(∃t′∈\dd{τ}(H))$ $δ(c)⋅∈⋅F′(t′)$, which by inspection is equivalent to $δ(c)⋅∈⋅\dF′○\dd{τ}(H)$.

\lstep{(\rf{8597})}.  Take $H⋅∈⋅\HH$.  (\rf{8488}) and the bijectivity of $δ$ imply $(∀c∈C)$ $c⋅∈⋅\dF(H)$ $⟺$ $c⋅∈⋅\dd{δ}^{-1}○\dF′○\dd{τ}(H)$.  Hence $\dF(H) = \dd{δ}^{-1}○\dF′○\dd{τ}(H)$.  Hence the bijectivity of $δ$ implies $\dd{δ}○\dF(H) = \dF′○\dd{τ}(H)$. 

\lstep{(\rf{8592})}.  As noted at the outset, $δ$ is a bijection with domain $C$.  Thus, since $\dF(H)⋅⊆⋅C$, it suffices to prove that $\dd{δ}○\dF(H)⋅=⋅\dF′○\dd{τ}(H)$.  This is part (\rf{8597}).

\lstep{(\rf{9297})}. As noted at the outset, \lic{8486} $δ{:}C→C′$ is a bijection.  Thus, since $\Ss⋅⊆⋅\PP(C)$, it suffices to show $⎨\,\dd{δ}(S)\,|\,S∈\Ss\,⎬ = \Ss′$.  It will be argued that  \begin{align}
\zz
⎨\,\dd{δ}(S)\,|\,S∈\Ss\,⎬
=&⋅⎨⋅\dd{δ}(A)⋅|⋅A⊆C⋅\text{and}⋅(∀H∈\HH)⋅|A∩\dF(H)|{=}1⋅⎬ \nt
=&⋅⎨⋅\dd{δ}(A)⋅|⋅A⊆C⋅\text{and}⋅(∀H∈\HH)⋅|\dd{δ}(A)∩\dd{δ}○\dF(H)|{=}1⋅⎬ \nt
=&⋅⎨⋅\dd{δ}(A)⋅|⋅A⊆C⋅\text{and}⋅(∀H∈\HH)⋅|\dd{δ}(A)∩\dF′(\dtau(H))|{=}1⋅⎬ \nt 
=&⋅⎨⋅\dd{δ}(A)⋅|⋅A⊆C⋅\text{and}⋅(∀H′∈\HH′)⋅|\dd{δ}(A)∩\dF′(H′)|{=}1⋅⎬ \nt
=&⋅⎨⋅A′⊆C′⋅|⋅(∀H′∈\HH′)⋅|A′∩\dF′(H′)|{=}1⋅⎬ \nt 
=&⋅\Ss′.\notag
\zz
\end{align} The first equality holds by the definition of $\Ss$, the second by \rf{8486}, the third by part (\rf{8597}), the fourth by SP Theorem 3.8(e), the fifth by \rf{8486} (again), and the last by the definition of $\Ss′$.

\lstep{(\rf{8480})}.  This paragraph checks domains and codomains.  $\Ss$ is both the domain of $ζ$ by definition, and the domain of $\dd{δ}\sj_{\Ss}$ by inspection.  On the lefthand side only, $\ZZ$ is both the codomain of $ζ$ by definition, and the domain of $\dtau\sj_{\ZZ}$ by inspection.  On the righthand side only, $\Ss′$ is both the codomain of $\dd{δ}\sj_{\Ss}$ by part (\rf{9297}), and the domain of $ζ′$ by definition.  Finally, $\ZZ′$ is both the codomain of $\dtau\sj_{\ZZ}$ by SP Proposition~2.7(h,i) and SP Theorem 3.9, and the codomain of $ζ′$ by definition.

Next take $S⋅∈⋅\Ss$. This paragraph shows \begin{center}
\zz
$(∀\,t′\,∈\,(\dtau\sj_{\ZZ})○ζ(S)⧷⎨t\po⎬)$ $q′(t′)\,∈\,\dd{δ}(S)$.
\zz
\end{center}  Toward that end, take \ilc{8501} $t′⋅∈⋅(\dtau\sj_{\ZZ})○ζ(S)⧷⎨t\po⎬$.  \rf{8501} implies there is \il{8502} $t⋅∈⋅ζ(S)$ such that \il{8503} $τ(t) = t′$.  \rf{8501} and \rf{8503} imply \il{9462} $τ(t)⋅≠⋅t\po$.  Since $τ{:}T→T′$ is a bijection (as noted at outset), and since $τ(t^o) = t\po$ (by SP Proposition 2.7(c)), \rf{9462} implies \il{8504} $t⋅≠⋅t^o$.  \rf{8502} and \rf{8504} imply $t⋅∈⋅ζ(S)⧷⎨t^o⎬$, and thus the definition of $ζ$ implies $q(t)⋅∈⋅S$.  Thus SP Proposition 3.8(d) implies $q′(τ(t))⋅∈⋅\dd{δ}(S)$.  Thus \rf{8503} implies $q′(t′)⋅∈⋅\dd{δ}(S)$.

Finally, the definition of $ζ′$ states that $ζ′○(\dd{δ}\sj_{\Ss})(S)$ is the unique member of \begin{center}
\zz
 $⎨⋅Z′∈\ZZ′⋅|⋅(∀t′∈Z′⧷⎨t\po⎬)\,q′(t′)∈\dd{δ}(S)⋅⎬$.  
\zz 
\end{center} The previous paragraph implies $(\dtau\sj_{\ZZ})○ζ(S)$ is a member of this collection.  Thus $(\dtau\sj_{\ZZ})○ζ(S) = ζ′○(\dd{δ}\sj_{\Ss})(S)$. \end{pf}

\section{For \ct{NCF}}\label{9290}\markb{\sc \rf{9290}. For \ct{NCF}}

\begin{lemma}\label{8584} Suppose $Φ$ is an \ct{NCF} form.  Then $(∀i∈I)$ $(\dF(H))_{H∈\HH_i}$ is pairwise disjoint and $∪_{H∈\HH_i}\dF(H) = C_i$. \end{lemma}

\begin{pf} Take $i⋅∈⋅I$.  SF Proposition 3.2(16b) implies that $(\dF(H))_{H∈\HH}$ is pairwise disjoint.  This and $\HH_i⋅⊆⋅\HH$ imply $(\dF(H))_{H∈\HH_i}$ is pairwise disjoint.  Thus it remains to show $∪_{H∈\HH_i}\dF(H) = C_i$.  

For the forward direction, suppose $c⋅∈⋅∪_{H∈\HH_i}\dF(H)$.  Then there is \ilc{8585} $H⋅∈⋅\HH_i$ such that \il{8586} $c⋅∈⋅\dF(H)$.  \rf{8586} implies there is \il{8587} $t⋅∈⋅H$ such that \il{8588} $c⋅∈⋅F(t)$.  Meanwhile, \rf{8585} implies there is \il{8589} $c^*⋅∈⋅C_i$ such that \il{8590} $H = F\T(c^*)$.  \rf{8590} and SF Lemma~A.1(b) imply \il{8591} $c^*⋅∈⋅\dF(H)$.  \rf{8591}, \rf{8587}, and SP Proposition 3.2(16a) imply $c^*⋅∈⋅F(t)$.  This and \rf{8589} imply $c^*⋅∈⋅F(t)∩C_i$.  This, \rf{F2}, and \rf{F3} imply $F(t)⋅⊆⋅C_i$.  This and \rf{8588} imply $c⋅∈⋅C_i$.  

For the reverse direction, take $c⋅∈⋅C_i$.  \rf{P3} implies $F\T(c)⋅≠⋅∅$.  So $c⋅∈⋅\dF○F\T(c)$.  Also note $F\T(c)⋅∈⋅\HH_i$ by the definition of $\HH_i$.  Thus $c⋅∈⋅∪_{H∈\HH_i}\dF(H)$.  \end{pf}

\begin{npf}[for Proposition~\rf{9072}]\label{9072p} It suffices to show [A] that the function followed by the purported inverse is the identity on $\Ss$ and [B] that the purported inverse followed by the function is the identity on $∏_{i∈I}\Ss_i$.

[A] Take $S⋅∈⋅\Ss$.  It will be argued that\begin{gather}
\zz
S⋅\mapsto⋅(S∩C_i)_{i∈I}⋅\mapsto⋅∪_{i∈I}(S∩C_i)_{i∈I} = S⋅∩⋅∪_{i∈I}C_i = S⋅∩⋅C = S.\notag
\zz
\end{gather} By inspection, the first arrow applies the function itself.  To show that the second arrow applies the purported inverse, it suffices to show $(S∩C_i)_{i∈I}⋅∈⋅∏_{i∈I}\Ss_i$.  For this it suffices to show $(∀i∈I)$ $S∩C_i⋅∈⋅\Ss_i$.  Toward that end, take $i⋅∈⋅I$.  By the definition of $\Ss_i$, it must be shown that [a] $S∩C_i⋅⊆⋅C_i$ and [b] $(∀H∈\HH_i)⋅|(S∩C_i)∩\dF(H)|{=}1$.  [a] is immediate.  For [b], take $H⋅∈⋅\HH_i$.  By Lemma~\rf{8584}, $\dF(H)⋅⊆⋅C_i$.  Thus $(S∩C_i)∩\dF(H)$ equals $S∩\dF(H)$, which is a singleton because $S⋅∈⋅\Ss$.  In the main argument, the first two equalities holds by manipulation, and the final equality holds because $S⋅∈⋅\Ss$ implies $S⋅⊆⋅C$.

[B] Take $(S_i)_{i∈I}⋅∈⋅∏_{i∈I}\Ss_i$.  Note that \lic{9299} $(∀k∈I)$ $S_k⋅∈⋅\Ss_k$, and that this implies \li{9300} $(∀k∈I)$ $S_k⋅⊆⋅C_k$.  It will be argued that\begin{gather}
\zz
(S_i)_{i∈I}⋅\mapsto⋅∪_{i∈I}S_i⋅\mapsto⋅((∪_{i∈I}S_i)∩C_j)_{j∈I} = (S_j∩C_j)_{j∈I} = (S_j)_{j∈I}.\notag
\zz
\end{gather} By inspection, the first arrow applies the purported inverse.  To show that the second arrow applies the function itself, it suffices to show $∪_{i∈I}S_i⋅∈⋅\Ss$.  By the definition of $\Ss$, it suffices to show [a] $∪_{i∈I}S_i⋅⊆⋅C$ and [b] $(∀H∈\HH)⋅|(∪_{i∈I}S_i)∩\dF(H)| = 1$.  For [a], it suffices to show $(∀k∈I)$ $S_k⋅⊆⋅C$.  This follows from \rf{9300}.  For [b], take $H⋅∈⋅\HH$.  SF Proposition~2.1(c) implies there is $j⋅∈⋅I$ such that $H⋅∈⋅\HH_j$.  Thus Lemma~\rf{8584} implies $\dF(H)⋅⊆⋅C_j$, which by \rf{F2} implies $(∀i∈I⧷⎨j⎬)$ $\dF(H)∩C_i = ∅$, which by \rf{9300} implies $(∀i∈I⧷⎨j⎬)$ $\dF(H)∩S_i = ∅$, which by inspection implies $(∪_{i∈I}S_i)∩\dF(H) = S_j∩\dF(H)$.  The righthand side is a singleton by \rf{9299} and the definition of $\Ss_j$.  Thus the lefthand side is a singleton.  Thus [b] holds.  To see the first equality in the main argument, take $j⋅∈⋅I$.  \rf{9300} implies $(∀i∈I⧷⎨j⎬)$ $S_i⋅⊆⋅C_i$, which by \rf{F2} implies $(∀i∈I⧷⎨j⎬)$ $S_i∩C_j = ∅$, which by inspection implies $(∪_{i∈I}S_i)∩C_j = S_j∩C_j$.  To see the second equality in the main argument, again take $j⋅∈⋅I$.  \rf{9300} implies $S_j⋅⊆⋅C_j$, which implies $S_j∩C_j = S_j$. \end{npf}

\begin{lemma}\label{9174} Suppose $Φ$ and $Φ′$ are \ct{NCF} forms.  Then the following are equivalent. \begin{tlist}
\yl{9249} $[Φ,Φ′,\inc_{I,I′},\inc_{T,T′},\inc_{C,C′}]$ is a subform inclusion.
\yl{9251} $[Π,Π′,\inc_{I,I′},\inc_{T,T′}]$ is a subpreform inclusion, $I\,⊆\,I′$, and $(∀i∈I)\,C_i\,⊆\,C′_i$. 
\yl{9250} $I\,⊆\,I′$, $T\,{=}\,⎨t′∈T′|t^o≼′t′⎬$, $(∀i∈I)\,C_i\,⊆\,C′_i$, $⊗ = ⊗′\sj_{F\gr}$\,, and $\HH\,⊆\,\HH′$.
\end{tlist} \end{lemma}

\begin{pf} {\em (\rf{9249})$⇒$(\rf{9251})}.  Suppose (\rf{9249}).  Then, by definition, \lic{9306} $[Φ,Φ′,\inc_{I,I′},\inc_{T,T′},$ $\!\inc_{C,C′}]$ is an inclusion, \il{9308} $T = ⎨t′∈T′|t^o≼′t′⎬$, and \il{9309} $\HH⋅⊆⋅\HH′$.  \rf{9306} and the ``forgetful'' functor of SF Theorem 2.7 imply  $[Π,Π′,\inc_{T,T′},\inc_{C,C′}]$ is an inclusion, and thus \rf{9308}--\rf{9309} imply it is a subpreform inclusion.  Also, \rf{9306} implies $I⋅⊆⋅I′$.  In light of the last two sentences, it remains to show $(∀i∈I)$ $C_i⋅⊆⋅C′_i$.  This follows from \rf{f3} for $[Φ,Φ′,\inc_{I,I′},\inc_{T,T′},\inc_{C,C′}]$.

{\em (\rf{9251})$⇒$(\rf{9250})}.  Suppose (\rf{9251}).  Then (\rf{9250}) follows from Lemma~\rf{9173}(\rf{9244})$⇒$(\rf{9245}). 

{\em (\rf{9249})$⇐$(\rf{9250})}. Suppose (\rf{9250}).  Then \lic{9207} $I\,⊆\,I′$, \li{9209} $T\,{=}\,⎨t′∈T′|t^o≼′t′⎬$, \li{9210} $(∀i∈I)\,C_i\,⊆\,C′_i$, \li{9211} $⊗ = ⊗′\sj_{F\gr}$\,, and \li{9212} $\HH\,⊆\,\HH′$.  \rf{9209} implies \li{9213} $T⋅⊆⋅T′$.  \rf{9210} implies \il{9214} $C⋅⊆⋅C′$.  By the definition of a subform inclusion, it suffices to show that [i] $I⋅⊆⋅I′$, that [ii] $T⋅⊆⋅T′$, that [iii] $C⋅⊆⋅C′$, and that [iv] $[Φ,Φ′,\inc_{I,I′},\inc_{T,T′},\inc_{C,C′}]$ is a morphism satisfying $T\,{=}\,⎨t′∈T′|t^o≼′t′⎬$ and $\HH\,⊆\,\HH′$.  By inspection, [i] is implied by \rf{9207}, [ii] is implied by \rf{9213}, [iii] is implied by \rf{9214}, and the two conditions in [iv] are implied by \rf{9209} and \rf{9212}.  Thus it remains to show that the quintuple $[Π,Π′,\inc_{I,I′},\inc_{T,T′},\inc_{C,C′}]$ is a morphism.  It will be argued that the quintuple satisfies \rf{f1}--\rf{f3}.  For \rf{f1}, note \rf{9207}, \rf{9213}, and \rf{9214} imply that $\inc_{I,I′}{:}I→I′$, $\inc_{T,T′}{:}T→T′$, and $\inc_{C,C'}{:}C→C′$ are well-defined.  For \rf{f2}, it suffices to show that $⊗\gr⋅⊆⋅⊗\pgr$.  This holds by \rf{9211}.  Finally, \rf{f3} is implied by \rf{9210}.  

\end{pf}

\begin{lemma}\label{8477} Suppose $[Φ,Φ′,τ,δ,ι]$ is an \ct{NCF} isomorphism.  Then $(∀i∈I)⋅\dd{δ}\sj_{\Ss_i}$ is a bijection from $\Ss_i$ onto $\Ss′_{ι(i)}$. \end{lemma} 

\begin{pf} Take $i⋅∈⋅I$.  SF Proposition~2.6(d) implies \lic{9301} $δ\sj_{C_i}$ is a bijection from $C_i$ onto $C′_{ι(i)}$.  Thus, since $\Ss_i⋅⊆⋅\PP(C_i)$, it suffices to show that \begin{align}
\zz
⎨\,\dd{δ}(S_i)\,|\,S_i∈\Ss_i\,⎬
=&⋅⎨⋅\dd{δ}(A)⋅|⋅A⊆C_i⋅\text{and}⋅(∀H∈\HH_i)⋅|A∩\dF(H)|{=}1⋅⎬ \nt
=&⋅⎨⋅\dd{δ}(A)⋅|⋅A⊆C_i⋅\text{and}⋅(∀H∈\HH_i)⋅|\dd{δ}(A)∩\dd{δ}○\dF(H)|{=}1⋅⎬ \nt
=&⋅⎨⋅\dd{δ}(A)⋅|⋅A⊆C_i⋅\text{and}⋅(∀H∈\HH_i)⋅|\dd{δ}(A)∩\dF′(\dtau(H))|{=}1⋅⎬ \nt
=&⋅⎨⋅\dd{δ}(A)⋅|⋅A⊆C_i⋅\text{and}⋅(∀H′∈\HH′_{ι(i)})⋅|\dd{δ}(A)∩\dF′(H′)|{=}1⋅⎬ \nt
=&⋅⎨⋅A′⊆C′_{ι(i)}⋅|⋅(∀H′∈\HH′_{ι(i)})⋅|A′∩\dF′(H′)|{=}1⋅⎬ \nt 
=&⋅\Ss′_{ι(i)}.\notag
\zz
\end{align} The first equality holds by the definition of $\Ss_i$, the second by \rf{9301}, the third by Lemma~\rf{8598}(\rf{8597}), the fourth by SF Proposition 2.6(m), the fifth by \rf{9301} (again), and the last by the definition of $\Ss′_{ι(i)}$.  \end{pf}

\section{For \ct{NCG}}\label{9311} \markb{\sc \rf{9311}. For \ct{NCG}}

\ssec{Basics}\label{9448}

\begin{lemma}\label{9350} Suppose $γ = [Γ,Γ′,ι,τ,δ,\vb]$ and $γ′ = [Γ′,Γ″,ι′,τ′,δ′,\vb′]$ are morphisms, with their $θ = [(T,p),(T′,p′),τ]$ and $θ′ = [(T′,p′),(T″,p″),τ′]$.  Then the following hold.
\par(a) Take $i⋅∈⋅I$.  Then $\dU′_{ι(i)}(\ZZ^{\prime\,θ′}) = \dU′_{ι(i)}(\ZZ′)$ implies $β′_{ι(i)}{*}β_i = β′_{ι(i)}○β_i$.
\par(b) $\ZZ^{\prime\,θ′} = \ZZ′$ implies $(∀i∈I)$ $β′_{ι(i)}{*}β_i|_{\dU_i(\ZZ^{θ′○θ})} = β′_{ι(i)}○β_i$. \end{lemma}

\begin{pf} {\em (a)}. Assume [1] $\dU′_{ι(i)}(\ZZ^{\prime\,θ′}) = \dU′_{ι(i)}(\ZZ′)$.  By the definition of $β′_{ι(i)}{*}β_i$, it suffices to show that [i] the domain of $β′_{ι(i)}{*}β_i$ is equal to the domain of $β_i$, [ii] the codomain of $β_i$ is equal to the domain of $β′_{ι(i)}$, and [iii] the codomain of $β′_{ι(i)}{*}β_i$ is equal to the codomain of $β′_{ι(i)}$.  

For [i], the domain of $β′_{ι(i)}{*}β_i$ by definition is equal to $\overline{β_i\T}○\dU′_{ι(i)}(\ZZ^{\prime\,θ′})$, which by assumption [1] is equal to $\overline{β_i\T}○\dU′_{ι(i)}(\ZZ′)$.  This is equal to the domain of $β_i$ because the codomain of $β_i$ is $\dU′_{ι(i)}(\ZZ′)$ by \rf{g2}.

For [ii], the codomain of $β_i$ by \rf{g2} is equal to $U′_{ι(i)}(\ZZ′)$, which by assumption [1] is equal to $U′_{ι(i)}(\ZZ^{\prime\,θ′})$, which by \rf{g2} for $γ′$ is equal to the domain of $β′_{ι(i)}$. 

For [iii], the codomain of $β′_{ι(i)}{*}β_i$ by definition is equal to $U″_{ι′○ι(i)}(\ZZ″)$, which by \rf{g2} for $γ′$ is equal to the codomain of $β′_{ι(i)}$. 

{\em (b)}.  Assume [2] $\ZZ^{\prime\,θ′} = \ZZ′$.  Then take $i⋅∈⋅I$.  [2] implies $\dU′_{ι(i)}(\ZZ^{\prime\,θ′}) = \dU′_{ι(i)}(\ZZ′)$.  Thus part (a) implies $β′_{ι(i)}{*}β_i = β′_{ι(i)}○β_i$.  Thus it suffices to show that $\dU_i(\ZZ^{θ′○θ})$ equals the domain of $β_i$.  Note that [2] and Lemma~\rf{9313}(\rf{9352}) imply $\ZZ^{θ′○θ} = \ZZ^θ$.  Thus $\dU_i(\ZZ^{θ′○θ})$ equals $\dU_i(\ZZ^θ)$, which by \rf{g2} equals the domain of $β_i$. \end{pf}

\begin{lemma}\label{9416} Suppose $γ = [Γ,Γ′,ι,τ,δ,\vb]$ and $γ′ = [Γ′,Γ″,ι′,τ′,δ′,\vb′]$ are morphisms.  Then $γ′○γ$ is a morphism. \end{lemma}

\begin{pf} Let $θ = [(T,p),(T′,p′),τ]$ and $θ′ = [(T′,p′),(T″,p″),τ′]$. \begin{cllist}

\yl{9417} $(∀i∈I,Z∈\ZZ^{θ′○θ})⋅β_i(U_i(Z))⋅∈⋅\dU′_{ι(i)}(\ZZ^{\prime\,θ′})$.  Take $i⋅∈I$ and \ilc{9418} $Z⋅∈⋅\ZZ^{θ′○θ}$.  \rf{9418} and Lemma~\rf{9313}(\rf{9314}) imply $Z⋅∈⋅\ZZ^θ$.  Thus \rf{g4} for $γ$ implies \il{9419} $β_i(U_i(Z)) = U′_{ι(i)}(P′○τ(t^o)∪\dtau(Z))$.  Also, \rf{9418} and Lemma~\rf{9313}(\rf{9315}) imply $P′○τ(t^o)∪\dtau(Z)⋅∈⋅\ZZ^{\prime\,θ′}$.  Thus $U′_{ι(i)}(P′○τ(t^o)∪\dtau(Z))$ $∈$ $U′_{ι(i)}(\ZZ^{\prime\,θ′})$.  This and \rf{9419} imply the result.

\yl{9421} $(∀i∈I,Z∈\ZZ^{θ′○θ})⋅β′_{ι(i)}(β_i(U_i(Z))) = U″_{ι′○ι(i)}(P″○τ′○τ(t^o)∪\dtau′○\dtau(Z))$.\linebreak Take $i⋅∈⋅I$ and \ilc{9356} $Z⋅∈⋅\ZZ^{θ′○θ}$.  \rf{9356} and Lemma~\rf{9313}(\rf{9314}) imply $Z⋅∈⋅\ZZ^θ$.  Thus \rf{g4} for $γ$ implies \il{9360} $β_i(U_i(Z)) = U′_{ι(i)}(P′○τ(t^o)∪\dtau(Z))$.  Also, \rf{9356} and Lemma~\rf{9313}(\rf{9315}) imply \il{9361} $P′○τ(t^o)∪\dtau(Z)⋅∈⋅\ZZ^{\prime\,θ′}$.  It will be argued that \pagebreak \begin{align}
\zz
&⋅β′_{ι(i)}(β_i(U_i(⋅Z⋅))) \nt
=&⋅β′_{ι(i)}(U'_{ι(i)}(⋅P′○τ(t^o)∪\dtau(Z)⋅)) \nt
=&⋅U″_{ι′○ι(i)}(⋅P″○τ′(t\po)⋅∪⋅\dtau′(P′○τ(t^o)∪\dtau(Z))⋅) \nt 
=&⋅U″_{ι′○ι(i)}(⋅P″○τ′○τ(t^o)⋅∪⋅\dtau′○\dtau(Z)⋅). \notag
\zz
\end{align} The first equality follows from \rf{9360}, Claim~\rf{9417}, and the fact that the domain of $β′_{ι(i)}$ is $U′_{ι(i)}(\ZZ^{\prime\,θ′})$ by \rf{g2} for $γ′$ at $i′ = ι(i)$.  The second equality follows from \rf{9361} and \rf{g4} for $γ′$ at $i′ = ι(i)$ and $Z′ = P′○τ(t^o)∪\dtau(Z)$.  The third equality follows from Lemma~\rf{8335}.

\end{cllist}

{\em Conclusion}. By the definition of a game morphism, it suffices to show that $γ′○γ$ $=$ $[Γ,Γ″,ι′○ι,τ′○τ,δ′○δ,$ $\!(β′_{ι(i)}{*}β_i|_{\dU_i(\ZZ^{θ′○θ})})_{i∈I}]$ satisfies the following:\begin{gather}
\zz
\mqi \text{[\r{g}1]} & [Φ,Φ′,ι′○ι,τ′○τ,δ′○δ]⋅\text{is an \ct{NCF} morphism}, \mqo \nt
\mqi \text{[\r{g}2]} & (∀i∈I)⋅β′_{ι(i)}{*}β_i|_{\dU_i(\ZZ^{θ′○θ})}\,{:}\,\dU_i(\ZZ^{θ′○θ})\,→\,\dU″_{ι′○ι(i)}(\ZZ″), \mqo \nt
\mqi \text{[\r{g}3]} & (∀i∈I)⋅β′_{ι(i)}{*}β_i|_{\dU_i(\ZZ^{θ′○θ})}⋅\text{is weakly increasing, and} \mqo \nt
\mqi \text{[\r{g}4]} & (∀i∈I,Z∈\ZZ^{θ′○θ})⋅β′_{ι(i)}{*}β_i|_{\dU_i(\ZZ^{θ′○θ})}(U_i(Z)) = U″_{ι′○ι(i)}(P″○τ′○τ(t^o)∪\dtau′○\dtau(Z)). \mqo \notag
\zz
\end{gather} 

[\r{g}1].  By \rf{g1} for $γ$, $[Φ,Φ′,ι,τ,δ]$ is an \ct{NCF} morphism.  Similarly, by \rf{g1} for $γ′$, $[Φ′,Φ″,ι′,τ′,δ′]$ is an \ct{NCF} morphism.  Thus $[Φ′,Φ″,ι′,τ′,δ′]○[Φ,Φ′,ι,τ,δ] = [Φ,Φ″,ι′○ι,τ′○τ,δ′○δ]$ is an \ct{NCF} morphism.

[\r{g}2].  Take $i⋅∈⋅I$.  By the definition of $β′_{ι(i)}{*}β_i$, it suffices to show that $\dU_i(\ZZ^{θ′○θ})$ is included in the domain of $β′_{ι(i)}{*}β_i$, which is $\overline{β_i\T}○\dU′_{ι(i)}(Z^{\prime\,θ′})$.  Toward that end, take $u_i⋅∈⋅\dU_i(\ZZ^{θ′○θ})$.  Then there is \ilc{9422} $Z⋅∈⋅\ZZ^{θ′○θ}$ such that \il{9423} $U_i(Z) = u_i$.  \rf{9422} and Claim~\rf{9417} imply $β_i(U_i(Z))⋅∈⋅\dU′_{ι(i)}(\ZZ^{\prime\,θ′})$.  Thus $U_i(Z)⋅∈⋅\overline{β_i\T}○\dU′_{ι(i)}(\ZZ^{\prime\,θ′})$.  This and \rf{9423} imply $u_i⋅∈⋅\overline{β_i\T}○\dU′_{ι(i)}(\ZZ^{\prime\,θ′})$.

[\r{g}3]. This follows from the definition of $(β_{ι(i)}{*}β_i)_{i∈I}$, from \rf{g3} for $γ$, and from \rf{g3} for $γ′$.

[\r{g}4]. Take $i⋅∈⋅I$ and $Z⋅∈⋅\ZZ^{θ′○θ}$.  In steps, $β′_{ι(i)}{*}β_i|_{\dU_i(\ZZ^{θ′○θ})}(U_i(Z))$ by the definition of $β′_{ι(i)}{*}β_i$ is equal to $β′_{ι(i)}(β_i(U_i(Z)))$, which by Claim~\rf{9421} is equal to\linebreak $U″_{ι′○ι(i)}(P″○τ′○τ(t^o)∪\dtau′○\dtau(Z))$. \end{pf}

\begin{npf}[for Theorem~\rf{8231}]\label{8231p}⋅ \begin{cllist}

\yl{8515} {\em If $Γ$ is a game, $\id_Γ$ is a morphism.}  Take $Γ$.  It must be shown that the sextuple $[Γ,Γ,\id_I,\id_T,\id_C,(\id_{\dU_i(\ZZ)})_{i∈I}]$ satisfies \rf{g1}--\rf{g4}.  For \rf{g1}, note that $[Φ,Φ,\id_I,\id_T,\id_C]$ is an \ct{NCF} identity, and thus a fortiori it is an \ct{NCF} morphism.  Further,  $\ZZ = \ZZ^θ$ by Proposition~\rf{9043}(\rf{9046}) and the fact that $θ = [(T,p),(T,p),\id_T]$ is a subtree inclusion.  Thus $(\id_{\dU_i(\ZZ)})_{i∈I}$ satisfies \rf{g2}--\rf{g3} by inspection.  For \rf{g4}, 
take $i⋅∈⋅I$ and $Z⋅∈⋅\ZZ$ (this suffices since $\ZZ^θ⋅⊆⋅\ZZ$, or alternatively, since $\ZZ^θ = \ZZ$ has been shown).  In steps, $β_i○U_i(Z)$ by $β_i = \id_{\dU_i(\ZZ)}$ is equal to $U_i(Z)$, which by $Γ = Γ′$ and $ι = \id_I$ is equal to $U′_{ι(i)}(Z)$, which by Lemma~\rf{8327}(\rf{8452}) is equal to $U′_{ι(i)}(P′(t\po)∪Z)$, which by $Γ = Γ′$ and $τ = \id_T$ is equal to $U′_{ι(i)}(P′○τ(t^o)∪\dtau(Z))$. \end{cllist}

{\em Conclusion}.  Claim \rf{8515} has established the well-definition of identity, and Lemma \rf{9416} has established the well-definition of composition.  The unit and associative laws are straightforward.  Thus \ct{NCG} is a category. \end{npf}

\begin{npf}[for Theorem~\rf{8647}]\label{8647p} By \rf{G1}, $\FO$ maps any game into a form.  By \rf{g1}, $\FA$ maps any game morphism into a form morphism.  Thus is suffices to show that $\FB$ preserves source, target, identity, and composition (Mac Lane 1998, page 13).  This is done in the following four claims. \begin{cllist}

\yl{9327} $\FA([Γ,Γ′,ι,τ,δ,\vb])^\src = \FO([Γ,Γ′,ι,τ,δ,\vb]^\src)$.  In steps,\linebreak $\FA([Γ,Γ′,ι,τ,δ,\vb])^\src$ by the definition of $\FA$ is equal to $[\FO(Γ),\FO(Γ′),ι,τ,δ]^\src$ which by the definition of $\src$ in \ct{NCF} is equal to $\FO(Γ)$, which by the definition of $\src$ in \ct{NCG} is equal to $\FO([Γ,Γ′,ι,τ,δ,\vb]^\src)$.

\yl{9328} $\FA([Γ,Γ′,ι,τ,δ,\vb])^\trg = \FO([Γ,Γ′,ι,τ,δ,\vb]^\trg)$.  This is very similar to Claim~\rf{9327}.  Simply change $\src$ to $\trg$. 

\yl{9329} $\FA(\id_Γ) = \id_{\FO(Γ)}$.  In steps, $\FA(\id_Γ)$ by the definition of $\id$ in \ct{NCG} is equal to $\FA([Γ,Γ,\id_I,\id_T,\id_C,(\id_{\dU_i(\ZZ)})_{i∈I}])$, which by the definition of $\FA$ is equal to $[\FO(Γ),\FO(Γ),\id_I,\id_T,\id_C]$, which by the definition of $\id$ in \ct{NCF} is equal to $\id_{\FO(Γ)}$. 

\yl{9330} $\FA([Γ′,Γ″,ι′,τ′,δ′,\vb′]○[Γ,Γ′,ι,τ,δ,\vb])$ $=$ $\FA([Γ′,Γ″,ι′,τ′,δ′,\vb′])$ $○$ \linebreak $\FA([Γ,Γ′,ι,τ,δ,\vb])$.  It will be argued that\begin{align}
\zz
&⋅\FA([Γ′,Γ″,ι′,τ′,δ′,\vb′]○[Γ,Γ′,ι,τ,δ,\vb]) \nt
=&⋅\FA([Γ,Γ″,ι′○ι,τ′○τ,δ′○δ,(β′_{ι(i)}{*}β_i|_{\dU_i(\ZZ^{θ′○θ})})_{i∈I}]) \nt
=&⋅[\FO(Γ),\FO(Γ″),ι′○ι,τ′○τ,δ′○δ] \nt
=&⋅[\FO(Γ′),\FO(Γ″),ι′,τ′,δ′]⋅○⋅[\FO(Γ),\FO(Γ′),ι,τ,δ] \nt
=&⋅\FA([Γ′,Γ″,ι′,τ′,δ′,\vb′])⋅○⋅\FA([Γ,Γ′,ι,τ,δ,\vb]), \notag
\zz
\end{align} where $θ = [(T,p),(T′,p′),τ]$ and $θ′ = [(T′,p′),(T″,p″),τ′]$.  The first equality holds by the definition of $○$ in \ct{NCG}, the second by the definition of $\FA$, the third by the definition of $○$ in \ct{NCF}, and the fourth by two applications of the definition of $\FA$. \end{cllist}
\end{npf}

\ssec{Subgames}\label{9449}

\begin{lemma}\label{9221} Suppose $Γ$ and $Γ′$ are \ct{NCG} games.  Then the following are equivalent. \begin{tlist}
\yl{9222} $[Γ,Γ′,\inc_{I,I′},\inc_{T,T′},\inc_{C,C′},(\inc_{\dU_i(\ZZ),\dU′_i(\ZZ′)})_{i∈I}]$ is a subgame inclusion.
\yl{9223} [b1] $[Φ,Φ′,\inc_{I,I′},\inc_{T,T′},\inc_{C,C′}]$ is a subform inclusion and{\linebreak} [b2] $(∀i∈I,Z∈\ZZ)$ $U_i(Z) = U′_i(P′(t^o)∪Z)$.
\yl{9224} [c1] $I\,⊆\,I′$, [c2] $T\,{=}\,⎨t′∈T′|t^o≼′t′⎬$, [c3] $(∀i∈I)\,C_i\,⊆\,C′_i$, [c4] $⊗\,{=}\,⊗′\sj_{F\gr}$, {\linebreak} [c5]~$\HH\,⊆\,\HH′$, and [c6] $(∀i∈I,Z∈\ZZ)$ $U_i(Z) = U′_i(P′(t^o)∪Z)$.
\end{tlist}\end{lemma}

\begin{pf} {\em (a)$⇒$(b)}. Suppose (a).  Then, by definition, \lic{9364} $γ$ $=$ $[Γ,Γ′,\inc_{I,I′},\inc_{T,T′},$ $\!\inc_{C,C′},(\inc_{\dU_i(\ZZ),\dU′_i(\ZZ′)})_{i∈I}]$ is an inclusion, \li{9252} $T = ⎨t′∈T′|t^o≼′t′⎬$, and \li{9365} $\HH′⋅⊆⋅\HH$.  \rf{9364} and the ``forgetful'' functor of Theorem~\rf{8647} imply $[Φ,Φ′,\inc_{I,I′},\inc_{T,T′},\inc_{C,C′}]$ is a form inclusion, and thus \rf{9252} and \rf{9365} imply it is a subform inclusion.  Hence [b1] holds.  Further, [b1], Lemma~\rf{9174}(a$⇒$b), and Lemma~\rf{9173}(a$⇒$b) imply $θ = [(T,p),(T′,p′),τ]$ is a subtree inclusion.  Hence Proposition~\rf{9043}(\rf{9046}) implies \li{9366} $\ZZ^θ = \ZZ$.  Also, \rf{9364} implies $γ$ is a morphism.  Thus \rf{g4} and \rf{9366} imply $(∀i∈I,Z∈\ZZ)$ $U_i(Z) = U′_i(P′(t^o)∪Z)$.  Thus [b2] holds.

{\em (a)$⇐$(b)}. Suppose (b).  [b1], Lemma~\rf{9174}(a$⇒$b), and Lemma~\rf{9173}(a$⇒$b) imply $θ = [(T,p),(T′,p′),τ]$ is a subtree inclusion.  Hence Proposition~\rf{9043}(\rf{9046}) implies \lic{9371} $\ZZ^θ = \ZZ$.  Further, by definition, [b1] implies \li{9367} $[Φ,Φ′,\inc_{I,I′},\inc_{T,T′},\inc_{C,C′}]$ is an inclusion, \li{9369} $T = ⎨t′∈T′|t^o≼′t′⎬$, and \li{9370} $\HH′⋅⊆⋅\HH$.  Now consider the sextuple $γ = [Γ,Γ′,\inc_{I,I′},$ \!$\inc_{T,T′}$, \!$\inc_{C,C′},$ $(\inc_{\dU_i(\ZZ),\dU′_i(\ZZ′)})_{i∈I}]$.  By \rf{9367}, $γ$ satisfies \rf{g1}.  By inspection and \rf{9371}, $γ$ satisfies \rf{g2}--\rf{g3}.  By [b2] and \rf{9371}, $γ$ satisfies \rf{g4}.  Thus $γ$ is a morphism.  Thus by inspection and \rf{9371}, $γ$ is a game inclusion.  Thus by \rf{9369} and \rf{9370}, $γ$ is a subgame inclusion.

{\em (b)$⟺$(c)}. [b1] is equivalent to [c1]--[c5] by Lemma~\rf{9174}(a$⟺$c).  [b2] and [c6] are identical. \end{pf}

\ssec{Isomorphisms}\label{9450}

\newcommand{\h}{^{\scriptscriptstyle{\#}}}

\begin{lemma}\label{9166} Suppose $γ = [Γ,Γ′,ι,τ,δ,\vb]$ is an isomorphism.  Then the following hold.
\par(a) Each member of $⎨ι,τ,δ⎬∪⎨β_i|i∈I⎬$ is a bijection.
\par(b) $γ^{-1} = [Γ′,Γ,ι^{-1},τ^{-1},δ^{-1},$ $\!(β^{-1}_{ι^{-1}(i′)})_{i′∈I′}]$. 
\end{lemma}

\begin{pf} Since $γ$ is an isomorphism, there are $ι\h$, $τ\h$, $δ\h$, and $\vb\h$ such that \begin{center}
\zz
$γ^{-1} = [Γ′,Γ,ι\h,τ\h,δ\h,\vb\h]$.
\zz
\end{center} Let $θ\h = [(T′,p′),(T,p),τ\h]$. \begin{cllist}

\yl{9378} {\em (a) $ι$ is a bijection, (b) $ι\h = ι^{-1}$, (c) $τ$ is a bijection, (d) $τ\h = τ^{-1}$, (e) $δ$ is a bijection, and (f) $δ\h = δ^{-1}$.}  Since $γ$ and $γ^{-1}$ are an inverse pair, the ``forgetful'' functor of Theorem~\rf{8647} implies that $[Φ,Φ′,ι,τ,δ]$ and $[Φ′,Φ,ι\h,τ\h,δ\h]$ are an inverse pair.  Thus the claim follows from SF Theorem~2.4.

\yl{9379} {\em (a) $θ$ is an isomorphism and (b) $θ\h = θ^{-1}$.}  (a) follows from the definition of $θ$, Claim~\rf{9378}(c), and the second sentence of SP Theorem 2.6.  For (b), in steps, $θ^{-1}$ by the third sentence of SP Theorem 2.6 is equal to $[(T,p),(T′,p′),τ^{-1}]$, which by Claim~\rf{9378}(d) is equal to $[(T,p),(T′,p′),τ\h]$, which by definition equals $θ\h$. 

\yl{9386} {\em (a) $\ZZ^θ = \ZZ$. (b) $\ZZ^{\prime\,θ^{-1}} = \ZZ′$. (c) $\ZZ^{\prime\,θ\h} = \ZZ′$.}   Claim~\rf{9379}(a) implies $θ$ and $θ^{-1}$ are isomorphisms.  Thus (a) follows from Proposition~\rf{9043}(\rf{9045})[1], and (b) follows from the same proposition with its $θ$ set equal to $θ^{-1}$.  (c) follows from (b) and Claim~\rf{9379}(b).

\yl{9383} {\em $(∀i∈I)$ $β\h_{ι(i)}○β_i = \id_{\dU_i(\ZZ)}$}.  Take $i⋅∈⋅I$.  Since $γ$ and $γ^{-1}$ are an inverse pair, $γ^{-1}○γ = \id_Γ$.  Thus the definitions of composition and identity imply \begin{center}
\zz
[i] $β\h_{ι(i)}{*}β_i|_{\dU_i(\ZZ^{θ\h○θ})} = \id_{\dU_i(\ZZ)}$.
\zz
\end{center} Meanwhile, consider Lemma~\rf{9350}(b) with its $γ′$ equal to $γ^{-1}$.  This and Claim~\rf{9386}(c) imply \begin{center}
\zz
[ii] $⋅β\h_{ι(i)}{*}β_i|_{\dU_i(\ZZ^{θ\h○θ})} = β\h_{ι(i)}○β_i$.
\zz
\end{center} The claim follows from [i] and [ii]. 

\yl{9380} {\em $(∀i′∈I′)$ $β_{ι\h(i′)}○β\h_{i′} = \id_{\dU′_{i′}(\ZZ′)}$.}  Take $i′⋅∈⋅I′$.  Since $γ$ and $γ^{-1}$ are an inverse pair, $γ○γ^{-1} = \id_{Γ′}$.  Thus the definitions of composition and identity imply \begin{center}
\zz
[i] $β_{ι\h(i′)}{*}β\h_{i′}|_{\dU′_{i′}(\ZZ^{\prime\,θ○θ\h})} = \id_{\dU′_{i′}(\ZZ′)}$.
\zz
\end{center} Meanwhile, consider Lemma~\rf{9350}(b), with its $γ$ equal to $γ^{-1}$ here, and its $γ′$ equal to $γ$ here.  This and Claim~\rf{9386}(a) imply \begin{center}
\zz
[ii] $β_{ι\h(i′)}{*}β\h_{i′}|_{\dU′_{i′}(\ZZ^{\prime\,θ○θ\h})} = β_{ι\h(i′)}○β\h_{i′}$.
\zz
\end{center} The claim follows from [i] and [ii].

\yl{8527} {\em $(∀i∈I)$ $β_i$ is a bijection, and $β\h_{ι(i)} = β^{-1}_i$.}  Take $i⋅∈⋅I$.  Claim~\rf{9383} implies \ilc{9384} $β\h_{ι(i)}○β_i = \id_{\dU_i(\ZZ)}$.  Meanwhile, Claim~\rf{9380} at $i′ = ι(i)$ implies $β_{ι\h○ι(i)}○β\h_{ι(i)} = \id_{\dU′_{ι(i)}(\ZZ′)}$.  Thus Claim~\rf{9378}(b) implies $β_{ι^{-1}○ι(i)}○β\h_{ι(i)} = \id_{\dU′_{ι(i)}(\ZZ′)}$.  Thus \il{9385} $β_i○β\h_{ι(i)} = \id_{\dU′_{ι(i)}(\ZZ′)}$.  The claim follows from \rf{9384} and \rf{9385}.

\yl{8530} {\em $(β\h_{i′})_{i′∈I′} = (β^{-1}_{ι^{-1}(i′)})_{i′∈I′}$.}  Claim~\rf{8527} implies $(β\h_{ι(i)})_{i∈I} = (β^{-1}_i)_{i∈I}$.  Thus the result follows from Claim~\rf{9378}(a). \end{cllist}

{\em Conclusion}.  Part (a) follows from Claims~\rf{9378}(a,c,e) and \rf{8527}.  Part (b) follows from Claims~\rf{9378}(b,d,f) and \rf{8530}. \end{pf}

\begin{lemma}\label{8532} Suppose $γ = [Γ,Γ′,ι,τ,δ,(β_i)_{i∈I}]$ is a morphism.  Also suppose that each member of $⎨ι,τ,δ⎬∪⎨β_i|i∈I⎬$ is a bijection.  Then $γ$ is an isomorphism. \end{lemma}

\begin{pf}  Let $θ = [(T,p),(T′,p′),τ]$.  Further, let \begin{center}
\zz
$\ddot{γ} = [Γ′,Γ,ι^{-1},τ^{-1},δ^{-1},$ $\!(β^{-1}_{ι^{-1}(i′)})_{i′∈I′}]$
\zz
\end{center} and let $\ddot{θ} = [(T′,p′),(T,p),τ^{-1}]$. \begin{cllist} 

\yl{9331} {\em (a) $θ$ is an isomorphism, and (b) $\ddot{θ} = θ^{-1}$.}  (a) follows from the definition of $θ$, the bijectivity of $τ$, and second sentence of SP Theorem 2.6.  For (b), in steps, $\ddot{θ}$ by definition equals $[(T′,p′),(T,p),τ^{-1}]$ which by the third sentence of SP Theorem 2.6 equals $θ^{-1}$.

\yl{9387} {\em (a) $\ZZ^θ = \ZZ$, (b) $\ZZ^{\prime\,θ^{-1}} = \ZZ′$, and (c) $\ZZ^{\prime\,\ddot{θ}} = \ZZ′$.}  Claim~\rf{9331}(a) implies $θ$ and $θ^{-1}$ are isomorphisms.  Thus (a) follows from Proposition~\rf{9043}(\rf{9045})[1], and (b) follows from the same proposition with its $θ$ set equal to $θ^{-1}$.  (c) follows from (b) and Claim~\rf{9331}(b).

\yl{9388} {\em (a) $P′○τ(t^o) = ∅$, and (b) $P○τ^{-1}(t\po) = ∅$.}  Claim~\rf{9331}(a) implies $θ$ and $θ^{-1}$ are isomorphisms.  Thus (a) follows from Proposition~\rf{9043}(\rf{9045})[2], and (b) follows from the same proposition with its $θ$ set equal to $θ^{-1}$.

\yl{9424} {\em $(∀i∈I)⋅β_i\,{:}\,\dU_i(\ZZ)\,→\,\dU′_{ι(i)}(\ZZ′)$ is a bijection.}  Take $i⋅∈⋅I$.  \rf{g2} for $γ$ implies $β_i\,{:}\,\dU_i(\ZZ^θ)→\dU′_{ι(i)}(\ZZ′)$.  Thus Claim~\rf{9387}(a) implies $β_i\,{:}\,\dU_i(\ZZ)→\dU′_{ι(i)}(\ZZ′)$.  $β_i$ is a bijection by assumption.

\yl{8535} {\em $\ddot{γ}$ is a morphism.}  It suffices to show \begin{gather}
\zz
\mqi \text{[\"{g}1]} & [Φ′,Φ,ι^{-1},τ^{-1},δ^{-1}]⋅\text{is an \ct{NCF} morphism,} \mqo \nt
\mqi \text{[\"{g}2]} & (∀i′∈I′)⋅β^{-1}_{ι^{-1}(i′)} : \dU′_{i′}(\ZZ^{\prime\,\ddot{θ}})⋅→⋅\dU_{ι^{-1}(i′)}(\ZZ), \mqo \nt
\mqi \text{[\"{g}3]} & (∀i′∈I′)⋅β^{-1}_{ι^{-1}(i′)}⋅\text{is weakly increasing, and} \mqo \nt
\mqi \text{[\"{g}4]} & (∀i′∈I′,Z′∈\ZZ^{\prime\,\ddot{θ}})⋅β^{-1}_{ι^{-1}(i′)}(U′_{i′}(Z′)) = U_{ι^{-1}(i′)}(P○\dtau^{-1}(t\po)∪\dtau^{-1}(Z′)). \mqo \notag
\zz
\end{gather}

For [\"{g}1], note that SF Theorem 2.4(b), and the bijectivity of $ι$, $τ$, and $δ$, imply that $[Φ′,Φ,ι^{-1},τ^{-1},δ^{-1}]$ is an \ct{NCF} isomorphism.  Thus, a fortiori, [\"{g}1] holds.

For [\"{g}2], take $i′⋅∈⋅I′$.  \rf{g2} for $γ$ at $i = ι^{-1}(i′)$ implies that $β_{ι^{-1}(i′)}$ $:$ $\dU_{ι^{-1}(i′)}(\ZZ^θ)$ $→$ $\dU′_{ι○ι^{-1}(i′)}(\ZZ′)$.  Thus, by applying Claim~\rf{9387}(a) to the domain, and by simplifying the codomain, $β_{ι^{-1}(i′)}$ $:$ $\dU_{ι^{-1}(i′)}(\ZZ)$ $→$ $\dU′_{i′}(\ZZ′)$.  Thus bijectivity implies $β^{-1}_{ι^{-1}(i′)}$ : $\dU′_{i′}(\ZZ′)$ $→$ $\dU_{ι^{-1}(i′)}(\ZZ)$.  This and Claim~\rf{9387}(c) imply 
[\"{g}2].

For [\"{g}3], take $i′⋅∈⋅I′$.  By \rf{g3} for $γ$ at $i = ι^{-1}(i′)$, $β_{ι^{-1}(i′)}$ is weakly increasing.  Thus, since this function is bijective by assumption, it is strictly increasing.  Thus its inverse $β^{-1}_{ι^{-1}(i′)}$ is also strictly increasing.  

For [\"{g}4], consider the following statements.\begin{gather}
\zz
\mqi \text{[1]} & (∀i∈I,Z∈\ZZ)⋅U′_{ι(i)}(\dtau(Z)) = β_i(U_i(Z)). \mqo \nt
\mqi \text{[2]} & (∀i∈I,Z∈\ZZ)⋅β_i^{-1}(U′_{ι(i)}(\dtau(Z))) = U_i(Z). \mqo \nt
\mqi \text{[3]} & (∀i∈I,Z′∈\ZZ′)⋅β_i^{-1}(U′_{ι(i)}(Z′)) = U_i(\dtau^{-1}(Z′)). \mqo \nt 
\mqi \text{[4]} & (∀i′∈I′,Z′∈\ZZ′)⋅β_{ι^{-1}(i′)}^{-1}(U′_{i′}(Z′)) = U_{ι^{-1}(i′)}(\dtau^{-1}(Z′)). \mqo \notag
\zz
\end{gather} \rf{g4} for $γ$ is equivalent to [1] because of Claims~\rf{9387}(a) and \rf{9388}(a).  [1]$⟺$[2] because of Claim~\rf{9424}.  [2]$⟺$[3] because of Claim~\rf{9331}(a) and Proposition~\rf{9050}.  [3]$⟺$[4] because $ι$ is bijective.  Finally, [4] is equivalent to [\"{g}4] because of Claims~\rf{9387}(c) and \rf{9388}(b).

\yl{9389} $(∀i∈I)$ $β^{-1}_{ι^{-1}○ι(i)}{*}β_i|_{\dU_i(\ZZ^{\ddot{θ}○θ})} = β^{-1}_{ι^{-1}○ι(i)}○β_i$.  Consider Lemma~\rf{9350}(b) with its $(γ,θ)$ equal to $(γ,θ)$, and its $(γ′,θ′)$ equal to $(\ddot{γ},\ddot{θ})$.  Lemma~\rf{9350}'s first sentence holds by assumption and Claim~\rf{8535}.  Lemma~\rf{9350}(b)'s condition $\ZZ^{\prime\,θ′} = \ZZ′$ becomes $\ZZ^{\prime\,\ddot{θ}} = \ZZ′$, which holds by Claim~\rf{9387}(c).  Thus the lemma implies the result.

\yl{9390} $(∀i′∈I′)$ $β_{ι^{-1}(i′)}{*}β^{-1}_{ι^{-1}(i′)}|_{\dU′_{i′}(\ZZ^{\prime\,θ○\ddot{θ}})} = β_{ι^{-1}(i′)}○β^{-1}_{ι^{-1}(i′)}$.  Consider Lemma \rf{9350}(b) with its $(γ,θ)$ equal to $(\ddot{γ},\ddot{θ})$, and its $(γ′,θ′)$ equal to $(γ,θ)$.  Lemma~\rf{9350}'s first sentence holds by assumption and Claim~\rf{8535}.  Lemma~\rf{9350}(b)'s condition $Z^{\prime\,θ′} = \ZZ′$ becomes $\ZZ^θ = \ZZ$, which holds by Claim~\rf{9387}(a).  Thus the lemma implies the result.

\yl{8533} $\ddot{γ}○γ = \id_{Γ}$.  It is argued that\begin{align}
\zz
\ddot{γ}○γ =&⋅[Γ′,Γ,ι^{-1},τ^{-1},δ^{-1},(β^{-1}_{ι^{-1}(i′)})_{i′∈I′}]○[Γ,Γ′,ι,τ,δ,(β_i)_{i∈I}] \nt
=&⋅[Γ,Γ,ι^{-1}○ι,τ^{-1}○τ,δ^{-1}○δ,(β^{-1}_{ι^{-1}○ι(i)}{*}β_i|_{\dU_i(\ZZ^{\ddot{θ}○θ})})_{i∈I}] \nt
=&⋅[Γ,Γ,ι^{-1}○ι,τ^{-1}○τ,δ^{-1}○δ,(β^{-1}_{ι^{-1}○ι(i)}○β_i)_{i∈I}] \nt
=&⋅[Γ,Γ,ι^{-1}○ι,τ^{-1}○τ,δ^{-1}○δ,(β^{-1}_i○β_i)_{i∈I}] \nt
=&⋅[Γ,Γ,\id_I,\id_T,\id_C,(\id_{\dU_i(\ZZ)})_{i∈I}] \nt
=&⋅\id_Γ\,,\notag
\zz
\end{align} In the righthand side of the second equality, $ι(i)$ appears due to the definition of  composition.  The third equality holds by Claim~\rf{9389}.  The last component of the fifth equality holds by Claim~\rf{9424}.  The remainder is immediate.

\yl{9333} $γ○\ddot{γ} = \id_{Γ′}$.  It is argued that \begin{align}
\zz
γ○\ddot{γ} =&⋅[Γ,Γ′,ι,τ,δ,(β_i)_{i∈I}]○[Γ′,Γ,ι^{-1},τ^{-1},δ^{-1},(β^{-1}_{ι^{-1}(i′)})_{i′∈I′}] \nt 
=&⋅[Γ′,Γ′,ι○ι^{-1},τ○τ^{-1},δ○δ^{-1},(β_{ι^{-1}(i')}{*}β^{-1}_{ι^{-1}(i′)}|_{\dU_{i′}(\ZZ^{\prime\,θ○\ddot{θ}})})_{i′∈I′}] \nt
=&⋅[Γ′,Γ′,ι○ι^{-1},τ○τ^{-1},δ○δ^{-1},(β_{ι^{-1}(i')}○β^{-1}_{ι^{-1}(i′)})_{i′∈I′}] \nt
=&⋅[Γ′,Γ′,\id_{I′},\id_{T′},\id_{C′},(\id_{\dU′_{ι○ι^{-1}(i′)}(\ZZ′)})_{i′∈I′}] \nt
=&⋅[Γ′,Γ′,\id_{I′},\id_{T′},\id_{C′},(\id_{\dU′_{i′}(\ZZ′)})_{i′∈I′}] \nt
=&⋅\id_{Γ′}\,.
\notag
\zz
\end{align} In the righthand side of the second equality, $β$ carries the subscript $ι^{-1}(i′)$ because of the definition of composition.  The third equality holds by Claim~\rf{9390}.  The last component of the fourth equality holds by Claim~\rf{9424} at $i = ι^{-1}(i′)$.  The remainder is immediate. \end{cllist} 

{\em Conclusion}.  Claims \rf{8535}, \rf{8533}, and \rf{9333} imply that $γ$ and $\ddot{γ}$ are an inverse pair.  Thus $γ$ is an isomorphism.  \end{pf}

\begin{npf}[for Theorem~\rf{8248}]\label{8248p} The forward direction of (a) and all of (b) hold by Lemma~\rf{9166}.  The reverse direction of (a) holds by Lemma~\rf{8532}. \end{npf}

\begin{npf}[for Proposition~\rf{9334}]\label{9334p} Figure~\rf{8307}'s functors imply that \lic{9435} $[Φ,Φ′,ι,τ,δ]$ is a form isomorphism, \li{9436} $[Π,Π′,τ,δ]$ is a preform isomorphism, and \li{9437} $θ = [(T,p),(T′,p′),τ]$ is a tree isomorphism.  

(\rf{9434}) follows from \rf{9437} and Proposition~\rf{9050}.  (\rf{9425}) follows from \rf{9437} and Proposition~\rf{9043}(\rf{9045}).  (\rf{9429}) follows from \rf{9436} and Lemma~\rf{8598}(\rf{8592}).  (\rf{9433}) follows from \rf{9435} and Lemma~\rf{8477}.  (\rf{9430}) follows from \rf{9436} and Lemma~\rf{8598}(\rf{9297}).  (\rf{9432}) follows from \rf{9436} and Lemma~\rf{8598}(\rf{8480}).  

(\rf{9427}). Take $i⋅∈⋅I$.  Theorem~\rf{8248}(a) implies that $β_i$ is a bijection from $\dU_i(\ZZ^θ)$ onto $\dU_i(\ZZ)$.  Part (\rf{9425})[1] implies $\ZZ^θ = \ZZ$.  Finally, \rf{g3} and bijectivity imply $β_i$ is strictly increasing.

(\rf{9428}). Take $i⋅∈⋅I$.  \rf{g4} and both halves of part (b) imply $(∀Z∈\ZZ)$ $β_i(U_i(Z)) = U′_{ι(i)}(\dtau(Z))$.  Thus it remains to check domains and codomains.  $\ZZ$ is both the domain of $U_i$ by \rf{G2}, and the domain of $\dtau\sj_{\ZZ}$ by inspection.  On the lefthand side only, $\dU_i(\ZZ)$ is both the codomain of $U_i$ by \rf{G2}, and the domain of $β_i$ by part (\rf{9427}).  On the righthand side only, $\ZZ′$ is both the codomain of $\dtau\sj_{\ZZ}$ by part (\rf{9434}), and the domain of $U′_{ι(i)}$ by \rf{G2} for $Γ′$ at $i′ = ι(i)$.  Finally, $\dU′_{ι(i)}(\ZZ′)$ is both the codomain of $β_i$ by \rf{g2}, and the codomain of $\dU′_{ι(i′)}$ by \rf{G2} for $Γ′$ at $i′ = ι(i)$. \end{npf}

\ssec{Nash-equilibria}\label{9451}

\begin{npf}[for Proposition~\rf{8261}]\label{8261p} Note Theorem~\rf{8248}(a) implies \ilc{9463} $ι{:}I→I′$ is a bijection and \il{9464} $δ{:}C→C′$ is a bijection.  Also, SF Proposition 2.6(d) and the ``forgetful'' functor of Theorem~\rf{8647} imply that \il{9465} $(∀i∈I)$ $δ_i{:}C_i→C′_{ι(i)}$ is a bijection.

Take $S⋅⊆⋅\Ss$, and consider the following.  \begin{gather}
\zz
\mqi \text{[1]} & (∀i∈I,S^+_i∈\Ss_i)⋅
U_i○ζ(S)⋅≥⋅U_i○ζ(\,(S⧷C_i)∪S^+_i\,). \mqo\nt
\mqi \text{[2]} & (∀i∈I,S^+_i∈\Ss_i)⋅
β_i○U_i○ζ(S)⋅≥⋅β_i○U_i○ζ(\,(S⧷C_i)∪S^+_i\,). \mqo\nt\notag
\mqi \text{[3]} & (∀i∈I,S^+_i∈\Ss_i)⋅
U′_{ι(i)}○\dtau\sj_{\ZZ}○ζ(S)⋅≥⋅U′_{ι(i)}○\dtau\sj_{\ZZ}○ζ(\,(S⧷C_i)∪S^+_i\,). \mqo\nt
\mqi \text{[4]} & (∀i∈I,S^+_i∈\Ss_i)⋅
U′_{ι(i)}○ζ′(⋅\dd{δ}(S)⋅)⋅≥⋅U′_{ι(i)}○ζ′(⋅\dd{δ}((S⧷C_i)∪S^+_i)⋅). \mqo\nt
\mqi \text{[5]} & (∀i∈I,S^+_i∈\Ss_i)⋅
U′_{ι(i)}○ζ′(⋅\dd{δ}(S)⋅)⋅≥⋅U′_{ι(i)}○ζ′(⋅(\dd{δ}(S)⧷C′_{ι(i)})∪\dd{δ}(S^+_i)\,). \mqo\nt
\mqi \text{[6]} & (∀i∈I,S^{\prime +}_{ι(i)}∈\Ss_{ι(i)})⋅
U′_{ι(i)}○ζ′(⋅\dd{δ}(S)⋅)⋅≥⋅U′_{ι(i)}○ζ′(⋅(\dd{δ}(S)⧷C′_{ι(i)})∪S^{\prime +}_{ι(i)}\,). \mqo\nt 
\mqi \text{[7]} & (∀i′∈I′,S^{\prime +}_{i′}∈\Ss_{i′})⋅
U′_{i′}○ζ′(⋅\dd{δ}(S)⋅)⋅≥⋅U′_{i′}○ζ′(⋅(\dd{δ}(S)⧷C′_{i′})∪S^{\prime +}_{i′}⋅). \mqo\notag
\zz
\end{gather} $S$ being a Nash-equilibrium in $Γ$ is equivalent to [1], by definition.  [1]$⟺$[2] by Proposition~\rf{9334}(\rf{9427}).  [2]$⟺$[3] by Proposition~\rf{9334}(\rf{9428}).  [3]$⟺$[4] by Proposition~\rf{9334}(\rf{9432}).  [4]$⟺$[5] by \rf{9464} and \rf{9465}.  [5]$⟺$[6] by Proposition~\rf{9334}(\rf{9433}).   [6]$⟺$[7] by \rf{9463}.  Finally, [7] is equivalent to  $\dd{δ}(S)$ being a Nash-equilibrium in $Γ′$, by definition.  \end{npf}

\ssec{Subcategories}\label{9452} 

\begin{lemma}\label{9391} Suppose $Γ$ is an \ct{NCG} game (with \ct{NCF} form $Φ$).  Further suppose $Φ′$ is an \ct{NCF} form and $[Φ,Φ′,ι,τ,δ]$ is an \ct{NCF} isomorphism.  Then (a) $ι{:}I→I′$ and $\dtau\sj_{\ZZ}{:}\ZZ→\ZZ′$ are bijections.  Further, define \begin{gather}
\zz
Γ′ = (I′,T′,(C′_{i′∈I′})_{i′∈I′},⊗′,(U_{ι^{-1}(i′)}○(\dtau\sj_{\ZZ})^{-1})_{i′∈I′}).\notag
\zz
\end{gather} Then (b) $Γ′$ is an \ct{NCG} game and (c) $[Γ,Γ′,ι,τ,δ,(\id_{\dU_i(\ZZ)})_{i∈I}]$ is an \ct{NCG} isomorphism. \end{lemma}

\begin{pf} Let $θ = [(T,p),(T′,p′),τ]$.  \begin{cllist}

\yl{9392} {\em (a) $\dtau\sj_{\ZZ}{:}\ZZ→\ZZ′$ is a bijection, (b) $\ZZ^θ = \ZZ$, and (c) $P′○τ(t^o) = ∅$.}  $θ$ is a tree isomorphism because of [i] the assumption that $[Φ,Φ′,ι,τ,δ]$ is an \ct{NCF} isomorphism and [ii] the ``forgetful'' functors of Figure~\rf{8307}.  Thus (a) follows from Proposition~\rf{9050}, and (b) and (c) follow from Proposition~\rf{9043}(\rf{9045}).

\yl{9395} {\em (a) $ι{:}I→I′$ is a bijection, and (b) $(∀i∈I)$ $U′_{ι(i)}○\dtau\sj_{\ZZ} = U_i$.}  (a) This follows from SF Theorem 2.4(a) and the assumption that $[Φ,Φ′,ι,τ,δ]$ is an \ct{NCF} isomorphism.  (b) The definition of $Γ′$ implies $(∀i′∈I′)$ $U′_{i′} = U_{ι^{-1}(i′)}○(\dtau\sj_{\ZZ})^{-1}$.  Thus Claim~\rf{9392}(a) implies $(∀i′∈I′)$ $U′_{i′}○\dtau\sj_{\ZZ} = U_{ι^{-1}(i′)}$.  This and part (a) imply part (b).  

\yl{9393} {\em $Γ′$ is an \ct{NCG} game.}  \rf{G1} for $Γ′$ holds because $Φ′$ is an \ct{NCF} form by assumption.  For \rf{G2} take $i′⋅∈⋅I′$.  Claim~\rf{9395}(a), and \rf{G2} for $Γ$ at $i = ι^{-1}(i′)$, imply $U_{ι^{-1}(i′)}$ is a surjective real-valued function from $\ZZ$.  Thus Claim~\rf{9392}(a) implies $U_{ι^{-1}(i′)}○(\dtau\sj_{\ZZ})^{-1}$ is a surjective real-valued function from $\ZZ′$.  Thus the definition of $Γ′$ implies $U′_{i′}$ is a surjective real-valued function from $\ZZ′$.  This is \rf{G2} for $Γ′$ at $i′$.

\yl{9394} {\em The sextuple $[Γ,Γ′,ι,τ,δ,(\id_{\dU_i(\ZZ)})_{i∈I}]$ is an \ct{NCG} morphism.}  $Γ$ is an \ct{NCG} game by assumption.  $Γ′$ is an \ct{NCG} game by Claim~\rf{9393}.  It remains to show\begin{gather}
\zz
\mqi \text{[\~{g}1]} & [Φ,Φ′,ι,τ,δ]⋅\text{is an \ct{NCF} morphism}, \mqo \nt
\mqi \text{[\~{g}2]} & (∀i∈I)⋅\id_{\dU_i(\ZZ)}\,{:}\,\dU_i(\ZZ^θ)\,→\,\dU′_{ι(i)}(\ZZ′), \mqo \nt
\mqi \text{[\~{g}3]} & (∀i∈I)⋅\id_{\dU_i(\ZZ)}⋅\text{is weakly increasing, and} \mqo \nt
\mqi \text{[\~{g}4]} & (∀i∈I,Z∈\ZZ^θ)⋅\id_{\dU_i(\ZZ)}(U_i(Z)) = \dU′_{ι(i)}(P′○τ(t^o)∪\dtau(Z)). \mqo \notag
\zz
\end{gather} [\~{g}1] is implied, a fortiori, by the assumption that $[Φ,Φ′,ι,τ,δ]$ is an \ct{NCF} isomorphism.  [\~{g}3] is obvious.  

For [\~{g}2], it suffices to show that [i] the domain of $\id_{\dU_i(\ZZ)}$, which is $\dU_i(\ZZ)$, is equal to $\dU_i(\ZZ^θ)$, and that [ii] the codomain of $\id_{\dU_i(\ZZ)}$, which is $\dU_i(\ZZ)$, is equal to $\dU′_{ι(i)}(\ZZ′)$.  [i] follows from Claim~\rf{9392}(b).  For [ii], in steps, $\dU′_{ι(i)}(\ZZ′)$ by definition equals $\dU_i○(\dtau\sj_{\ZZ})^{-1}(\ZZ′)$, which by Claim~\rf{9392}(a) equals $\dU_i(\ZZ)$.

For [\~{g}4], take $i⋅∈⋅I$ and $Z⋅∈⋅\ZZ^θ$.  Claim~\rf{9392}(b) [or $\ZZ^θ⋅⊆⋅\ZZ$] implies $Z⋅∈⋅\ZZ$, which implies [1] $U_i(Z)⋅∈⋅\dU_i(\ZZ)$.  Then, in steps, $\id_{\dU_i(\ZZ)}(U_i(Z))$ by [1] is equal to $U_i(Z)$, which by Claim~\rf{9395}(b) is equal to $U′_{ι(i)}(\dtau(Z))$, which by Claim~\rf{9392}(c) is equal to $U′_{ι(i)}(P′○τ(t^o)∪\dtau(Z))$.

\yl{9396} {\em $[Γ,Γ′,ι,τ,δ,(\id_{\dU_i(\ZZ)})_{i∈I}]$ is an \ct{NCG} isomorphism}.  This follows from Claim~\rf{9394} and Corollary~\rf{8636}(\rf{9126}$⇐$\rf{9127}) because [i] $[Φ,Φ′,ι,τ,δ]$ is an \ct{NCF} isomorphism by assumption and [ii] each $\id_{\dU_i(\ZZ)}$ is a bijection by inspection. \end{cllist}

{\em Conclusion}.  Part (a) follows from Claims \rf{9395}(a) and \rf{9392}(a).  Part (b) follows from Claim~\rf{9393}.  Part (c) follows from Claim~\rf{9396}.  \end{pf} 

\markb{\sc References}
\eput